\documentclass[manuscript,screen, bookmarksnumbered,unicode,hidelinks]{acmart}

\usepackage{appendix}

\usepackage[ruled,vlined,linesnumbered]{algorithm2e} 
\SetAlFnt{\small}       
\SetAlCapFnt{\small}    
\SetAlCapNameFnt{\small}
\LinesNumberedHidden    
\DontPrintSemicolon    
\SetAlgoSkip{0.5em}    
\usepackage{listings}
\usepackage{array}
\usepackage{graphicx}
\usepackage{placeins}
\usepackage[braket, qm]{qcircuit}
\usepackage{amsmath}
\usepackage{tikz}
\usepackage{placeins}
\usepackage{pgfplots}
\SetKw{Throw}{Throw}
\usepackage[percent]{overpic}
\usepackage{amsthm}
\usepackage{makecell}
\usepackage{pgfplots}
\pgfplotsset{compat=1.18}
\usepackage{subcaption} 
\usepackage{enumitem} 
\usepackage{braket}
\usepackage{threeparttable}

\usetikzlibrary{shapes, arrows.meta, positioning, fit, calc, backgrounds}
\usepackage{pifont} 
\tikzset{
  process/.style={rectangle, rounded corners, minimum width=2.5cm, minimum height=1cm, text centered, draw=black, fill=gray!10},
  io/.style={trapezium, trapezium left angle=70, trapezium right angle=110, minimum width=2.5cm, minimum height=1cm, text centered, draw=black, fill=blue!10},
  decision/.style={diamond, minimum width=2cm, minimum height=1cm, text centered, draw=black, fill=orange!20},
  arrow/.style={thick,->,>=stealth}
}

\newcommand{\ibmfoot}{\footnote{\url{https://quantum-computing.ibm.com/}}}

\newcommand{\helmifoot}{\footnote{\url{https://docs.csc.fi/computing/quantum-computing/overview/}}}

\newcommand{\cmark}{\ding{51}}%
\newcommand{\xmark}{\ding{55}}%
\usepackage{tcolorbox}
\usepackage{booktabs}
\usepackage{pgfplots}
\usepackage{pgfplotstable}
\usepackage{siunitx}
\usepackage{xcolor}
\pgfplotsset{compat=1.18}
\usetikzlibrary{mindmap,trees}   
\usepackage{mathtools}  

\usepackage{amssymb}
\usepackage{tcolorbox}
\definecolor{mygreen}{rgb}{0,0.6,0}
\definecolor{mygray}{rgb}{0.5,0.5,0.5}
\definecolor{mymauve}{rgb}{0.58,0,0.82}
\newtcolorbox{rqsummary}[1][]{%
  colback=gray!5!white,
  colframe=gray!75!black,
  boxrule=0.5pt,
  arc=2pt,
  width=\textwidth,
  fonttitle=\bfseries,
  before skip=10pt,
  after skip=10pt,
  float,                 
  floatplacement=htbp,   
  #1}
\lstdefinelanguage{json}{
    morestring=[b]",
    morestring=[s]{'}{'},
    morecomment=[l]{//},
    morecomment=[s]{/*}{*/},
    morekeywords={true,false,null},
    sensitive=false,
}

\AtBeginDocument{%
  }

\begin{document}
\title{\textit{C2$\ket{\rm{Q}}$}: A Robust Framework for Bridging Classical and Quantum Software Development}



\author{Boshuai Ye}
\email{boshuai.ye@oulu.fi}
\orcid{0009-0000-3480-1234}
\affiliation{%
  \institution{M3S Research Group, SEIS Unit, University of Oulu}
  \city{Oulu}
  \country{Finland}
}

\author{Arif Ali Khan*}
\email{arif.khan@oulu.fi}
\orcid{0000-0002-8479-1481}
\affiliation{%
  \institution{M3S Research Group, SEIS Unit, University of Oulu}
  \city{Oulu}
  \country{Finland}
}

\author{Teemu Pihkakoski}
\email{teemu.pihkakoski@oulu.fi}
\orcid{0009-0008-4598-7271}
\affiliation{%
  \institution{Nano and Molecular Systems Research Unit, University of Oulu}
  \city{Oulu}
  \country{Finland}
}

\author{Peng Liang}
\email{liangp@whu.edu.cn}
\orcid{0000-0002-2056-5346}
\affiliation{%
  \institution{School of Computer Science, Wuhan University}
  \city{Wuhan}
  \country{China}
}

\author{Muhammad Azeem Akbar}
\email{azeem.akbar@lut.fi}
\orcid{0000-0002-4906-6495}
\affiliation{%
  \institution{Software Engineering Department, Lappeenranta-Lahti University of Technology}
  \city{Lappeenranta}
  \country{Finland}
}

\author{Matti Silveri}
\email{matti.silveri@oulu.fi}
\orcid{0000-0002-6319-2789}
\affiliation{%
  \institution{Nano and Molecular Systems Research Unit, University of Oulu}
  \city{Oulu}
  \country{Finland}
}

\author{Lauri Malmi}
\email{lauri.malmi@aalto.fi}
\orcid{0000-0003-1064-796X}
\affiliation{%
  \institution{Department of Computer Science, Aalto University}
  \city{Espoo}
  \country{Finland}
}

\renewcommand{\shortauthors}{Ye et al.}

\begin{abstract}
Quantum Software Engineering (QSE) is emerging as a critical discipline to make quantum computing accessible to a broader developer community; however, most quantum development environments still require developers to engage with low-level details across the software stack—including problem encoding, circuit construction, algorithm configuration, hardware selection, and result interpretation—making them difficult for classical software engineers to use. To bridge this gap, we present \textit{C2$\ket{\rm{Q}}$}, a hardware-agnostic quantum software development framework that translates specific types of classical specifications into quantum-executable programs while preserving methodological rigor. The framework applies modular software engineering principles by classifying the workflow into three core modules: an \textit{encoder} that classifies problems, produces Quantum-Compatible Formats (QCFs), and constructs quantum circuits, a \textit{deployment} module that generates circuits and recommends hardware based on fidelity, runtime, and cost, and a \textit{decoder} that interprets quantum outputs into classical solutions. This architecture supports systematic evaluation across simulators and Noisy Intermediate-Scale Quantum (NISQ) quantum devices, remaining scalable to new problem classes and algorithms. 
In evaluation, the \textit{encoder} module achieved a 93.8\% completion rate, the hardware recommendation module consistently selected the appropriate quantum devices for workloads scaling up to 56 qubits. End-to-end experiments on 434 Python programs and 100 JSON problem instances demonstrate that the full \textit{C2$\ket{\rm{Q}}$} workflow executes reliably on simulators and can be deployed successfully on representative real quantum hardware, with empirical runs limited to small- and medium-sized instances consistent with current NISQ capabilities. A proxy-based usability analysis further indicates substantial reductions in handwritten lines of code and explicit configuration steps compared to conventional quantum SDK workflows. These results indicate that \textit{C2$\ket{\rm{Q}}$} lowers the entry barrier to quantum software development by providing a reproducible, extensible toolchain that connects classical specifications to quantum execution. The open-source implementation of \textit{C2$\ket{\rm{Q}}$} is available at \url{https://github.com/C2-Q/C2Q} and as a ready-to-use Python package at \url{https://pypi.org/project/c2q-framework/}.
\end{abstract}



\maketitle
\section{Introduction} \label{sec:intro}
Quantum Computing (QC) is gaining increasing global attention, offering unique opportunities across both high-tech and traditional industries. At its core, QC is powered by quantum-mechanical principles such as superposition, entanglement, and interference, which in combination allow quantum systems to represent and process many possible quantum states of information simultaneously~\cite{nielsen2010quantum}. This enables quantum computers to tackle problems that are intractable for classical methods, including integer factorisation and combinatorial optimisation tasks like the Travelling Salesman Problem (TSP). In certain cases, quantum algorithms can provide polynomial speedups, such as quadratic improvements achieved through amplitude amplification~\cite{preskill}. Building on these foundations, QC has emerged as a transformative paradigm with applications spanning cryptography, optimisation, machine learning, finance, and chemistry~\cite{shor1999polynomial, moll2018quantum, biamonte2017quantum, egger2020quantum, mcardle2020quantum}.

Over the past five years, QC has progressed from laboratory prototypes to a cloud-based ecosystem accessible to both researchers and industry.  The leading vendors now provide cloud access to superconducting or trapped-ion processors (IBM, Quantinuum, IonQ, Rigetti) via services such as IBM Quantum~\cite{ibm_quantum}, Amazon Braket~\cite{braket}, and Azure Quantum~\cite{Microsoft_Azure_Quantum_Development}. In parallel, to realise the full potential of QC, the software stack has matured rapidly. Open source software development kits (SDKs) such as \emph{Qiskit} (IBM)~\cite{javadiabhari2024quantumcomputingqiskit}, \emph{Cirq} (Google)~\cite{cirq}, and \emph{PennyLane} (Xanadu)~\cite{pennylane} provide high-level Python interfaces for circuit construction, transpilation, and hybrid quantum–classical workflows.

Despite these advances, developing practical quantum software applications remains challenging. Existing SDKs still require developers to engage with low-level quantum software programming abstractions that require specialised expertise that most classical programmers lack \cite{zhao2021quantumsoftwareengineeringlandscapes}. Surveys and empirical studies confirm that even experienced users face steep learning curves and fragmented workflows when transitioning from classical code bases to runnable quantum implementations \cite{Murillo_2025, alsalman2024quantum}. Furthermore, the diversity of quantum hardware platforms and their cloud access environments introduces significant variability, making it difficult to execute high-level problem specifications in a platform-independent manner \cite{Murillo_2025}. Collectively, these issues underline a persistent gap between classical and quantum software development.

Given these challenges, this research aims to design a framework that makes quantum software development more accessible. The framework serves as a bridge between classical and quantum computing software development: it takes problems written in classical form, prepares these problems automatically for quantum processing (encoded into a quantum-compatible format), and then translates the quantum results back into classical solutions. In this way, users can focus on solving their problems without having to deal with the complexity of quantum software development and the differences between hardware platforms. To realize this vision, this research addresses \textbf{three main tasks}:  
(i) identifying how problems expressed in classical specifications (source code snippets or structured inputs) can be recognized and transformed into representations suitable for execution on a quantum device;  
(ii) determining how these representations can guide the automated selection and configuration of appropriate quantum algorithms; and  
(iii) ensuring that the resulting workloads can be executed seamlessly on different quantum processors without requiring users to handle platform-specific details.
Therefore, to achieve the above objectives, we formulate the following research questions (RQs):

\textbf{RQ1}: How can a framework be designed to bridge the gap between classical and quantum software development?
\begin{itemize}
   \item \textbf{Rationale}: Quantum software development remains challenging and largely inaccessible to classical (traditional) software engineers, who are accustomed to formulating problems in classical programming languages or structured data formats. Addressing this gap requires a framework capable of automatically interpreting classical code inputs and generating executable quantum programs. Such a framework should abstract low-level quantum complexities, support diverse problem domains, and remain compatible with a variety of algorithmic and hardware backends. Automating this translation pipeline is critical for lowering the entry barrier and enabling the broader adoption of quantum computing in practice. This RQ can be further decomposed into two sub-questions, focusing respectively on problem encoding and quantum circuit generation.
\end{itemize}

 \textbf{RQ1.1:} What is an efficient approach for analysing classical code snippets and transforming them into quantum-compatible formats (QCFs) that can serve as input to quantum algorithms?
    \begin{itemize}
        \item \textbf{Rationale}: Classical code snippets are a natural and familiar way for developers to describe computational tasks. However, enabling their execution on quantum hardware requires both structural analysis and systematic translation. This process involves identifying the underlying problem type and reformulating it into a quantum-compatible format (QCF)—a canonical representation that can be directly mapped to quantum algorithms and circuits. Automating this analysis and transformation is essential to integrate classical specifications seamlessly into quantum workflows.
    \end{itemize}

\textbf{RQ1.2}: How can suitable quantum algorithms be automatically selected and configured based on the translated QCFs to produce executable quantum programs? 
    \begin{itemize}
     \item \textbf{Rationale}:  Selecting and configuring quantum algorithms is a central challenge in quantum software development. A single problem type may correspond to multiple algorithms, each with unique input requirements and parameters, and choosing among them currently demands expert knowledge. Without automation, this step becomes a major barrier for most software developers and limits the practical use of QC. Automating this phase, which includes algorithm selection and circuit construction, ensures correctness, consistency, and usability of the proposed framework for non-experts.
    \end{itemize}

\textbf{RQ2}: How can the framework automatically recommend and deploy quantum hardware that best matches a program’s requirements while hiding platform-specific complexity?
    \begin{itemize}
        \item \textbf{Rationale}: Quantum hardware platforms vary widely in architecture, qubit connectivity, noise characteristics, and access models. This heterogeneity presents a challenge for selecting an appropriate execution quantum device, particularly when balancing performance, cost, and reliability. Automating this selection and deployment process enhances portability across devices and reduces platform-specific complexity, enabling seamless execution across diverse quantum hardware.
    \end{itemize}
   
   \textbf{RQ3}: How effective and generalisable is the proposed framework?  

    \begin{itemize}
        
        \item \textbf{Rationale}: To establish the effectiveness and generalisability of the proposed framework, a comprehensive end-to-end evaluation is essential. We conducted this evaluation in two complementary ways: (i) simulation-based experiments, to analyse performance and scalability under controlled conditions, and (ii) execution on real quantum computers, to validate practical applicability and robustness in real-world settings. Together, these assessments provide a balanced and reliable demonstration of the framework capabilities.

    \end{itemize}

To address these RQs, this article presents the \textit{C2$\ket{\rm{Q}}$} framework, an automated and modular pipeline designed to bridge classical and quantum software development. The framework transforms high-level problem descriptions expressed in classical form into executable quantum workloads, thereby enabling seamless integration of classical problem specifications with quantum computation. 

Conceptually, the framework follows an \textbf{encoder–deployment–decoder} modular pattern: the \emph{encoder} (parser, translator, generator, transpiler) transforms classical inputs into quantum compatible format, the \emph{deployment layer} (recommender, execution) selects and executes the transformed code on suitable hardware, and the \emph{decoder} (interpreter) maps raw quantum outputs back into classical solutions. The pipeline begins with a \textit{parser} module that analyses user-supplied classical code, identifies the underlying problem, and extracts relevant structured data. The \textit{translator} sub-module then maps problem instances into \textit{Quantum-Compatible Formats} (QCFs). Next, the sub-module \textit{generator} selects and configures appropriate quantum algorithms for each QCF, producing logical quantum circuits. These are adapted to hardware constraints by the \textit{transpiler}. Next, the \textit{recommender} sub-module evaluates quantum circuits based on the problem type, expected execution time, and the cost of the required hardware. Using these criteria, it recommends the most suitable available quantum device. Finally, the \textit{execution} runs the quantum circuit on the selected backend, either a simulator or accessible real quantum hardware, and the \textit{interpreter} translates the outputs into a classical format for ease of understanding. This multi-stage design directly addresses the RQs: RQ~1 (overall framework design), RQ~1.1 (input analysis), RQ~1.2 (algorithm selection and configuration), RQ~2 (hardware deployment), and RQ~3 (system-level evaluation), collectively fulfilling the research objective.

We validated the entire pipeline or modules of the proposed framework through three complementary experiments. First, the input–analysis module is benchmarked on 434 synthetic Python code snippets. The \textit{encoder} correctly classifies problem type with a weighted-average \(F_{1}\) score of 98.2\% and extracts relevant data with a 93.8\% completion rate. Second, the hardware \textit{recommender} module is evaluated on a representative problem class, with workloads scaling up to 56 qubits, which consistently selected trapped-ion devices (\textit{Quantinuum H1}, \textit{Quantinuum H2}), reflecting the dominance of fidelity in the decision process. 
While not exercised in this experiment, the scoring function also allows superconducting devices such as \textit{IBM Sherbrooke} to become preferable when accuracy is weighted less heavily, owing to their faster runtimes and lower costs. Third, the full pipeline is executed end-to-end on all collected inputs, comprising 434 Python programs and 100 structured JSON specifications, automatically generating and running quantum jobs 
across both simulators and quantum hardware platforms natively supported by the framework, including \textit{IBM Brisbane\ibmfoot} and Finnish \textit{Helmi\helmifoot} quantum computer by VTT and IQM. Across this evaluation, the workflow achieved a 93.8\% \textit{completion rate}, and demonstrated substantial reductions in handwritten lines of code and explicit configuration steps compared to standard quantum SDK workflows, as indicated by proxy-based workload measures.

To conclude, the principal \textbf{contributions} of this work are:
\begin{enumerate}
  \item \textit{C2$\ket{\rm{Q}}$}, a novel fully automated framework that translates specific types of classical specifications into executable quantum programs for a defined set of canonical problem families, including combinatorial optimisation problems (e.g., TSP), reversible-arithmetic (e.g., addition), and number-theoretic (e.g., integer factorisation) tasks;
  \item a hardware \emph{recommender} module that balances fidelity, latency, and cost, demonstrated across superconducting and trapped-ion processors;
  \item a comprehensive evaluation across quantum simulators and real hardware platforms (e.g., \textit{IBM Brisbane} and Finland’s \textit{Helmi}), validating the framework end to end;
  \item empirical evidence of high completion rates of the workflow and measurable reductions in handwritten code volume and explicit configuration steps, demonstrating the practical usability of the full classical-to-quantum pipeline; and
  \item A public \emph{repository} containing source code, experiment scripts, and extended results, available via GitHub~\cite{C2Q_github_2025}.
\end{enumerate}
The remainder of this article is organised as follows.  
Section~\ref{sec:background} provides the essential background on quantum computing and quantum software development.  
Section~\ref{sec:relatedwork} reviews the prior work.  
Section~\ref{sec:framework} presents the design and implementation of the \textit{C2$\ket{\rm{Q}}$} framework.  
Section~\ref{sec:evaluation} reports evaluation results.  
Section~\ref{sec:discussion} reflects on key insights from the evaluation.  
Section~\ref{sec:implications} discusses the implications of our work.  
Section~\ref{sec:validity} describes the threats to the validity of this study.  
Section~\ref{sec:conclusions} concludes this work with future work directions.

\section{Background} \label{sec:background}
We provide the background of this study by outlining the core concepts of quantum computing (Section~\ref{qc}), the representative quantum algorithms employed in the proposed framework together with their associated problem domains (Section~\ref{sec:quantum_algorithms}), and the foundations of quantum software engineering (Section~\ref{sec:qse}).

\subsection{Quantum Computing}\label{qc}

Quantum computing (QC) is a computational paradigm that leverages quantum mechanical phenomena to solve certain types of problems significantly faster than classical computing~\cite{boixo, preskill, nasem, nielsen2010quantum}. Classical computers represent information using bits that exist in one of two states: 0 or 1. In contrast, quantum computers encode information in quantum bits (\textit{qubits}), which can exist in a superposition of both 0 and 1 simultaneously. This property, known as \textit{superposition}, enables qubits to represent and process multiple states simultaneously~\cite {preskill, nasem}. A single qubit can be represented as:
\begin{equation}
   |\psi\rangle = \alpha|0\rangle + \beta|1\rangle, 
\end{equation}   
where $\alpha$ and $\beta$ are complex numbers, satisfying $|\alpha|^2 + |\beta|^2 = 1$. These amplitudes represent the probabilities of observing the qubit in the $|0\rangle$ or $|1\rangle$ state upon measurement~\cite{nielsen2010quantum, preskill}. An $n$-qubit system exists in a superposition of $2^n$ basis states:
\begin{equation}
    |\psi\rangle = \sum_{j=0}^{2^n - 1} \alpha_j |j\rangle,
\end{equation}
where each $|j\rangle$ corresponds to an $n$-bit binary string (e.g., $|000\rangle$, $|101\rangle$), and $\alpha_j$ are complex amplitudes such that $\sum_j |\alpha_j|^2 = 1$. These states are constructed using tensor products of individual qubit states:
\begin{equation}
    |j\rangle = |j_1\rangle \otimes |j_2\rangle \otimes \dots \otimes |j_n\rangle.
\end{equation}
As the number of qubits increases, the number of states that a quantum system can represent grows exponentially. An $n$-qubit register can encode $2^n$ basis states simultaneously, a capacity far beyond what is feasible in classical systems. 
This exponential scaling together with linearity of quantum operations, and interference provide the foundation for the potential computational advantage of quantum computing over classical approaches in certain tasks~\cite{nielsen2010quantum, preskill}.

Another important property in quantum computing is \textit{entanglement}, a phenomenon in which two or more qubits become correlated in such a way that the state of each qubit cannot be described independently~\cite{nielsen2010quantum}. Once entangled, the measurement outcome of one qubit is immediately related to the outcome of the others, regardless of spatial separation~\cite{preskill}. A common example of entanglement is the \textit{Bell state}, which is a specific two-qubit quantum state exhibiting perfect correlation. One of the four canonical Bell states is:
\begin{equation}
    |\Phi^+\rangle = \frac{1}{\sqrt{2}}(|00\rangle + |11\rangle).
\end{equation}
In this state, the qubits are equally likely to be measured as $|00\rangle$ or $|11\rangle$, but never as $|01\rangle$ or $|10\rangle$. The outcome of measuring one qubit immediately determines the result of the other. Bell states are widely used in quantum communication and serve as a minimal example of maximal entanglement~\cite{nielsen2010quantum}. 

Entanglement and superposition are generated and controlled through quantum logic gates. In QC, the quantum logic gates operate on qubits to implement quantum operations and drive the evolution of quantum states. These gates are analogous to classical logic gates (e.g., AND, OR, NOT)~\cite{nielsen2010quantum}, but they act on quantum states and can produce effects unique to quantum systems, such as superposition and entanglement. Quantum gates must be \textit{unitary}—a mathematical property that guarantees probability conservation and reversibility of operations~\cite{preskill}. These gates enable the creation and manipulation of quantum phenomena such as superposition and entanglement, which are essential for quantum computation~\cite{nielsen2010quantum, preskill, Barenco_1995}. Common examples of quantum gates include:

\begin{itemize}
    \item \textbf{Hadamard (H)}: Creates an equal superposition of the basis states \( |0\rangle \) and \( |1\rangle \).
    \item \textbf{Pauli gates} $(X, Y, Z)$: Apply discrete quantum transformations. 
    \begin{itemize}
        \item $X$ performs a bit flip, transforming \( |0\rangle \leftrightarrow |1\rangle \).
        \item $Z$ performs a phase flip, changing the sign of the \( |1\rangle \) component: \( |0\rangle \rightarrow |0\rangle, |1\rangle \rightarrow -|1\rangle \).
        \item $Y$ combines a bit flip and a phase flip; it maps $|0\rangle$ to $i|1\rangle$ and $|1\rangle$ to $-i|0\rangle$.

    \end{itemize}
    
    \item \textbf{Control gates}: Multi-qubit gates that apply an operation to a \emph{target} qubit depending on the state of a \emph{control} qubit.

    \begin{itemize}
        \item CNOT (Controlled-NOT) applies an $X$ gate to the target qubit if the control qubit is in state \( |1\rangle \); commonly used to generate entanglement.
        \item CZ (Controlled-Z) applies a $Z$ gate to the target qubit when the control qubit is \( |1\rangle \); useful in introducing conditional phase shifts.
        \item Other controlled gates (e.g., \texttt{CRX}, \texttt{CRZ}, \texttt{CU3}) operate similarly by conditionally applying specific single-qubit operations based on the control qubit's state.
    \end{itemize}
\end{itemize}


By composing quantum gates in sequence, developers build \textit{quantum circuits}, the core abstraction for programming gate-based quantum computers. Quantum circuits typically start with initialized qubits, apply unitary gate operations, and end with measurements that produce classical outcomes~\cite{nielsen2010quantum, javadiabhari2024quantumcomputingqiskit}. While logical circuits are hardware-agnostic, they must be adapted to device constraints through \textit{transpilation}, which maps qubits and gates to hardware-native forms and applies basic optimisations~\cite{cross2015quantum}. This step, handled automatically by SDKs such as Qiskit~\cite{javadiabhari2024quantumcomputingqiskit}, Cirq~\cite{cirq}, and PennyLane~\cite{pennylane}, is critical for efficient use of today’s NISQ devices. Figure~\ref{fig:circuit_demo} illustrates a basic quantum circuit that prepares a Bell state using a Hadamard and a CNOT gate, followed by measurement:

\begin{figure}[h]
    \centering
    \[
    \Qcircuit @C=1em @R=1em {
        \lstick{q_0\  |0\rangle} & \gate{H} & \ctrl{1} & \meter \cwx[2] & \qw  \\
        \lstick{q_1 \  |0\rangle} & \qw      & \targ    & \qw    & \qw \\
        \lstick{c}   & \cw      & \cw      & \cw    & \cw
    }
    \]
    \caption{A quantum circuit that prepares a Bell state.}
    \label{fig:circuit_demo}
\end{figure}

\subsection{Quantum Algorithms} \label{sec:quantum_algorithms}
\textit{Quantum algorithms} are computational procedures that leverage quantum mechanical principles to solve problems that are classically hard ~\cite{farhi2014, peruzzo2014}. Quantum algorithms are typically represented as circuits consisting of sequences of quantum gates applied to qubits, followed by measurements. The measurement process projects the quantum state onto a basis state, and the outcome is recorded as a classical bit.
Well-known examples include Shor's algorithm~\cite{shor1999polynomial} for integer factorization, Grover's algorithm~\cite{grover1996} for unstructured search, and the Deutsch–Jozsa algorithm~\cite{deutsch1992rapid} for function property testing. Quantum algorithms have found applications in diverse domains, including optimization~\cite{lucas2014}, finance~\cite{egger2020quantum}, chemistry~\cite{mcardle2020quantum}, and cryptography~\cite{shor1999polynomial}. This section introduces the core algorithms used in the proposed framework, focusing on software-relevant abstractions, encoding strategies, and modular integration.

Quantum algorithms are not universally applicable but tend to be effective for a limited set of problem domains where they offer potential computational advantages over classical approaches~\cite{preskill, lucas2014, shor1999polynomial}. In this work, we focus on three representative domain problem categories that align with the objectives of the proposed framework. First, \textit{combinatorial optimisation problems}~\cite{lucas2014}, such as Maximum Independent Set, MaxCut, and the Traveling Salesman Problem (TSP). Second, \textit{arithmetic operations}, particularly in the form of quantum arithmetic~\cite{draper2000additionquantumcomputer, Ruiz_Perez_2017}, such as addition, multiplication, and related operations. Third, \textit{number-theoretic tasks}~\cite{shor1999polynomial}, including integer factorization. Many problems in the first category belong to the NP-complete class~\cite{garey1979computers}, meaning they are computationally hard and no polynomial-time exact algorithms are currently known. These targeted domains represent typical scenarios where quantum algorithms can potentially offer performance improvements over classical approaches.

In this work, the proposed framework targets representative problems from each of these categories and integrates algorithms aligned with the \textit{Quantum-Compatible Formats (QCFs)} they operate on. A QCF is a standardized problem representation that allows seamless mapping to quantum circuits 
(see Appendix~\ref{app:qcf_data}). Once defined, a QCF can be directly instantiated as a logical circuit for the corresponding algorithm. Quantum Approximate Optimization Algorithm (QAOA)~\cite{farhi2014} and Variational Quantum Eigensolver (VQE)~\cite{peruzzo2014} are employed for finding an eigenstate of a problem matrix (for example, QUBO-based optimisation formulations or physics\&chemistry based Hamiltonias), Grover’s search~\cite{grover1996} for oracle-based encodings, and QFT-based arithmetic~\cite{draper2000additionquantumcomputer} for structured numerical operations. These algorithms were selected for their practical applicability and theoretical maturity, ensuring they can address a broad spectrum of problem instances within the targeted domains. Within the framework, they serve as automated, modular subroutines for solving classical problems expressed in QCFs, enabling broad coverage across the selected problem categories. The selected algorithms are defined below, along with their relevance to the framework:

\begin{itemize}
    \item \textbf{Quantum Approximate Optimization Algorithm (QAOA)} \\
    QAOA is a hybrid quantum-classical algorithm designed to solve combinatorial optimization problems. It encodes the objective (the function to be minimized or maximized) into a \textit{cost Hamiltonian} \( H_C \), a quantum operator whose lowest energy state corresponds to the optimal solution. These problems are typically expressed in \textit{Quadratic Unconstrained Binary Optimization (QUBO)} format, which is widely used in classical optimization and can be transformed into \( H_C \) through standard mappings (see Appendix~\ref{subapp:qubo})~\cite{lucas2014}. The algorithm begins by initializing the quantum system in a uniform superposition \( |+ \rangle^{\otimes n} \) and applies alternating sequences of the cost Hamiltonian \( H_C \) and a \textit{mixer Hamiltonian} \( H_M \), parameterized by angle vectors \( \vec{\gamma} \) and \( \vec{\beta} \), with \( \gamma_k \) and \( \beta_k \) denoting the cost and mixer angles at each layer, respectively. This yields the variational state \( |\psi(\vec{\gamma}, \vec{\beta})\rangle \). A classical optimizer iteratively adjusts the parameters to minimize the expected energy:
    \begin{equation}
    E(\vec{\gamma}, \vec{\beta}) = \langle \psi(\vec{\gamma}, \vec{\beta}) | H_C | \psi(\vec{\gamma}, \vec{\beta}) \rangle.
    \end{equation}
    This variational structure allows QAOA to support a broad class of problems that can be encoded into QUBO form~\cite{farhi2014}. In the proposed \textit{C2$\ket{\rm{Q}}$} framework, QAOA is used to solve combinatorial optimisation problems such as Maximum Independent Set (MIS), once the problem is transformed into a QUBO-based QCF. It is automatically selected and configured by the \textit{generator} module based on the problem tag (See Section~\ref{subsec:gen}).

    \item \textbf{Variational Quantum Eigensolver (VQE)} \\
    VQE is a hybrid quantum-classical algorithm originally developed for estimating the ground-state energy of quantum systems~\cite{peruzzo2014}. It constructs a parameterized quantum state \(|\psi(\theta)\rangle\) using an \textit{ansatz circuit}—a parameterized quantum circuit designed to generate trial states that approximate the ground state of a given Hamiltonian~\cite{McClean_2016}. The algorithm evaluates the expectation value of the Hamiltonian \( H \), and a classical optimizer iteratively updates the parameters \( \theta \) to minimize the expected energy:
    \begin{equation}
       E(\theta) = \langle \psi(\theta) | H | \psi(\theta) \rangle.
    \end{equation}
    Although widely adopted in quantum chemistry~\cite{mcardle2020quantum}, VQE can also be adapted for optimization tasks by encoding the objective into a Hamiltonian—analogous to QAOA~\cite{farhi2014}—making it a flexible and extensible component within the broader class of variational quantum algorithms.

    In the proposed \textit{C2$\ket{\rm{Q}}$} framework, VQE serves as an alternative to QAOA for solving QUBO-encoded problems. The \textit{generator} submodule selects VQE when this preference is explicitly specified in the Quantum-Compatible Format (QCF), either by the user or via configuration settings (see Section~\ref{subsec:gen}).

    \item \textbf{Grover's Algorithm} \\
    Grover's algorithm is a quantum search algorithm that achieves a quadratic speedup over classical methods for unstructured search problems~\cite{grover1996}. It is particularly effective for decision tasks where candidate solutions can be efficiently verified but lack inherent structure~\cite{nielsen2010quantum}. This includes problems such as constraint satisfaction, Boolean formula evaluation, and decision variants of combinatorial problems—all of which are relevant targets within the proposed framework. The algorithm operates by preparing a uniform superposition and repeatedly applying the \textit{Grover operator}, which amplifies the amplitudes of marked solutions. The Grover operator \( G \) is defined as:
    \begin{equation}
        G = H^{\otimes n} \cdot (2|0\rangle\langle 0| - I) \cdot H^{\otimes n} \cdot O_f
    \end{equation}
    Here, \( H^{\otimes n} \) denotes Hadamard gates applied to all qubits, \( (2|0\rangle\langle 0| - I) \) reflects about the initial state, and \( O_f \) is a problem-specific oracle that flips the phase of marked states~\cite{grover1996, nielsen2010quantum}. Additional details and formal oracle definitions are provided in Appendix~\ref{subapp:oracle}.

    Grover’s algorithm is employed in the proposed \textit{C2$\ket{\rm{Q}}$} framework to solve oracle-based problems, including decision variants of combinatorial optimisation and factorisation. These problems are reduced to oracle queries, enabling efficient search over marked solutions. The \textit{parser} submodule detects Grover-compatible tasks and triggers automatic integration via the \textit{generator} submodule.


    \item \textbf{Quantum Fourier Transform (QFT)–based arithmetic algorithms} \\
    The QFT (quantum analogue of the Discrete Fourier Transform) is widely used as a subroutine in phase estimation~\cite{nielsen2010quantum} and Shor’s factoring~\cite{kitaev1995quantum,nielsen2010quantum,shor1999polynomial}. 
    In this work, we employ QFT-based constructions to implement arithmetic on basis-encoded integers: Draper’s Fourier-basis adder using controlled phase rotations~\cite{draper2000additionquantumcomputer} and a multiplier formed by composing QFT additions~\cite{Ruiz_Perez_2017}. 
    Subtraction is realized as two’s-complement addition of the right operand.

    The QCF records (i) operand bit-widths ($w_A$, $w_B$; optional product width $w_C$), (ii) numeric interpretation (unsigned or two’s-complement; optional fixed-point scale $2^{-f}$), and (iii) operation semantics on computational-basis encodings:
\[
\textsf{Add}:\; |a\rangle\,|b\rangle \;\mapsto\; |a\rangle\,\big|(b{+}a)\bmod 2^{w_B}\big\rangle,\qquad
\textsf{Sub}:\; |a\rangle\,|b\rangle \;\mapsto\; |a\rangle\,\big|(b{-}a)\bmod 2^{w_B}\big\rangle,
\]
\[
\textsf{Mul}:\; |a\rangle\,|b\rangle\,|0\rangle \;\mapsto\; |a\rangle\,|b\rangle\,\big|(a\!\cdot\! b)\bmod 2^{w_C}\big\rangle.
\]
Temporary qubits (if any) are reset to $|0\rangle$ at the end, and very small controlled-phase rotations may be truncated to meet hardware constraints. These QFT-based arithmetic routines are encapsulated as reusable circuit blocks within the \textit{generator} and instantiated automatically when the input QCF specifies an arithmetic task.

\end{itemize}

\subsection{Quantum Software Engineering} \label{sec:qse}
Quantum software refers not only to a set of executable quantum programs but also to the supporting libraries, documentation, and other artifacts necessary for the development, operation, and maintenance of these programs \cite{zhao2021quantumsoftwareengineeringlandscapes}. 
\textit{Quantum Software Engineering (QSE)} refers to a set of methodologies and tools used for the design, development, testing, and maintenance of quantum software systems ~\cite{zhao2021quantumsoftwareengineeringlandscapes, Akbar}. Like classical software engineering, QSE follows a structured life cycle, including requirements analysis, design, implementation, testing, and maintenance~\cite{zhao2021quantumsoftwareengineeringlandscapes}. However, the unique properties of quantum software makes it difficult to directly apply classical software engineering principles. Murillo et al.’s roadmap highlights that QSE must evolve beyond classical practices to address unique quantum realities—such as entanglement, noise, and hybrid execution workflows—and calls for new life-cycle models, debugging methodologies, and developer tools tailored to quantum environments \cite{Murillo_2025}. As quantum hardware continues to evolve, QSE plays a critical role in harnessing the full potential of quantum computing by enabling scalable, reliable, and efficient quantum software development on real quantum computers~\cite{zhao2021quantumsoftwareengineeringlandscapes, khan2023softwarearchitecturequantumcomputing}.

Despite rapid advancements in quantum hardware and algorithm design, \textit{Quantum Software Engineering (QSE)} remains a nascent field with critical challenges. Prior surveys highlight that existing approaches provide limited support for quantum software development workflows, including high-level modeling constructs, hybrid classical–quantum integration, and scalable software architectures~\cite{zhao2021quantumsoftwareengineeringlandscapes, Murillo_2025, khan2023softwarearchitecturequantumcomputing}. Toolchains are still fragmented, with a lack of standardized abstractions, reusable patterns, and mature infrastructure to support portability across heterogeneous hardware backends~\cite{Murillo_2025, Akbar}. Moreover, quantum software developers face a steep entry barrier due to immature programming environments and non-standardized automation support~\cite{zhao2021quantumsoftwareengineeringlandscapes, Murillo_2025}. These limitations underline the need for frameworks that explicitly target developer productivity by providing modular abstractions, cross-platform compatibility, and automated workflow support.


\subsubsection{Hardware Diversity and Deployment Challenges} \label{sec:platform_diff}
In addition to the QSE limitations, practical deployment introduces further challenges due to hardware heterogeneity. Current quantum devices, typically accessed through cloud-based services, differ significantly in architecture, supported intermediate representations (IRs), qubit connectivity, and performance metrics~\cite{Murillo_2025}. As a result, even when software abstractions are available, developers often need platform-specific optimisation or manual adaptation to execute programs efficiently on different devices, which undermines portability and maintainability.

Regarding hardware heterogeneity, architectures vary across superconducting circuits and trapped-ion processors~\cite{nielsen2010quantum, Linke_2017, Kjaergaard_2020, Bruzewicz_2019, preskill}, leading to differences in qubit connectivity and native gate sets, which directly affect the fidelity of compiled circuits. Scalability dictates the size of problems that can be encoded, with some large-scale systems, such as IBM's 127-qubit Eagle processor~\cite{Jurcevic2021, ibmroadmap}, pushing current hardware limits. Performance metrics such as gate fidelity, relaxation time ($T_1$), dephasing time ($T_2$), and overall quantum volume (QV)~\cite{Cross2019} are critical for device reliability and for ensuring correct execution on NISQ hardware. Cost and accessibility are also non-trivial concerns, as cloud-based services impose fees, queueing delays, calibration downtime, and variability in execution latencies~\cite{ibm_quantum, braket} and performance metrics. Consequently, selecting a quantum device for task execution is a complex process that must take into account both the circuit structure and hardware specifications. This consideration motivates the use of automated device recommendation modules, such as the recommender integrated into our \textit{C2$\ket{\rm{Q}}$} framework. We therefore treat any limitations caused by hardware heterogeneity, such as error rates and costs, as external constraints rather than limitations of the \textit{C2$\ket{\rm{Q}}$} architecture.

\subsubsection{Quantum Software Development and Its Challenges}

\textit{Quantum software development} refers to the process of designing and implementing executable programs that run on quantum hardware devices. Unlike classical software development, which is grounded in deterministic execution models and relies on well-established abstractions such as variables, control structures, and functions, quantum software development must account for quantum-mechanical principles. These include superposition, entanglement, projective measurement, and constraints such as the no-cloning theorem, which fundamentally alter how information is represented, manipulated, and observed
~\cite{zhao2021quantumsoftwareengineeringlandscapes, khan2023softwarearchitecturequantumcomputing}. Programs are typically expressed as quantum circuits composed of gate operations acting on qubits, and are developed using quantum software programming languages~\cite{selinger2004towards, heim2020quantum} or Software Development Kits (SDKs) such as Qiskit~\cite{javadiabhari2024quantumcomputingqiskit}, Cirq~\cite{cirq}, and PennyLane~\cite{pennylane}.

Quantum software development is still at an early stage and faces several critical challenges. 
As Zhao~\cite{zhao2021quantumsoftwareengineeringlandscapes} observes, there is currently no widely adopted and efficient quantum programming language, nor consensus on standard abstractions for low-level constructs such as parameterized quantum circuits, quantum gates, and quantum data structures. 
The absence of such standardization constrains both the portability and the reusability of quantum programs across heterogeneous platforms, a limitation emphasized in prior surveys as well
~\cite{destefano2022softwareengineeringquantumprogramming, di2024abstraction}. In addition, prior studies~\cite{alsalman2024quantum, zhao2021quantumsoftwareengineeringlandscapes, khan2023softwarearchitecturequantumcomputing, Murillo_2025} emphasize the shortage of skilled quantum developers and the need for engineering methodologies specifically tailored to the hybrid classical–quantum software programming paradigm. Standardized frameworks that abstract and encapsulate low-level quantum details are essential to bridge the gap between classical and quantum development, lowering the entry barrier for software engineers transitioning to quantum computing. Such frameworks should provide high-level abstractions, automate low-level component construction, enable seamless integration with classical software environments, support cross-platform execution and workflow automation, and offer developer-friendly APIs with built-in result interpretation and performance feedback.

These limitations motivate the development of the proposed \textit{C2$\ket{\rm{Q}}$} framework, a hybrid classical–quantum software framework designed to bridge the gap between classical problem definitions and executable quantum programs. \textit{C2$\ket{\rm{Q}}$} automates the transformation of high-level classical code or structured data into executable quantum workloads, while abstracting away platform-specific details. Its modular architecture enables problem parsing, quantum-compatible encoding, algorithm selection, hardware recommendation, and execution in a unified workflow, thereby lowering the barrier for classical developers and improving the portability, scalability, and maintainability of quantum software.

\section{Related Work} \label{sec:relatedwork}
This section reviews tools and frameworks that support quantum software development, including SDKs, quantum programming languages, and cloud platforms, and compares their capabilities to highlight key contributions that motivate the design of the proposed \textit{C2$\ket{\rm{Q}}$} framework.

\subsection{Existing Studies}
Despite promising advances in quantum hardware, the parallel development of software frameworks aimed at providing practical tools for \textit{Quantum Software Engineering} (QSE) has lagged behind~\cite{zhao2021quantumsoftwareengineeringlandscapes}. In response, major quantum technology companies and research groups have released software development kits (SDKs) and platforms that support circuit construction, optimisation, transpilation, and simulation~\cite{javadiabhari2024quantumcomputingqiskit, cross2022openqasm}. These toolkits aim to bridge the gap between high-level quantum algorithms and hardware execution by providing structured workflows for design, development, debugging, testing, deployment, and maintenance of quantum programs, thereby promoting quantum software engineering practices~\cite{khan2023softwarearchitecturequantumcomputing}. 

To understand the current state-of-the-art in quantum software development, we have presented an overview of major tools—both \textit{software development kits (SDKs)} and \textit{quantum programming languages}, as well as \textit{cloud platforms}—that support quantum software development. Table~\ref{tab:quantum-sdks} and Table~\ref{tab:quantum-cloud-hardware} summarise these development platforms and tools. Each row in Table~\ref{tab:quantum-sdks} is described by: (i)~the SDK or language name; (ii)~its type, distinguishing general-purpose SDKs from domain-specific quantum programming languages (QPLs); and (iii)~the primary implementation language, which determines its integration with classical development workflows. Table~\ref{tab:quantum-cloud-hardware} presents cloud-based quantum platforms, specifying: (i)~the platform name; (ii)~supported SDKs or languages for programming; and (iii)~the hardware providers and technology available through the platform.

Among the SDKs listed, \textit{Qiskit}, developed by IBM, provides a full-stack quantum software development environment, covering circuit design, simulation, and execution on IBM’s superconducting hardware. It offers a high-level abstraction layer with constructs such as quantum algorithms, circuits,  operators, and simulators, making it accessible to developers ~\cite{javadiabhari2024quantumcomputingqiskit}. In contrast, \textit{Cirq}, developed by Google, focuses on fine-grained, low-level control over quantum circuits, making it well-suited for research on Google’s Sycamore processors, with features such as custom gate definitions, calibration routines, and noise modeling, albeit with limited hardware diversity~\cite{cirq}. Complementing these circuit-oriented SDKs, \textit{PennyLane}, by Xanadu, adopts a hardware-agnostic, machine-learning–oriented design, integrating seamlessly with libraries such as PyTorch and TensorFlow, supporting variational quantum algorithms through differentiable programming, and enabling execution on multiple quantum hardware devices via a plugin system~\cite{pennylane}. Moreover, the \textit{Intel\textsuperscript{\textregistered} Quantum SDK} offers a C++ based environment with a low-level virtual machine (LLVM) compiler toolchain for hybrid quantum–classical programming, providing familiar syntax for C++ developers, built-in support for variational parameters, and high-performance quantum simulation capabilities~\cite{Khalate2022, intel_overview}. Finally, \textit{tket}~\cite{sivarajah2020t} is a platform-agnostic SDK and compiler framework that provides a Python interface backed by a high-performance C++ core. It is designed for advanced circuit transpilation and optimisation, and supports a wide range of quantum hardware through modular backend interfaces, making it a versatile choice for cross-platform quantum program deployment. More recently, Quantinuum has introduced \textit{Guppy}, a domain-specific quantum programming language embedded in Python, designed to express high-level hybrid quantum programs with measurement-dependent control flow~\cite{koch2025guppy}. Unlike \textit{tket}, which focuses on circuit-level compilation and optimisation, \textit{Guppy} targets algorithm-level program structure, supporting hybrid quantum-classical programs with measurement-dependent control flow.

Quantum programming languages (QPLs) operate at a higher level of abstraction than SDKs, enabling developers to describe algorithms in a more declarative form and often without directly specifying individual gate operations \cite{heim2020quantum, di2024abstraction}. \textit{Ket}~\cite{da2021ket} is an open-source, embedded quantum programming language supported by the Ket Quantum Programming Platform; it leverages the simplicity of Python syntax, integrates with quantum simulators, and allows execution on IBM Quantum devices. Microsoft’s \textit{Q\#}~\cite{Svore_2018} is a domain-specific, strongly typed language for quantum software programming, where the type system enforces strict separation between quantum and classical data (e.g., \texttt{Qubit}, \texttt{Result}, \texttt{Int}), reducing programming errors. It is designed to describe quantum subroutines abstractly and integrate seamlessly with Azure Quantum; it includes extensive standard libraries for algorithms, simulation, and resource estimation. \textit{Qrisp}~\cite{seidel2024qrispframeworkcompilablehighlevel} offers a high-level, Pythonic programming model in which developers manipulate \textit{QuantumVariables} instead of manually constructing gate sequences, thereby automating circuit bookkeeping while compiling to efficient low-level implementations—although a solid grounding in quantum mechanics and linear algebra remains essential. \textit{Quipper}~\cite{green2013introduction} is an embedded functional language in Haskell aimed at scalable, modular descriptions of quantum circuits, supporting features such as circuit families, hierarchical design, and resource estimation for very large-scale algorithms. \textit{Scaffold}~\cite{abhari2012scaffold} extends C/C++ like syntax to describe quantum operations alongside classical control, enabling modular, hybrid quantum–classical program design with support for compilation to hardware-specific formats. Finally, \textit{LIQUi\textbar\textgreater}~\cite{wecker2014liqui}, developed by Microsoft Research, is an F\#-based architecture and simulator framework that provides a rich library of quantum algorithms, supports error correction simulation, and enables detailed performance and resource analysis.

On the platform side, cloud-based services integrate programming interfaces with quantum hardware execution. \textit{IBM Quantum}~\cite{ibm_quantum} provides direct access to IBM’s superconducting hardware with device-aware transpilation and integrated development through Qiskit. \textit{IonQ Cloud}~\cite{ionq_devices} supports a broad range of SDKs including Qiskit, Cirq, tket, Q\#, PennyLane, and ProjectQ, targeting IonQ’s trapped-ion systems. \textit{Amazon Braket}~\cite{braket} offers multi-vendor access, supporting IonQ, Rigetti, OQC, and QuEra hardware within a unified job model. \textit{Azure Quantum}~\cite{Microsoft_Azure_Quantum_Development} provides integrated access to IonQ, Quantinuum, Rigetti, and Pasqal devices, with tight coupling to \textit{Q\#}~\cite{Svore_2018}. Other examples include \textit{Rigetti QCS}~\cite{rigetti_qcs} for superconducting hardware via Quil/pyQuil~\cite{smith2016practical}, \textit{D-Wave Leap}~\cite{dwave_leap} for quantum annealing, and \textit{Xanadu Cloud}~\cite{Killoran_2019} for photonic quantum computing.

\begin{table}[h!]
\centering
\caption{Representative quantum SDKs and programming languages.}
\label{tab:quantum-sdks}
\begin{tabular}{p{3.5cm} p{3.5cm} p{5.5cm}}
\toprule
\textbf{SDK / Language} & \textbf{Type} & \textbf{Primary Language} \\
\midrule
Qiskit~\cite{javadiabhari2024quantumcomputingqiskit} & SDK & Python \\
Cirq~\cite{cirq} & SDK & Python \\
PennyLane~\cite{pennylane} & SDK & Python \\
Q\#~\cite{Svore_2018} & QPL & Q\# (domain-specific) \\
pyQuil~\cite{smith2016practical} & SDK & Python \\
Qrisp~\cite{seidel2024qrispframeworkcompilablehighlevel} & QPL & Python \\
Strawberry Fields~\cite{Killoran_2019} & SDK/QPL & Python \\
Intel Quantum SDK~\cite{intel_overview} & SDK & C++ \\
Quipper~\cite{green2013introduction} & QPL & Haskell \\
Scaffold~\cite{abhari2012scaffold} & QPL & C/C++-like \\
LIQUi$\rangle$~\cite{wecker2014liqui} & QPL & F\# \\
tket~\cite{sivarajah2020t} & SDK/Compiler & Python/C++ bindings \\
Guppy~\cite{koch2025guppy} & QPL & Python \\
\bottomrule
\end{tabular}
\end{table}

\begin{table}[h!]
\centering
\caption{Representative quantum cloud platforms and their hardware providers.}
\label{tab:quantum-cloud-hardware}
\begin{tabular}{p{3.5cm} p{3.5cm} p{5.5cm}}
\toprule
\textbf{Platform} & \textbf{Supported SDKs / Languages} & \textbf{Hardware Providers / Technology} \\
\midrule
IBM Quantum~\cite{ibm_quantum} & Qiskit (primary), bridges & IBM (superconducting qubits) \\
IonQ Cloud~\cite{ionq_devices} & Qiskit, Cirq, tket, Q\#, PennyLane, ProjectQ & IonQ (trapped-ion qubits) \\
Amazon Braket~\cite{braket} & Braket SDK, Qiskit, PennyLane, Cirq & IonQ (trapped-ion), Rigetti (superconducting), OQC (superconducting), QuEra (neutral atoms) \\
Azure Quantum~\cite{Microsoft_Azure_Quantum_Development} & Q\#, Qiskit, Cirq & IonQ (trapped-ion), Quantinuum (trapped-ion), Rigetti (superconducting), Pasqal (neutral atoms) \\
Rigetti QCS~\cite{rigetti_qcs} & Quil/pyQuil & Rigetti (superconducting qubits) \\
D-Wave Leap~\cite{dwave_leap} & Ocean SDK & D-Wave (quantum annealing) \\
Xanadu Cloud~\cite{Killoran_2019} & Strawberry Fields, PennyLane & Xanadu (photonic qubits) \\
\bottomrule
\end{tabular}
\end{table}

\subsection{Comparative Analysis}
Building on the related work review in the preceding sections, this subsection compares selected tools— including SDKs, quantum programming languages, and cloud platforms with the proposed \textit{C2$\ket{\rm{Q}}$} framework, highlighting differences in capabilities and limitations (see Table~\ref{tab:tool_comparison}). To ensure clarity and focus, we selected a representative subset of tools from each category based on their popularity, maturity, and inclusion of functionalities that are broadly characteristic of their respective tool classes. During this review, we found that while these tools provide essential functionalities for quantum software development, however, a significant methodological gap remains between the practices familiar to classical software developers and the specialized requirements of current quantum software development environments.

Across SDKs and quantum programming languages, even though multiple layers of abstraction are provided, the classical-to-quantum encoding step still relies heavily on the developer’s expertise in quantum software programming and quantum mechanics. Developers remain responsible for formulating problem-specific quantum logic, translating it into executable circuits, selecting appropriate algorithms, and adapting implementations to the constraints of available hardware~\cite{destefano2022softwareengineeringquantumprogramming, zhao2021quantumsoftwareengineeringlandscapes}. While higher-level languages can automate parts of circuit construction, they do not eliminate the need for substantial knowledge of quantum mechanics, algorithm design, and hardware characteristics since problem encoding, algorithm parameterization, and hardware-aware circuit design remain developer-dependent tasks~\cite{zhao2021quantumsoftwareengineeringlandscapes, destefano2022softwareengineeringquantumprogramming}. Furthermore, the post-execution stage is typically under-supported: quantum programs often return raw measurement findings—such as probability distributions over bitstrings—leaving it to the user to interpret these results, extract valid solutions, and map them back to the original problem domain~\cite{Murillo_2025, destefano2022softwareengineeringquantumprogramming}.

Cloud platforms such as \textit{Amazon Braket}~\cite{braket} and \textit{Azure Quantum}~\cite{Microsoft_Azure_Quantum_Development} integrate selected SDKs and quantum programming languages with backend execution, providing unified access to heterogeneous hardware—including superconducting, trapped-ion, photonic, and neutral-atom devices. While this integration increases flexibility and enables cross-hardware development, it does not remove the complexity of matching algorithms to appropriate hardware or the challenge of interpreting quantum outputs in a domain-specific context. Selecting an optimal device remains non-trivial due to substantial variations in architecture, qubit count, connectivity, error rates, and suitability for specific problem types (see Section~\ref{sec:platform_diff}).

As synthesised in Table~\ref{tab:tool_comparison}, these tools—despite differing in scope and abstraction—share three persistent challenges:
\begin{enumerate}
  \item \emph{Knowledge gap in problem encoding:} Translating high-level problem descriptions into executable quantum circuits still relies heavily on manual, expert-driven design.
  \item \emph{Lack of hardware recommendation:} While some platforms offer cross-hardware execution, selecting the most suitable device for a given problem remains the user’s responsibility.
  \item \emph{Insufficient result interpretability:} Existing SDKs provide raw measurement data but lack interfaces for automated post-processing, leaving developers to manually map outputs into classical solution formats.
\end{enumerate}

\begin{table}[h!]
\centering
\caption{Comparison of selected SDKs, QPLs, and cloud platforms by functionality. Abbreviations: PE = Problem Encoding, GC = Gates \& Circuits, QA = Quantum Algorithms, AS = Algorithm Selection, CP = Cross-Platform, HR = Hardware Recommendation, RI = Result Interpretation, SIM = Simulator Support. A checkmark (\cmark) indicates that the functionality is supported, while a cross (\xmark) indicates it is not.}
\label{tab:tool_comparison}
\small
\begin{tabular}{lcccccccc}
\toprule
\textbf{Tool} & \textbf{PE} & \textbf{GC} & \textbf{QA} & \textbf{AS} & \textbf{CP} & \textbf{HR} & \textbf{RI} & \textbf{SIM} \\
\midrule
Qiskit          & \xmark & \cmark & \cmark & \xmark & \xmark & \xmark & \xmark & \cmark \\
Cirq            & \xmark & \cmark & \xmark & \xmark & \xmark & \xmark & \xmark & \cmark \\
PennyLane       & \xmark & \cmark & \cmark & \xmark & \cmark & \xmark & \xmark & \cmark \\
Qrisp           & \xmark & \cmark & \cmark & \xmark & \xmark & \xmark & \xmark & \cmark \\
Amazon Braket   & \xmark & \cmark & \xmark & \xmark & \cmark & \xmark & \xmark & \cmark \\
Azure Quantum   & \xmark & \cmark & \xmark & \xmark & \cmark & \xmark & \xmark & \cmark \\
Proposed Framework (\textit{C2$\ket{\rm{Q}}$}) & \cmark & \cmark & \cmark & \cmark & \cmark & \cmark & \cmark & \cmark \\
\bottomrule
\end{tabular}
\end{table}

The proposed \textit{C2$\ket{\rm{Q}}$} framework addresses all three challenges by automating the pipeline from parsing classical problem descriptions to producing quantum-compatible representations, selecting suitable algorithms, and recommending optimal execution hardware—while including an interpretation layer that delivers results in a domain-relevant form. This integration reduces reliance on deep quantum expertise and enables practical deployment on NISQ devices.

\section{Proposed Approach} \label{sec:framework}
This section presents the architecture and workflow of the proposed \textit{C2$\ket{\rm{Q}}$} framework, detailing how its modular components enable the automated transformation of classical computational inputs into executable quantum programs.

\subsection{Framework Overview} \label{subsec:overview}
\begin{figure}[h]
    \centering
    \includegraphics[width=1.1\linewidth]{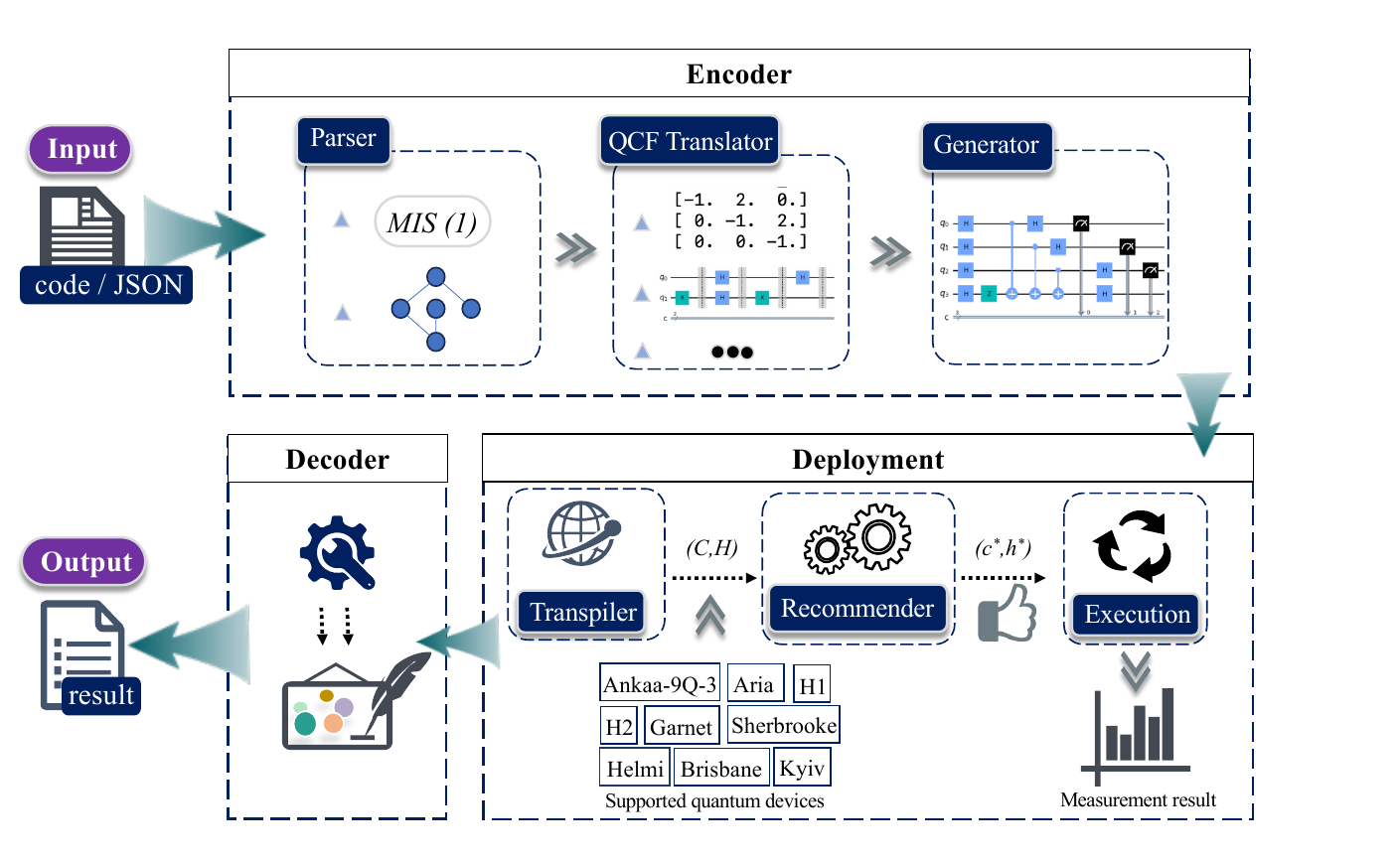}
    \caption{\textit{C2$\ket{\rm{Q}}$} pipeline with encoder–deployment–decoder architecture.}
    \label{fig:architecture}
\end{figure}

The framework is organised into modular components, each responsible for performing a specific task within the development pipeline. Building on this modular structure, the overall workflow can be understood as a sequential transformation pipeline that follows an \textit{encoder–deployment–decoder} architecture. The \emph{encoder} comprises the \textit{parser}, \textit{QCF translator}, and \textit{generator}, which together transform high-level classical problem specifications (source code or structured inputs) into hardware-agnostic quantum circuits. The \emph{deployment} module consists of the \textit{transpiler}, \textit{recommender}, and \textit{execution} components, responsible for selecting appropriate quantum devices and running the transpiled circuits. Finally, the \emph{decoder} module translates the quantum measurement outputs into user-readable classical solutions. The complete workflow is illustrated in Figure~\ref{fig:architecture}. 

The pipeline flow begins with user-provided inputs, such as high-level code snippets or structured JSON problem specifications, and then proceeds through the transformation pipeline described above. Each core module (\textit{encoder}, \textit{deployment}, \textit{decoder}) is implemented by dedicated sub-modules that ensures modularity, verifiability, and hardware-agnostic intermediate representations until device-specific deployment. 

We now discuss each module and its sub-modules, providing a technical understanding of the proposed framework and its overall input–output flow, as shown in Figure \ref{fig:architecture}.
As mentioned above, the \textit{encoder} module consists of \textit{parser}, which processes classical specifications \( C \in \mathcal{L} \), where \( \mathcal{L} \) denotes the set of supported \textit{High-Level Programming Languages (HLPLs)} or structured JSON specifications. It identifies the target problem \( \mathcal{P} \) and extracts the problem input data \( \mathcal{D} \):
\begin{equation}
\text{Parser:} \quad C \mapsto (\mathcal{P}, \mathcal{D}), \quad 
\mathcal{P} \in \mathcal{T},\;
\mathcal{D} \in \mathbb{S},
\end{equation}
Here, \( \mathcal{T} \) denotes supported problem types (e.g., MIS, MaxCut, TSP, arithmetic operations), and \( \mathbb{S} \) represents structured data (e.g., edge lists, adjacency matrices, integers). The \textit{QCF translator} then standardises these heterogeneous representations into a \emph{Quantum-Compatible Format (QCF)} \(Q\):
\begin{equation}
Q = \langle \mathcal{P}, \mathcal{D}, \epsilon \rangle,
\end{equation}
where $\mathcal{P}$ denotes the identified problem type (task tag), $\mathcal{D}$ the structured input data, and $\varepsilon$ the encoding:
\[
  \varepsilon \in \{\text{QUBO},\ \text{Oracle},\ \text{QFT--Arithmetic},\ \text{Reversible--Arithmetic},\ \text{Other}\}.
\]
This abstraction is hardware-agnostic and extensible, allowing multiple quantum algorithms to operate on the same task and data representation (see Appendix~\ref{app:qcf_data}).

Next, the \textit{generator} selects an algorithm from a selected list \( \mathbb{A} = \{\text{QAOA}, \text{VQE}, \text{Grover}, \dots\} \) (see Section~\ref{sec:quantum_algorithms}) and produces logical circuits \(C\):
\begin{equation}
\text{Generator:} \quad (\mathcal{Q}, \mathbb{A}) \mapsto \mathcal{C} = \{c_1, \dots, c_n\}.
\end{equation}
These quantum logical circuits \(C\) are refined by the \textit{transpiler}, which adapts \(C\) to hardware constraints:
\begin{equation}
\text{Transpiler:} \quad \mathcal{C} \mapsto \mathcal{C'}
\end{equation} 
The \textit{recommender} evaluates the circuits \(\mathcal{C'}\) across supported quantum devices \( \mathbb{H} \), selecting an optimal device–circuit pair \((c^*, h^*)\) by minimising a cost metric \(\mathcal{M}(c, h)\), which integrates error rates, execution time, and financial cost (provider pricing):
\begin{equation}
\text{Recommender:} \quad (\mathcal{C'}, \mathbb{H}) \mapsto (c^*, h^*) = \arg\min_{c', h} \mathcal{M}(c', h).
\end{equation}
The \textit{execution} runs the chosen configuration and returns quantum measurement outcomes \(\mathcal{O}\):
\begin{equation}
\text{Execution:} \quad (c^*, h^*) \mapsto \mathcal{O}, \quad \mathcal{O} \in \{\text{bitstrings}, \text{distribution}\}.
\end{equation}
Finally, the \textit{decoder} module translates quantum outcomes into a classical solution \(\mathcal{S}\):
\begin{equation} \label{equation:decoder}
\text{decode}_{\mathcal{P}}(\mathcal{O}; \epsilon) \mapsto \mathcal{S}.
\end{equation}
%
%
In the following subsections, we present the design and implementation details of each mentioned module and their sub-modules: \textit{encoder} module (\textit{parser}, \textit{QCF translator}, \textit{generator}), the \textit{deployment} module (\textit{transpiler, recommender, execution}), and the \textit{decoder} module. We further describe how they integrate into the end-to-end workflow of \textit{C2$\ket{\rm{Q}}$} (see Figure \ref{fig:architecture}). 
These modules address RQ1 (RQ1.1–RQ1.2) and RQ2.

\subsection{Scope, Inputs, and Boundaries} \label{subsec:scope}
This section clarifies the intended scope of \textit{C2$\ket{\rm{Q}}$}, the types of inputs it supports, and the boundaries of problems and abstractions addressed in this work. The \textit{C2$\ket{\rm{Q}}$} framework is designed as a quantum software development workflow for automating the translation of selected classes of classical problem specifications into executable quantum workloads.

The framework accepts high-level problem specifications expressed in two complementary forms. First, users can provide classical Python programs that implement a computational problem, reflecting common software engineering workflows. These programs are intended to be concise problem-specification snippets rather than full-scale software applications; detailed assumptions and technical constraints of the \textit{parser}, including input size considerations, which are discussed in Section~\ref{section:parser}. Second, users can alternatively provide structured JSON inputs that explicitly specify problem instances. Both classical code and JSON inputs are processed by the same downstream pipeline, with JSON serving as a more declarative and unambiguous specification mechanism than imperative code when appropriate.

The decision to support classical code as a primary entry point is motivated by established software engineering practice. In practice, developers often encode requirements, constraints, and problem intent implicitly through executable program structure, such as constraint checks, cost aggregation, feasibility conditions, and control flow, rather than through explicit problem-type or mathematical specifications. As a result, program comprehension and maintenance often rely on source code itself when higher-level specifications are absent or incomplete~\cite{brooks1983towards, rajlich2002staged}.
While structured JSON inputs provide a declarative and unambiguous specification mechanism, classical code offers a natural and accessible starting point for users with existing implementations, enabling problem intent to be captured without requiring an upfront commitment to a dedicated specification language. 

As mentioned in Section~\ref{sec:quantum_algorithms}, to ensure broad coverage of quantum-applicable problems while maintaining a clearly defined scope, we categorise the supported problems into the following three high-level domains of canonical problem families, each reflecting distinct quantum encoding strategies:

\begin{enumerate}
    \item \textbf{Combinatorial optimisation} (e.g., MaxCut, Independent Set, Traveling Salesman, Clique, K-Coloring, Vertex Cover)
    \item \textbf{Arithmetic operations} (e.g., addition, subtraction, multiplication)
    \item \textbf{Number-theoretic tasks} (e.g., integer factorisation) 
\end{enumerate}

These domains span a wide range of computational tasks, from arithmetic operations to complex combinatorial optimisation tasks (see Table~\ref{table:problem_summary}), and are well established in quantum algorithm research. Importantly, each domain admits standardised and widely used quantum-compatible formats (QCFs), such as QUBO formulations for optimisation~\cite{lucas2014,glover2018tutorial}, reversible and QFT-based circuit constructions for arithmetic~\cite{Ruiz_Perez_2017,draper2000additionquantumcomputer}, and oracle-based encodings for search and decision problems~\cite{grover1996,brassard2000quantum}.

This design choice makes these problem families particularly suitable for evaluating and validating an automated classical-to-quantum software development workflow. The availability of well-defined QCFs provides clear abstraction boundaries between problem specification, algorithm instantiation, and hardware recommendation, which is essential for assessing the correctness, modularity, and automation properties of a quantum software development framework. Therefore, these problem families provide a concrete and well-scoped foundation for prototyping and for investigating potential extensions to more specialised application domains within a controlled software engineering setting.

Since these problems can be described at a relatively coarse level of abstraction, the structured JSON inputs in \textit{C2$\ket{\rm{Q}}$} support consistent problem specification and function as a lightweight domain-specific language (DSL) for these canonical problem families. 
In the DSL literature, DSLs are commonly defined as languages tailored to a restricted application domain, offering improved expressiveness and usability within that domain compared to general-purpose specifications~\cite{mernik2005and, kosar2008preliminary}. Following these established definitions, the JSON schema in \textit{C2$\ket{\rm{Q}}$} provides a declarative, schema-based representation of problem instances and parameters, bridging developer-specified objectives and the canonical representations required by downstream quantum algorithms. We emphasize that, while this JSON-based interface aligns with established notions of lightweight DSLs for restricted domains, it does not introduce a dedicated grammar or formal semantic/type system; instead, it provides a schema-driven specification format scoped to canonical problem families.

\begin{table}[h!]
\centering
\caption{Supported problem types in \textit{C2$\ket{\rm{Q}}$} with domains, inputs, algorithms, and required QCFs. Combinatorial optimisation problems are considered in their optimisation form; each admits a corresponding NP-complete decision variant~\cite{garey1979computers}. QCFs: QUBO = matrix-based QUBO formulation, Oracle = oracle-based encoding and corresponding circuits, QFT = Quantum Fourier Transform arithmetic subroutines, Reversible = reversible logic arithmetic circuits.}
\label{table:problem_summary}
\small
\begin{tabular}{c l l p{3.6cm} p{2cm}}
\toprule
\textbf{Tag (\(\mathcal{T}\))} & \textbf{Problem Type} & \textbf{Domain} & \textbf{Required Data / Algorithms} & \textbf{QCFs Required} \\
\midrule
0 & Maximum Cut (MaxCut)    & Combinatorial optimisation & Edges or adjacency matrix; QAOA, VQE & QUBO \\
1 & Maximum Independent Set (MIS) & Combinatorial optimisation & Edges / adjacency matrix with constraints; QAOA, VQE, Grover & QUBO, Oracle \\
2 & Traveling Salesman (TSP) & Combinatorial optimisation & Distance matrix or weighted edges; QAOA, VQE & QUBO \\
3 & Clique                  & Combinatorial optimisation & Edges / adjacency matrix, clique size $k$; QAOA, VQE, Grover & QUBO, Oracle \\
4 & K-Coloring (KColor)     & Combinatorial optimisation & Edges, color constraints; QAOA, VQE, Grover & QUBO, Oracle \\
5 & Vertex Cover (VC)       & Combinatorial optimisation & Edges / adjacency matrix, cover size $k$; QAOA, VQE, Grover & QUBO, Oracle \\
6 & Integer Factorisation (Factor)            & Number-theoretic task      & Integer $N$; Shor’s, Grover & Oracle, Reversible \\
7 & Addition (ADD)          & Arithmetic operation       & Integer operands $(a,b)$; QFT Adder, Full Adder & QFT, Reversible \\
8 & Multiplication (MUL)    & Arithmetic operation       & Integer operands $(a,b)$; QFT Multiplier & QFT \\
9 & Subtraction (SUB)       & Arithmetic operation       & Integer operands $(a,b)$; QFT Subtractor, Full Adder & QFT, Reversible \\
\bottomrule
\end{tabular}
\end{table}

\subsection{Encoder (RQ1)} \label{section:encoder}
The \textit{encoder} module of the \textit{C2$\ket{\rm{Q}}$} framework consists of three sub-modules as discussed in Section~\ref{subsec:overview}. We now describe each sub-module in detail.

\subsubsection{RQ 1.1: Parser \& QCF Translator} \label{section:parser}

Parser is responsible for analysing classical problem specifications, extracting relevant structured data, and preparing them for downstream quantum processing. The parser outputs a well-formed \emph{problem descriptor} comprising (i) a formal problem tag, (ii) required structured data elements, and (iii) an initial specification of the target Quantum-Compatible Format (QCF). This separation ensures that classical-to-quantum encoding is handled in a modular and verifiable stage.

\begin{figure}[h]
    \includegraphics[width=0.6\linewidth]{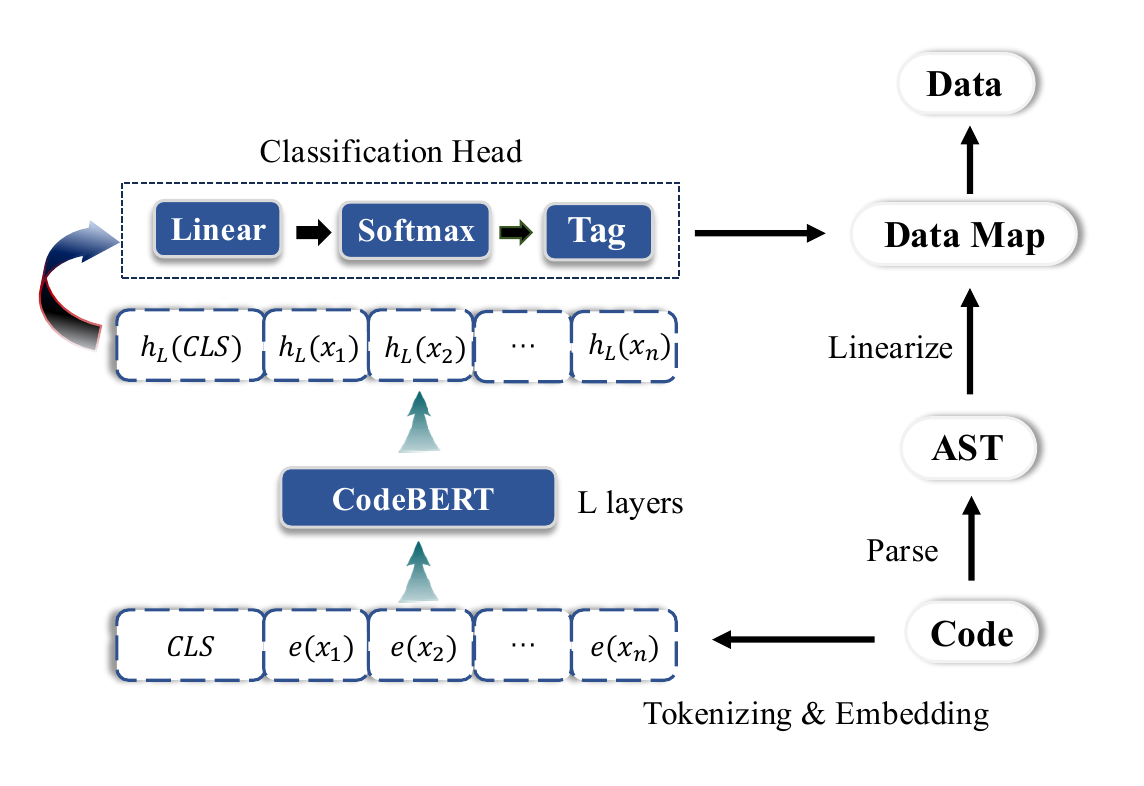}
    \caption{Architecture of Parser.}
    \label{fig:parser}
\end{figure}

\paragraph{Design of Parser}
As shown in Figure~\ref{fig:parser}, the parser adopts a hybrid design that maps classical code into one of the supported problem types (see Table~\ref{table:problem_summary}). It combines Abstract Syntax Tree (AST) analysis~\cite{ast} with a transformer-based semantic model, CodeBERT~\cite{feng2020codebertpretrainedmodelprogramming}, fine-tuned for quantum-relevant problem classification and structured data extraction. The AST layer ensures syntactic correctness and captures structural code patterns, while the CodeBERT layer provides robustness to variable naming diversity, idiomatic coding styles, and structural variations common in real-world codebases. Together, these mechanisms classify classical input code into one of the canonical problem types (e.g., MaxCut, MIS, TSP) and extract the corresponding problem-relevant data \((\mathcal{P}, \mathcal{D})\) for quantum encoding.




Given a code snippet, the parser first checks syntactic validity using Python's \texttt{ast} module; invalid inputs are assigned to \texttt{UNKNOWN} rather than forced into a class. Valid snippets are encoded with CodeBERT~\cite{feng2020codebertpretrainedmodelprogramming}, a transformer-based model with a maximum input length of 512 subword tokens produced by its tokenizer. When inputs exceed this limit, the tokenizer truncates the resulting sequence of tokens produced from the source code, and problem-type classification is performed on the leading portion of this token sequence. In contrast, AST-based parsing and QCF data extraction operate on the full program, ensuring that problem-relevant data is not lost due to token-length constraints. The [CLS] representation is passed through a linear classification head with softmax to predict the problem tag~\cite{devlin2019bertpretrainingdeepbidirectional}. Formally, the predicted tag \(\hat y\) is given by
\begin{equation}
    \hat{y}=\arg\max \mathrm{softmax}\!\left(W\,\mathbf{h}_{\text{[CLS]}}+b\right),
\end{equation}
where \(\hat{y}\) is mapped to a concrete problem tag \(\mathcal{P} \in \mathcal{T}\), where \(\mathcal{T}\) denotes the set of supported problem types defined in Table~\ref{table:problem_summary}. 

In parallel, AST traversal provides structural features that complement the semantic embedding, enabling recovery of problem-specific data \((\mathcal{D})\). The combined output forms a structured \emph{problem descriptor} \((\mathcal{P}, \mathcal{D})\), which serves as the interface to the next stage of the pipeline. 

In addition to classical code inputs, the parser also supports structured JSON specifications, which serve as a lightweight problem-family DSL as defined in Section~\ref{subsec:scope}. For JSON inputs, parsing is deterministic and schema-driven: the problem tag \(\mathcal{P}\) and structured data \(\mathcal{D}\) are read directly from the JSON fields without invoking AST analysis or neural inference. Each JSON specification follows a lightweight domain-specific schema and must include a \texttt{problem\_type} field identifying the canonical task (e.g., \texttt{MIS}, \texttt{MaxCut}) and a corresponding \texttt{data} field encoding the structured instance information (e.g., edge lists, distance matrices, or integers), conforming to the expected schema of that problem family. A complete example is provided in Appendix~\ref{app:json}.

\paragraph{Dataset and Model Training}
Due to the absence of large-scale, labeled corpora for quantum-relevant problem types, we bootstrap the initial parser (\textit{v0}) using a synthetic dataset generated with a large language model (LLM), specifically OpenAI’s GPT series \cite{gpt}. This approach is motivated by instruction-tuning pipelines such as Self-Instruct \cite{wang2023selfinstructaligninglanguagemodels} and Alpaca \cite{taori2023stanford}, which demonstrate the effectiveness of LLM-generated supervision when human annotations are scarce. Similar strategies have also been applied in recent studies, where synthetic datasets are created through systematic prompt design, validation mechanisms, and code-centric data generation pipelines \cite{xu2025kodcodediversechallengingverifiable, Nad__2025}. Building on these insights, we designed dataset prompts to ensure:  
\begin{itemize}
  \item \textbf{Problem type coverage}: balanced across targeted domains (Table~\ref{table:problem_summary});  
  \item \textbf{Structural diversity}: inclusion of recursive, iterative, and heuristic algorithm styles;  
  \item \textbf{I/O variety}: multiple semantically equivalent input–output encodings;  
  \item \textbf{Identifier variation}: both consistent and intentionally inconsistent naming schemes;  
  \item \textbf{Language constraint}: Python-only, simplifying pipeline integration for \textit{v0}.  
\end{itemize}

The resulting dataset comprises 434 Python code snippets, each containing both the problem specification and executable logic, spanning combinatorial optimisation, arithmetic, and number-theoretic problems. Quality control includes syntactic validation with Python’s \textit{ast} module, token-based duplicate filtering, and class balancing. Beyond syntactic correctness and class balance, we explicitly assess within-class diversity of the generated programs. At the implementation level, we compute static software metrics that are standard in empirical software engineering research \cite{ray2014large,nagappan2006mining}, including (i) program size measured by lines of code (LOC), (ii) control-flow usage (counts of \texttt{for} loops, \texttt{while} loops, and recursion), (iii) data-structure and external library usage, and (iv) cyclomatic complexity as a measure of control-flow complexity \cite{mccabe1976complexity}. In addition, we quantify algorithmic diversity via Shannon entropy computed over automatically detected algorithm families~\cite{shannon1948mathematical}, and structural instance diversity using exact fingerprints and Weisfeiler--Lehman graph hashing~\cite{shervashidze2011weisfeiler}.

Across the dataset, the analysis reveals measurable within-class variation. For example, LOC ranges span more than a factor of four in several families (e.g., K-Coloring from 10 to 44 LOC and TSP from 11 to 37 LOC), control-flow structures differ markedly (e.g., MIS programs combine iterative and recursive patterns, whereas MaxCut programs are predominantly iterative), and mean cyclomatic complexity varies from approximately 1.2 in arithmetic tasks to over 5.2 in MaxCut. Algorithmic entropy values for graph-based problems range from 1.35 to 1.90, indicating heterogeneous distributions of solution strategies. Structural instance diversity is assessed using two complementary measures: exact fingerprints (to detect identical instances) and Weisfeiler–Lehman (WL) graph hashing as a scalable approximation of structural non-isomorphism. For graph-based tasks, Exact Unique ratios range from 0.14 to 0.36, while WL Unique ratios range from 0.14 to 0.22. A detailed breakdown of all metrics is provided in Appendix~\ref{app:dataset_diversity}, and the full dataset together with validation scripts is available in the replication package~\cite{ye_khan_c2q_dataset_2025}.

\begin{table}[h!]
\centering
\caption{Training configuration for the parser model.}
\label{table:model_training_configurations}
\renewcommand{\arraystretch}{1.25}
\small
\begin{tabular}{l l}
\toprule
\textbf{Parameter} & \textbf{Value} \\
\midrule
Model & \texttt{microsoft/codebert-base} \\
Number of labels & 10 \\
Device & GPU (Apple M1 Max) \\
Evaluation strategy & Epoch \\
Number of epochs & 10 \\
Train batch size (per device) & 8 \\
Evaluation batch size (per device) & 16 \\
Train–test split & 80\% / 20\% \\
Shuffle & True \\
Optimizer & AdamW \\
Learning rate & $5 \times 10^{-5}$ \\
Data source & Preprocessed tokenized dataset (434 samples) \\
\bottomrule
\end{tabular}
\end{table}

The parser was fine-tuned from the \texttt{microsoft/codebert-base} checkpoint \cite{feng2020codebertpretrainedmodelprogramming} using the configuration summarised in Table~\ref{table:model_training_configurations}. Training progress was monitored via cross-entropy loss \cite{goodfellow2016deep}, defined as:
\begin{equation}
    \text{Loss} = - \frac{1}{N} \sum_{j=1}^{N} \sum_{i=1}^{C} y_{ji} \log({p}_{ji}),
\end{equation}
where \(y_{ji}\) is the ground-truth label and \(p_{ji}\) the predicted probability for class \(i\). On the evaluation split, the parser achieves a cross-entropy loss of 0.0674 and a weighted-average F1-score of 98.2\% (reported in Experiment~1; see Section~\ref{ex1}). To prevent misclassification, the framework applies a confidence threshold of 0.80 to the softmax output: if the predicted probability is below this threshold, the program is assigned an \texttt{UNKNOWN} tag and excluded from downstream processing. This threshold reflects a conservative design choice that prioritises precision over recall, ensuring that only high-confidence problem-type predictions are forwarded to subsequent stages that assume correct canonical classification. In addition, syntactically valid inputs that pass this check may still be rejected at the data extraction stage if required structured elements cannot be reliably retrieved. In rare cases, an input program that does not truly conform to any supported problem type may nevertheless (i) pass the softmax confidence threshold for a supported class and (ii) successfully complete structured data extraction due to partial structural similarity; such cases are considered failures, though they did not appear in our evaluation dataset.

\paragraph{Data Extraction}
Within the encoder, a key step is the extraction of problem-relevant data, which is subsequently transformed into QCF. This process combines Abstract Syntax Tree (AST) traversal, which captures structural cues, with a semantic map that records variable assignments, function calls, and arguments. Together, these representations allow the system to resolve user-defined identifiers—such as adjacency lists, weight matrices, and integer operands—that would otherwise remain opaque to purely syntactic analysis. Each problem type has explicitly defined data requirements (see Table~\ref{table:problem_summary}), which guide the extraction process. To ensure that this data extraction operates reliably, we employ declarative, rule-based strategies over the combined AST–semantic map. This approach ensures interpretability while keeping the module extensible for future integration of learning-based extraction methods. Supported cases include:
\begin{itemize}
    \item \textbf{Combinatorial optimisation problems (e.g., MaxCut, TSP):} Detection of edge lists or adjacency matrices, with default weights assigned where missing. 
    \item \textbf{Arithmetic operations (e.g., addition, subtraction, multiplication):} Resolution of integer operands based on both syntactic roles and semantic usage, ensuring robustness to naming variation.  
    \item \textbf{Number-theoretic problems (e.g., factorisation):} Identification of integer inputs \(N\) for prime factorisation tasks.  
\end{itemize}

\paragraph{QCF Translation}
After the parser produces a structured \textit{problem descriptor}—comprising the problem tag and extracted problem-relevant data—the \textit{QCF translator} transforms this representation into a Quantum-Compatible Format (QCF) suitable for algorithm instantiation. By mapping heterogeneous descriptors into canonical encoding schemes, the translator ensures a consistent interface across downstream modules.

The translator currently implements three categories of encodings:  
\begin{itemize}
    \item \textbf{QUBO (Quadratic Unconstrained Binary Optimization):}  
    Combinatorial NP-complete problems are reduced to QUBO, enabling compatibility with variational solvers such as QAOA and VQE. Standard formulations, such as those surveyed by Lucas~\cite{lucas2014}, are adopted to map graph- and constraint-based problems into binary quadratic forms.  

    \item \textbf{Oracle-based encoding:}  
    Decision and search problems are expressed as Boolean formulas. The PySAT toolkit~\cite{imms-sat18} is used to construct and manipulate Conjunctive Normal Form (CNF), which is then compiled into Grover-style oracles for quantum search.  

    \item \textbf{Arithmetic and number-theoretic encodings:}  
    Problems that fall outside the combinatorial optimisation setting are encoded through:  
    \begin{itemize}
        \item \textbf{Arithmetic operations (addition, subtraction, multiplication):} Represented via reversible logic or Fourier-based constructions~\cite{Ruiz_Perez_2017,draper2000additionquantumcomputer}.
        \item \textbf{Integer factorisation:} Supported via two quantum algorithms:
        \begin{itemize}
            \item Shor’s algorithm, which encodes modular exponentiation and applies the Quantum Fourier Transform for period finding~\cite{shor1999polynomial}. 
            \item Grover’s algorithm, which places candidate solutions in superposition and marks valid ones through an oracle, offering a search-based alternative~\cite{grover1996, brassard2000quantum}.
        \end{itemize}
    \end{itemize}
\end{itemize}

\paragraph{Example.} Appendix~\ref{app:mis} illustrates this translation process using the Maximum Independent Set (MIS) problem. The appendix presents a general construction of MIS into two distinct QCFs: (i) a QUBO formulation, where adjacency constraints are encoded as quadratic penalties and vertex selection is rewarded, and (ii) a Boolean formula representation compiled into a Grover-style oracle that enforces independence and maximality conditions. This example highlights how a single high-level problem type can be systematically normalised into alternative encodings, enabling the framework to support both variational and oracle-based quantum algorithms.



    

\subsubsection{RQ 1.2: Generator} \label{subsec:gen}
To address RQ~1.2, the \textit{C2$\ket{\rm{Q}}$} generator module instantiates quantum algorithms from the problem's QCF representation. Supported algorithms include variational methods (QAOA~\cite{farhi2014}, VQE~\cite{peruzzo2014}), search-based methods (Grover’s algorithm~\cite{grover1996}), and arithmetic routines such as reversible full-adders and QFT-based constructions for addition, subtraction, and multiplication~\cite{Ruiz_Perez_2017,draper2000additionquantumcomputer}, all of which are described in detail in Section~\ref{sec:quantum_algorithms}. This design enables mapping of diverse problem types—combinatorial, number-theoretic, and arithmetic—into appropriate algorithmic back-ends (Table~\ref{table:problem_summary}). Within the \textit{encoder} module of the proposed framework’s workflow (Figure~\ref{fig:architecture}), the generator serves as the final sub-module. It uses QCFs and produces hardware-agnostic logical circuits. All circuits are implemented in Qiskit~\cite{javadiabhari2024quantumcomputingqiskit} using standard gates (e.g., Pauli, Hadamard, CNOT), ensuring platform portability and enabling automated device recommendation. With the generator defined as the bridge from QCFs to circuits, we now specify how algorithms are configured within this module.

\paragraph{Algorithm Configuration}
We define parameter settings for selected QAOA, VQE, and Grover’s algorithm. While QCFs specify the choice of algorithm (e.g., a QUBO formulation guides the circuit construction for VQE and QAOA), configuration parameters control how the circuits are instantiated in practice. The generator uses a set of fixed default configuration parameters for each supported quantum algorithm. These defaults are selected from commonly used, literature-supported baseline settings to ensure reproducibility, comparability across problem families, and compatibility with current NISQ hardware. The parameter settings and their supporting references are summarized in Table~\ref{table:algorithm_configurations}. Configuration parameters are intentionally held constant across inputs and are not adapted to individual problem instances. This design choice allows the evaluation to focus on the correctness and automation capabilities of the framework rather than instance-specific performance optimisation. For QAOA, we adopt a shallow fixed-depth setting (3 layers) as a hardware-agnostic default. This choice reflects a conservative design decision to limit circuit depth on current NISQ devices, where prior work has shown that increasing depth can quickly negate potential benefits due to noise and compilation overhead~\cite{alam2019analysis,zhou2020quantum}. The variational angles are randomly initialised per layer from symmetry-reduced periodic domains commonly used in QAOA implementations ($\gamma_k\!\sim\!\mathrm{Unif}[0,\pi]$, $\beta_k\!\sim\!\mathrm{Unif}[0,\pi/2]$), where $\gamma_k$ and $\beta_k$ denote the per-layer \emph{cost} and \emph{mixer} angles (see Section~\ref{sec:quantum_algorithms}). These parameters are then optimized using Simultaneous Perturbation Stochastic Approximation (SPSA), a stochastic gradient-approximation optimizer that estimates gradients using only two evaluations per iteration~\cite{spall1992spsa}. SPSA is executed for 500 iterations and is selected for its shot efficiency and robustness under on-device noise~\cite{pellow2021comparison}. For VQE, we employ a shallow two-local hardware-efficient ansatz composed of $R_y$ rotations and $CZ$ entangling gates, instantiated with three layers as a fixed default. Hardware-efficient and two-local ansätze are widely adopted in NISQ-era VQE implementations due to their compatibility with native gate sets and controllable expressibility at low depth~\cite{McClean_2016,Kandala_2017,sim2019expressibility}. As with QAOA, the chosen depth is intended to balance circuit executability and expressibility rather than to optimise instance-specific performance. Grover’s algorithm follows the randomized adaptive iteration schedule of Boyer \emph{et al.}~\cite{boyer1998}, with multiplicative growth ($\lambda = 8/7$) and termination upon success~\cite{grover1996}.

\begin{table}[htbp]
\centering
\caption{Configurations of supported algorithms in the \textit{C2$\ket{\rm{Q}}$} generator.}
\label{table:algorithm_configurations}
\sffamily
\renewcommand{\arraystretch}{1.1}
\begin{tabular}{p{3.2cm} p{10cm}}
\toprule
\textbf{Algorithm} & \textbf{Configuration} \\
\midrule
QAOA &
\begin{itemize}[leftmargin=*]
  \item Layers: $p=3$ (conservative fixed depth; deeper circuits can be quickly limited by NISQ noise and compilation overhead)~\cite{alam2019analysis,zhou2020quantum}
  \item Cost Hamiltonian: Ising form derived from QUBO (standard mapping, see Appendix~\ref{subapp:qubo})~\cite{lucas2014}
  \item Mixer: $H_M=\sum_i X_i$
  \item Initial parameters: for each layer $k$, sample independently
        $\gamma_k \sim \mathrm{Unif}[0,\pi]$ and $\beta_k \sim \mathrm{Unif}\!\big[0,\tfrac{\pi}{2}\big]$, which restricts parameters to a non-redundant fundamental domain and avoids trivial symmetric initializations.
  \item Classical optimiser (default): \textit{SPSA}, max 500 iterations (stochastic, shot-efficient)~\cite{spall1992spsa, pellow2021comparison}
\end{itemize} \\

VQE &
\begin{itemize}[leftmargin=*]
  \item Ansatz: Two-local hardware-efficient ansatz with $R_y$ rotations and $CZ$ entanglement (3 layers; fixed default following widely used hardware-efficient and two-local designs)~\cite{McClean_2016,Kandala_2017,sim2019expressibility}
  \item Cost Hamiltonian: Ising form
  \item Initial parameters: small random angles (independent draws per parameter)
  \item Classical optimiser (default): \textit{SPSA}, max 500 iterations (stochastic, shot-efficient)~\cite{spall1992spsa, pellow2021comparison}
\end{itemize} \\

Grover’s Algorithm &
\begin{itemize}[leftmargin=*]
  \item Initial state: uniform superposition $H^{\otimes n}\ket{0}^{\otimes n}$
  \item Iterations: randomized adaptive schedule~\cite{boyer1998}; 
        initialize $m{=}1$, each round sample $j\in\{0,\dots,m{-}1\}$ uniformly, 
        prepare the initial state, apply $j$ Grover iterations (oracle then diffusion), and measure.
        On failure, set $m\leftarrow\min(\lceil\lambda m\rceil,\lceil\sqrt{2^n}\rceil)$ with $1{<}\lambda{<}4/3$ (we use $\lambda{=}8/7$) and repeat~\cite{grover1996}.
\end{itemize} \\
\bottomrule
\end{tabular}

\begin{flushleft}
\footnotesize
\footnotesize\emph{Notation.} $\mathrm{Unif}[a,b]$ denotes a uniform distribution on $[a,b]$. “Independently” means each layer/parameter is sampled separately (no sample influences another), and all samples are taken from the same distribution.
\end{flushleft}
\end{table}

\paragraph{Algorithms for Arithmetic and Number-Theoretic Problems}
In addition to combinatorial optimisation problems, the \textit{C2$\ket{\rm{Q}}$} framework also supports arithmetic operations—addition, subtraction, and multiplication—as well as number-theoretic problems such as integer factorisation. These are implemented in the \textit{generator} using established quantum arithmetic primitives. For addition and subtraction, the framework supports both quantum full-adder circuits~\cite{draper2000additionquantumcomputer} and Quantum Fourier Transform (QFT)–based arithmetic methods~\cite{Ruiz_Perez_2017}, providing a balance between clarity of implementation and scalability. Subtraction is handled by encoding the right operand in two’s complement form and applying the same addition circuit, consistent with classical practice. For multiplication, a QFT-based multiplier~\cite{Ruiz_Perez_2017} is used, which reduces the required number of gates compared to repeated-addition approaches and is therefore better suited for NISQ devices. For factorisation, the framework integrates both Grover’s search (oracle-based)~\cite{grover1996} and Shor’s algorithm (period-finding)~\cite{shor1999polynomial}, demonstrating flexibility across different algorithmic paradigms. Grover’s method is effective for small instances with clearly marked solutions, while Shor’s algorithm provides a scalable option for larger problems but requires significantly more quantum resources.

\subsection{Deployment (RQ2)} \label{Sec:Deployment}

The \textit{deployment} module consists of sub-modules: \textit{transpiler}, \textit{recommender}, and \textit{execution}. These sub-modules work collaboratively to adapt logical circuits to hardware constraints and to assist users in selecting a suitable quantum device for a given computational task.

The \textit{transpiler} and \textit{recommender} sub-modules form the core of the deployment stage, working together to adapt logical circuits to hardware constraints and to select the most suitable quantum device that is available on the cloud. As part of this process, the framework leverages Qiskit SDK transpilers~\cite{javadiabhari2024quantumcomputingqiskit} to perform hardware-specific circuit transformations. This choice is motivated by the use of Qiskit’s standard gate set for logical circuit construction and by the transpiler’s broad backend compatibility, which enables the generation of physical circuits across heterogeneous quantum devices and supports effective coordination between the \textit{transpiler} and \textit{recommender} modules. Their workflow proceeds in three steps:   

\begin{enumerate}
    \item \textbf{Transpilation:} Circuits are mapped to each device’s \emph{coupling map}, which specifies its physical topology (see Figure~\ref{fig:helmi} for the Helmi device~\cite{csc_helmi_specs}). This ensures hardware compatibility and minimises gate overhead.  
    \item \textbf{Resource Calculation:} The recommender analyses transpiled circuits to extract resource metrics such as gate counts and circuit depth.  
    \item \textbf{Metric Evaluation:} Devices are compared using three criteria—\emph{estimated error}, \emph{execution time}, and \emph{cost}. The integration of these metrics into the selection process is formalised through an optimisation algorithm (Algorithm~\ref{alg:recommender}), which will be introduced later.  
\end{enumerate}




We currently support devices from three major platforms: \textbf{Azure Quantum} (IonQ, Quantinuum, Rigetti), \textbf{IBM Quantum}, and \textbf{Amazon Braket} (IonQ, IQM). Each device is characterised by calibration data, including coherence times ($T_1, T_2$), gate error rates, gate timings, and provider-specific pricing models (see Section~\ref{sec:platform_diff}). These parameters, summarised in Table~\ref{tab:quantum_devices}, form the basis for estimating the three core evaluation metrics—\emph{error}, \emph{execution time}, and \emph{cost}—which drive the recommender’s device selection process. 
It is worth noting that \textit{IonQ Aria} is accessible via both Azure and AWS; while the underlying hardware specifications are identical, the recommender treats them as distinct devices due to differences in calibration updates, pricing, and access policies.

\begin{table}[h!]
\centering
\caption{Representative quantum devices and their physical parameters (calibration snapshots as of December~2024). 
Values fluctuate with calibration; actual metrics should be retrieved at runtime from provider dashboards or APIs. 
Symbols correspond to parameters used in Section~\ref{Sec:Deployment}: 
$e_1$ (single-qubit gate error), $e_2$ (two-qubit gate error), $e_m$ (measurement error), 
$t_1$ (single-qubit gate time), $t_2$ (two-qubit gate time), and coherence times $T_1, T_2$.}
\label{tab:quantum_devices}
\renewcommand{\arraystretch}{1.25}
\small
\resizebox{\textwidth}{!}{%
\begin{tabular}{l c c c c c c c}
\toprule
\textbf{Device / Provider} & \textbf{Qubits} & \textbf{$e_1$} & \textbf{$e_2$} & \textbf{$e_m$} & \textbf{$t_1$} & \textbf{$t_2$} & \textbf{Coherence $T_1/T_2$} \\
\midrule
IBM Kyiv~\cite{ibm_quantum} & 127 & $2.8 \times 10^{-4}$ & $1.2 \times 10^{-2}$ & $7.0 \times 10^{-3}$ & $50$ ns & $500$ ns & $258\,\mu$s / $109\,\mu$s \\
IBM Sherbrooke~\cite{ibm_quantum} & 127 & $2.2 \times 10^{-4}$ & $7.8 \times 10^{-3}$ & $1.3 \times 10^{-2}$ & $57$ ns & $530$ ns & $261\,\mu$s / $168\,\mu$s \\
IBM Brisbane~\cite{ibm_quantum} & 127 & $2.5 \times 10^{-4}$ & $7.7 \times 10^{-3}$ & $1.3 \times 10^{-2}$ & $60$ ns & $660$ ns & $221\,\mu$s / $134\,\mu$s \\
Rigetti Ankaa-9Q-3~\cite{rigetti_qcs} & 9 & $1.0 \times 10^{-3}$ & $8.0 \times 10^{-3}$ & $6.6 \times 10^{-2}$ & $40$ ns & $70$ ns & $21\,\mu$s / $24\,\mu$s \\
IQM Garnet~\cite{iqm_devices} & 20  & $8.0 \times 10^{-4}$ & $4.9 \times 10^{-3}$ & $1.9 \times 10^{-2}$ & $20$ ns & $40$ ns & $50\,\mu$s / $8\,\mu$s \\
IQM Helmi~\cite{iqm_devices} & 5   & $3.8 \times 10^{-3}$ & $3.9 \times 10^{-2}$ & $4.8 \times 10^{-2}$ & $120$ ns & $120$ ns & $36\,\mu$s / $17\,\mu$s \\
IonQ Aria (AWS/Azure)~\cite{ionq_devices} & 25 & $6.0 \times 10^{-4}$ & $6.0 \times 10^{-3}$ & $4.8 \times 10^{-3}$ & $135\,\mu$s & $600\,\mu$s & $100$ s / $1$ s \\
Quantinuum H1~\cite{quantinuum_devices} & 20  & $2.0 \times 10^{-5}$ & $1.0 \times 10^{-3}$ & $3.0 \times 10^{-3}$ & $63\,\mu$s & $308\,\mu$s & $60$ s / $4$ s \\
Quantinuum H2~\cite{quantinuum_devices} & 56  & $3.0 \times 10^{-5}$ & $1.5 \times 10^{-3}$ & $1.5 \times 10^{-3}$ & $63\,\mu$s & $308\,\mu$s & $60$ s / $4$ s \\
\bottomrule
\end{tabular}
}
\end{table}
\begin{flushleft}
\footnotesize
\textit{Note.} Quantinuum gate durations were derived based on vendor-reported transport and gate operation times~\cite{pino2021demonstration}. The effective single-qubit gate time ($t_1$) is the sum of intrazone transport time ($58\,\mu$s) and native gate time ($5\,\mu$s), totaling $63\,\mu$s. The two-qubit gate time ($t_2$) includes interzone transport ($283\,\mu$s) plus entangling gate
duration ($25\,\mu$s), totaling $308\,\mu$s.
\end{flushleft}

\begin{figure}[ht]
\centering
\begin{tikzpicture}[
    qubit/.style={circle, draw, minimum size=6mm, inner sep=0pt, font=\small},
    node distance=2cm
]
  \node[qubit] (q3) at (0,0) {Q$_3$};

  \node[qubit, above=of q3] (q0) {Q$_0$};
  \node[qubit, right=of q3] (q1) {Q$_1$};
  \node[qubit, below=of q3] (q2) {Q$_2$};
  \node[qubit, left=of q3]  (q4) {Q$_4$};

  \draw (q0) -- (q3);
  \draw (q1) -- (q3);
  \draw (q2) -- (q3);
  \draw (q3) -- (q4);
\end{tikzpicture}
\caption{Topology of the Helmi quantum computer represented by the coupling map \texttt{coupling\_map = [(0, 3), (1, 3), (2, 3), (3, 4)]}.}
\label{fig:helmi}
\end{figure}

\paragraph{Error Estimation.}
In quantum computing, \textit{errors} arise from imperfections in gate operations, qubit decoherence, and measurement noise, which degrade circuit fidelity as circuit depth increases. Fidelity is typically defined as $F = 1 - E$, where $E$ denotes the total error probability. 
To assess reliability, we estimate the cumulative error of each transpiled circuit by counting the number of single-qubit, two-qubit, and measurement operations, and retrieving their error rates $(e_1,e_2,e_m)$ from device calibration data (Table~\ref{tab:quantum_devices}). We then compute the total error probability using a product-of-survivals model that treats operation errors as independent and stochastic~\cite{aseguinolaza2024error}:
\begin{equation} \label{eq:error}
    E \;=\; 1 - \Big[(1-e_1)^{N_1} (1-e_2)^{N_2} (1-e_m)^{N_m}\Big].
\end{equation}
Although this model abstracts away coherent and correlated noise, it provides a practical and interpretable estimate for cross-hardware comparison. Techniques such as randomized benchmarking~\cite{Magesan_2011, magesan2012efficient}, cycle benchmarking~\cite{erhard2019characterizing}, and randomized compiling~\cite{wallman2016noise} are commonly employed by hardware providers to characterize gate- and cycle-level error rates. These methods operate under stochastic noise assumptions similar to those used in our model, and their results inform the calibration data (e.g., $e_1$, $e_2$, $e_m$) we rely on to estimate circuit-level error.


\paragraph{Execution Time Estimation.}
Execution time is derived from circuit depth and device-specific gate durations (\(t_{1}, t_{2}\), see Table~\ref{tab:quantum_devices}). 
For most devices, the runtime per iteration is estimated as:
\begin{equation} \label{eq:time}
    T_{\text{exec}} = (D_{\text{1Q}} \cdot t_{1} + D_{\text{2Q}} \cdot t_{2}) \times S \times I,
\end{equation}
where \(D_{\text{1Q}}, D_{\text{2Q}}\) are the single- and two-qubit depths, and \(S, I\) are the number of shots and iterations. 
Special cases (e.g., Quantinuum) instead use hardware credit-based models also for estimating the execution time~\cite{quantinuum_devices}.

\paragraph{Cost Estimation.}
Cost estimation reflects the diversity of provider pricing schemes. 
IBM and Rigetti adopt time-based pricing, Amazon Braket applies a per-shot model, and IonQ and Quantinuum employ gate- or credit-based models. 
By normalising these heterogeneous schemes into comparable cost estimates, the recommender balances fidelity, execution time, and budget when suggesting devices.\footnote{See provider documentation for pricing models: IBM~\url{https://quantum.cloud.ibm.com/}, Rigetti~\url{https://www.rigetti.com/}, Amazon Braket~\url{https://docs.aws.amazon.com/braket/latest/developerguide/braket-pricing.html}, IonQ~\url{https://ionq.com/}, Quantinuum~\url{https://www.quantinuum.com/}.}

\paragraph{Device Selection Algorithm}
We define a feasibility threshold $\tau=0.5$, requiring circuits to achieve at least 50\% estimated fidelity in order to be considered executable. Having defined estimates for error, execution time, and cost, the recommender (Algorithm~\ref{alg:recommender}) ranks circuit–device pairs by first filtering out infeasible options based on qubit capacity and the fidelity constraint ($F \ge \tau$). It then normalises the feasible metrics and computes a weighted score using user preferences $(\lambda_1,\lambda_2,\lambda_3)$, with default values shown in Table~\ref{tab:parameter_defaults}. Ties are resolved deterministically by favouring lower error, then lower cost, then lower latency to ensure reproducibility. A small adjustment ($\delta{=}0.05$) is applied to trapped-ion devices when accuracy dominates or qubit counts exceed 10, reflecting their lower error accumulation while respecting user-defined weights. This policy provides a transparent and tunable balance among fidelity, runtime, and cost.

\begin{algorithm}[h] 
\small
\caption{Recommender Module: Select Optimal Circuit–Hardware Pair}
\label{alg:recommender}
\KwIn{
  $\mathcal{C'}$: transpiled circuits; 
  $\mathbb{H}$: candidate devices; 
  $(\lambda_1,\lambda_2,\lambda_3)$: weights; 
  $\tau=0.5$
}
\KwOut{Best pair $(c^*,h^*)$}

$\mathcal{P} \gets \{(c',h) \mid c'\!\in\!\mathcal{C'},\, h\!\in\!\mathbb{H},\, \textbf{Feasible}(c',h)\}$\;
\If{$\mathcal{P}=\varnothing$}{\Return{No compatible pair found}}

\BlankLine
\textit{Compute normalised metrics using Eqs.\,(\ref{eq:error})--(\ref{eq:time}) and the pricing model:}\\
$E_{\max}\!\gets\!\max\limits_{(c',h)\in\mathcal{P}} E(c',h)$,\quad
$T_{\max}\!\gets\!\max\limits_{(c',h)\in\mathcal{P}} T(c',h)$,\quad
$P_{\max}\!\gets\!\max\limits_{(c',h)\in\mathcal{P}} P(c',h)$\;
For each $(c',h)\!\in\!\mathcal{P}$ set
$\hat{E}\!\gets\!E/E_{\max}$,\;
$\hat{T}\!\gets\!T/T_{\max}$,\;
$\hat{P}\!\gets\!P/P_{\max}$\;

\BlankLine
\textit{Rank by weighted score with a soft bias (no hard zero) and deterministic tie-break:}\\
Define $\textbf{Score}(c',h)$ as
\[
\textbf{Score}(c',h)=
\begin{cases}
\lambda_1\hat{E}+\lambda_2\hat{T}+\lambda_3\hat{P} - \delta(h) & \text{if } \lambda_1\!\ge\!0.8 \text{ or } \text{qubitsReq}(c')\!>\!10,\\[2pt]
\lambda_1\hat{E}+\lambda_2\hat{T}+\lambda_3\hat{P} & \text{otherwise,}
\end{cases}
\]
where $\delta(h)=0.05$ if $h$ is trapped-ion and $0$ otherwise\;

$(c^*,h^*) \gets \arg\min\limits_{(c',h)\in\mathcal{P}} \textbf{Score}(c',h)$ with tie-break by lexicographic order on $(\hat{E}, \hat{P}, \hat{T})$\;
\Return{$(c^*,h^*)$}

\SetKwProg{Fn}{Function}{:}{}
\Fn{\textbf{Feasible}$(c',h)$}{
  \If{$\text{qubitsReq}(c') > \text{qubitsAvail}(h)$}{\Return False}
  \If{$\text{gateSetIncompatible}(c',h)$}{\Return False}
  \If{$(1 - E(c',h)) < \tau$}{\Return False \tcp*{reject if estimated fidelity $<\tau$}}
  \Return True
}
\end{algorithm}

\begin{table}[h!]
\centering
\caption{Default parameters used in recommender scoring model.}
\label{tab:parameter_defaults}
\sffamily
\begin{tabular}{l c l}
\toprule
\textbf{Symbol} & \textbf{Default Value} & \textbf{Rationale} \\
\midrule
$\tau$          & 0.5                   & Majority-vote threshold; below this outcomes are unreliable \\
$\lambda_1$     & 0.6                   & Accuracy prioritised in scientific prototyping \\
$\lambda_2$     & 0.3                   & Latency weighted moderately \\
$\lambda_3$     & 0.1                   & Cost weighted least in early-stage research \\
\bottomrule
\end{tabular}
\end{table}

\subsection{Decoder (RQ1)}
The \textit{decoder} converts quantum measurement outcomes \(\mathcal{O}\) (either bitstrings or probability distributions) from the \textit{Execution} stage into a classical solution \(\mathcal{S}\). The decoding strategy depends on the problem tag \(\mathcal{P} \in \mathcal{T}\) and the encoding \(\epsilon\) specified in the QCF (see Equation~\ref{equation:decoder}).

For \textit{combinatorial optimisation problems} encoded as QUBO, bitstrings are interpreted as candidate solutions, validated against feasibility constraints, and ranked by objective value. In oracle-based search problems, the most frequently measured marked state is selected as the solution. In \textit{number-theoretic problems} such as Shor’s algorithm, post-processing techniques (e.g., continued fractions for order finding) map measurement results to integer factors. For \textit{arithmetic operations}, quantum registers are read directly as integers. To ensure robustness, the decoder also reports a confidence score (derived from outcome frequencies) and applies lightweight post-selection (e.g., majority voting or constraint checks). This guarantees that the final output is expressed in the classical format natural to the problem domain—such as graph partitions, tours, integer factors, or arithmetic results—while abstracting away quantum-specific details from end-users. In the next section, we evaluated the discussed modules of the proposed \textit{C2$\ket{\rm{Q}}$} framework with example problems.
\section{Proposed Framework Evaluation} \label{sec:evaluation}
In this section, we evaluate the performance of the proposed \textit{C2$\ket{\rm{Q}}$} framework and how well it addresses the proposed research questions (RQs). The goal of the evaluation is not to benchmark algorithmic efficiency or hardware performance, but to demonstrate the feasibility of end-to-end automation of the proposed framework. Accordingly, we design controlled experiments that validate each module of the framework under standardized settings. This positions our results as \textit{proof-of-concept (POC)} rather than comparative performance benchmarking. 
\subsection{Experimentation Setup} 
We conducted three primary experiments to answer the three RQs (see Section~\ref{sec:intro}):
\begin{itemize}
    \item \textbf{Experiment 1: Encoder Benchmark (RQ1)} — This experiment evaluates the parser and QCF translator modules in terms of problem classification, data extraction, and QCF translation (RQ1.1). A synthetic, labeled dataset of 434 Python programs and 100 JSON specifications covering all supported problem types was used. This dataset has been publicly available in the replication package~\cite{ye_khan_c2q_dataset_2025}. Evaluation metrics include precision, recall, F1-score, and extraction accuracy, validating whether the encoder can reliably map diverse classical inputs into structured QCF representations suitable for circuit generation (RQ1.2).
    \item \textbf{Experiment 2: Deployment Evaluation (RQ2)} — This experiment assesses the hardware recommender by executing QAOA circuits for max-cut problem on 3-regular graphs of increasing size. Error, execution time, and cost are estimated across multiple quantum hardware devices (IBM, IonQ, Rigetti, IQM, Quantinuum) using calibration data and provider cost models. The recommender’s weighted scoring algorithm is applied to balance fidelity, latency, and cost, demonstrating how the framework selects quantum hardware under default and user-defined parameter settings.
    \item \textbf{Experiment 3: Full Workflow Validation (RQ3)} — This experiment evaluates the complete framework using the same dataset as mentioned above. The workflow consists of classical problem specification, data extraction, QCF translation, algorithm selection, circuit generation, deployment, and result interpretation. As a detailed case study, the Maximum Independent Set (MIS) problem is used to illustrate the end-to-end process. To establish a baseline, the framework’s programming interfaces are compared with manual Qiskit implementations, ensuring that the baseline reflects domain knowledge and best practices and enabling a rigorous assessment of correctness, algorithm selection effectiveness, and overall usability using reproducible proxy-based workload indicators.

\end{itemize}

A summary of the experimental tools used across the three experiments is provided in Table~\ref{tab:evaluation_tools}, including simulators, hardware devices, and datasets. Due to the high cost and limited accessibility of current NISQ devices, large-scale quantum program execution remains impractical for routine evaluation. Moreover, simulators offer controlled noise models and deterministic seeding, enabling ablation-style analyses and fair comparisons that are difficult to achieve on time-shared hardware. Accordingly, all benchmark inputs used for full workflow validation, comprising 434 Python code snippets and 100 JSON specifications, were conducted on Qiskit \textit{Aer}~\cite{javadiabhari2024quantumcomputingqiskit}, which provides a scalable, noise-configurable environment. To complement simulator-based testing, one selected representative instance (see Section~\ref{ex:ex3}) was also executed on Finland’s superconducting quantum computer, \textit{Helmi}~\cite{csc_helmi_specs}, and IBM’s \textit{ibm\_brisbane} backend~\cite{ibm_quantum}, providing external validity that the generated circuits are executable on real devices. Unless otherwise specified, all circuit executions in this evaluation are performed with 1024 shots.

\begin{table}[h!]
\centering
\caption{Summary of evaluation tools and components for the three experiments.}
\label{tab:evaluation_tools}
\sffamily
\begin{tabular}{ll}
\hline
\textbf{Component} & \textbf{Tool/Source} \\
\hline
Simulator (primary) & Qiskit \textit{Aer} \\
Hardware (representative) & \textit{Helmi} (5 qubits), IBM \textit{ibm\_brisbane} (127 qubits) \\
Dataset & 434 Python code snippets and 100 JSON inputs covering supported problems \\
\hline
\end{tabular}
\end{table}

The evaluation metrics employed to assess each experiment are summarised in Table~\ref{tab:eval_metrics_all}. These include standard classification measures such as precision, recall, F1 score~\cite{powers2020evaluationprecisionrecallfmeasure}, and encoder completion rate in Section~\ref{ex1}, device-related metrics such as error rates, qubit availability, execution time, and cost in Section~\ref{sec:ex2}, and system-level measures of correctness and implementation workload, assessed using proxy-based indicators (e.g., handwritten lines of code and explicit manual configuration decisions),
in Section~\ref{ex:ex3}.  



\begin{table}[h!]
\centering
\caption{Evaluation metrics across the three experiments.}
\label{tab:eval_metrics_all}
\renewcommand{\arraystretch}{1.2}
\begin{tabular}{p{0.28\textwidth} p{0.28\textwidth} p{0.35\textwidth}}

\multicolumn{1}{c}{\textbf{Experiment 1}} &
\multicolumn{1}{c}{\textbf{Experiment 2}} &
\multicolumn{1}{c}{\textbf{Experiment 3}} \\
\multicolumn{1}{c}{\textit{Encoder Evaluation}} &
\multicolumn{1}{c}{\textit{Deployment Evaluation}} &
\multicolumn{1}{c}{\textit{Full Workflow Validation}} \\ \hline

\textbf{Metrics} &
\textbf{Metrics} &
\textbf{Metrics} \\

Precision &
Execution Time &
Completion Rate \\

Recall &
Qubit Availability &
Lines-of-Code (LOC) \\

F1-Score &
Error Rate &
Configuration Decisions \\

Encoder Completion Rate &
Cost Estimation &
-- \\

\end{tabular}
\end{table}

\subsection{Experiment 1: Encoder Evaluation (RQ1)} \label{ex1}
To evaluate the effectiveness of the \textit{encoder}, we assessed its three core components: the \textit{parser}, the \textit{QCF translator}, and the \textit{generator}. The experiment measured the ability to (i) correctly classify classical problem types from source code, (ii) accurately extract and translate the associated data into quantum-compatible formats (QCFs) (RQ1.1), and (iii) automatically map these QCFs to appropriate quantum algorithms and generate the corresponding quantum circuits (RQ1.2). The evaluation used a dataset of 434 Python code snippets and 100 JSON inputs, labelled according to the supported problem types defined in Table~\ref{table:problem_summary}. Accordingly, Experiment~1 addresses RQ1.1 (problem classification, data extraction and QCF translation), RQ1.2 (algorithm selection), and the overarching RQ1 (bridging classical inputs to quantum execution).

Across all supported problem types, the encoder achieved consistently high precision, recall, and F1-scores on the Python dataset. We report a weighted-average F1 of 0.982, which accounts for the varying number of test cases per problem type. The most challenging cases were \textsc{MaxCut} and \textsc{MIS}, which both rely on graph inputs and share structural similarities, leading to occasional misclassifications. In contrast, arithmetic and number-theoretic problems (e.g., \textsc{ADD}, \textsc{Factor}) achieved near-perfect results due to simpler input structures. On the additional set of 100 JSON inputs, the encoder reached 100\% classification accuracy, as expected given the well-structured nature of JSON specifications (see Table~\ref{table:parser_eval_results}).

The QCF translator showed comparable accuracy, since correct translation is conditional on successful data extraction. Algorithm selection and circuit generation also performed reliably across all problem types, with variational, oracle-based, and arithmetic algorithms instantiated as expected. As summarised in Table~\ref{table:parser_eval_results}, the overall encoder completion rate—covering data extraction, QCF translation, and circuit generation—was 93.8\%, remaining consistent with or slightly below classification precision (weighted-average F1). Detailed results on QCF translation, algorithm–circuit generation, and end-to-end workflow execution are reported in Section~\ref{ex:ex3}.

\begin{table}[h]
\centering
\caption{Encoder evaluation on 434 Python programs. “Encoder Completion” counts a case as successful only if it was correctly recognised \emph{and} its data were extracted and translated to QCF without error.}
\label{table:parser_eval_results}
\sffamily
\begin{tabular}{lccccc}
\hline
\textbf{Problem Type} & \textbf{Cases} & \textbf{Precision} & \textbf{Recall} & \textbf{F1-Score} & \textbf{Encoder Completion (\%)} \\
\hline
MaxCut   & 68 & 0.942 & 0.956 & 0.949 & 94.1 \\
MIS      & 70 & 0.957 & 0.943 & 0.950 & 94.3 \\
TSP      & 44 & 1.000 & 1.000 & 1.000 & 93.2 \\
Clique   & 33 & 0.971 & 1.000 & 0.985 & 93.9 \\
KColor   & 58 & 1.000 & 1.000 & 1.000 & 94.8 \\
Factor   & 31 & 1.000 & 1.000 & 1.000 & 93.5 \\
ADD      & 28 & 1.000 & 1.000 & 1.000 & 92.9 \\
MUL      & 27 & 1.000 & 1.000 & 1.000 & 92.6 \\
SUB      & 28 & 1.000 & 1.000 & 1.000 & 92.9 \\
VC       & 47 & 1.000 & 0.979 & 0.989 & 93.6 \\
\hline
\textbf{Overall} & \textbf{434} & -- & -- & \textbf{0.982} & \textbf{93.8} \\
\hline
\end{tabular}
\end{table}

\begin{rqsummary}[title=Summary to RQ1 — Bridging classical and quantum programming]
\begin{itemize}[leftmargin=*]
    \item \textbf{Bridging classical and quantum software development:} A modular architecture is needed to translate classical problem specifications into formats directly usable by quantum algorithms. In \textit{C2$\ket{\rm{Q}}$}, this role is realised by the encoder, which combines classification, data extraction, QCF translation, and algorithm instantiation. This lowers the entry barrier for classical developers.

    \item \textbf{Encoder effectiveness (RQ1.1):} The encoder achieved a weighted-average F1 of 0.982 on 434 Python programs. Data extraction accuracy aligned with classification precision, confirming reliable QCF generation once the problem type was recognised. Misclassifications occurred mainly between \textsc{MaxCut} and \textsc{MIS}. On 100 JSON inputs, accuracy was 100\%.

    \item \textbf{Algorithm selection and configuration (RQ1.2):} Given valid QCFs, the generator consistently instantiated the expected algorithms. Combinatorial optimisation problems were mapped to QAOA and VQE, while oracle-based problems used Grover. Arithmetic and number-theoretic tasks were likewise mapped to their correct templates. This shows algorithm selection and circuit generation can be automated without manual intervention.
\end{itemize}
\end{rqsummary}

\subsection{Experiment 2: Deployment Evaluation (RQ2)} \label{sec:ex2}
To evaluate the deployment module of the framework, we executed Quantum Approximate Optimization Algorithm (QAOA) circuits on the Max-Cut problem, one of the supported problem types (see Table~\ref{table:problem_summary}), using 3-regular graphs as representative instances to provide scalable and realistic benchmarks. Max-Cut is a widely used benchmark in quantum optimisation~\cite{farhi2014} because (i) its qubit requirements scale directly with graph size, (ii) it naturally maps to QAOA formulations, and (iii) the generated circuits are sufficiently deep to expose realistic trade-offs in fidelity, runtime, and cost on NISQ devices~\cite{crooks2018performancequantumapproximateoptimization}. In this evaluation, we constructed QAOA circuits with depth $p=1$ \textit{(intentionally lower than the generator default of $p=3$ in Table~\ref{table:algorithm_configurations}, to reduce depth on NISQ hardware and isolate hardware-recommendation effects)}, executed $50$ rounds of classical parameter optimisation~\cite{farhi2014}, and performed $1000$ measurement shots per iteration. The tests began with 4-vertex graphs and scaled upward in steps of 2 (i.e., 4, 6, 8, …, 60) until the qubit requirements exceeded device capacities. This experimental setup ensured comparability across heterogeneous backends while exercising the recommender’s weighted evaluation pipeline.

\begin{figure}[h]
    \centering
    \begin{subfigure}{0.49\textwidth}
        \centering
        \includegraphics[width=\linewidth]{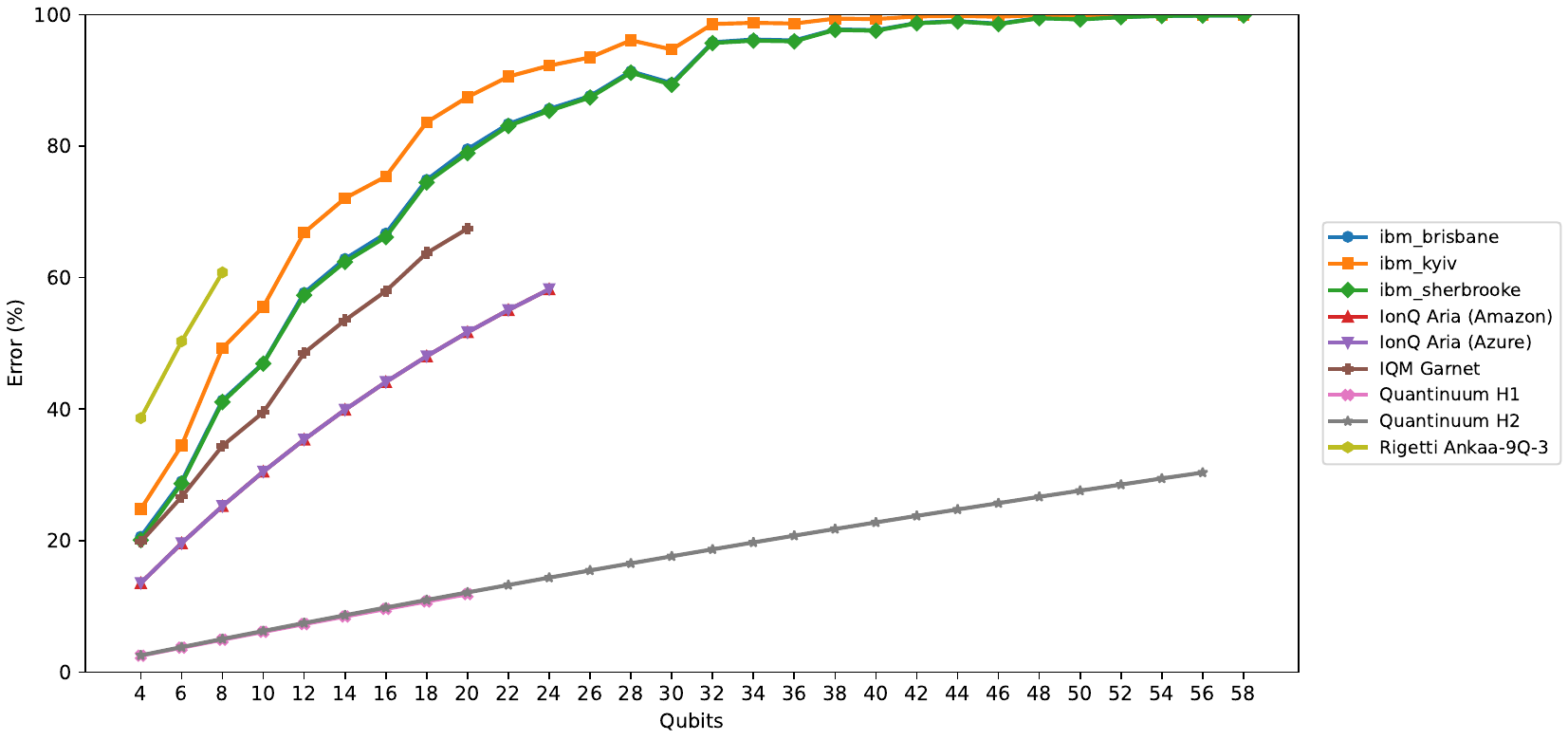}
        \caption{Estimated error rates.}
        \label{fig:rec_errors}
    \end{subfigure}
    \hfill
    \begin{subfigure}{0.49\textwidth}
        \centering
        \includegraphics[width=\linewidth]{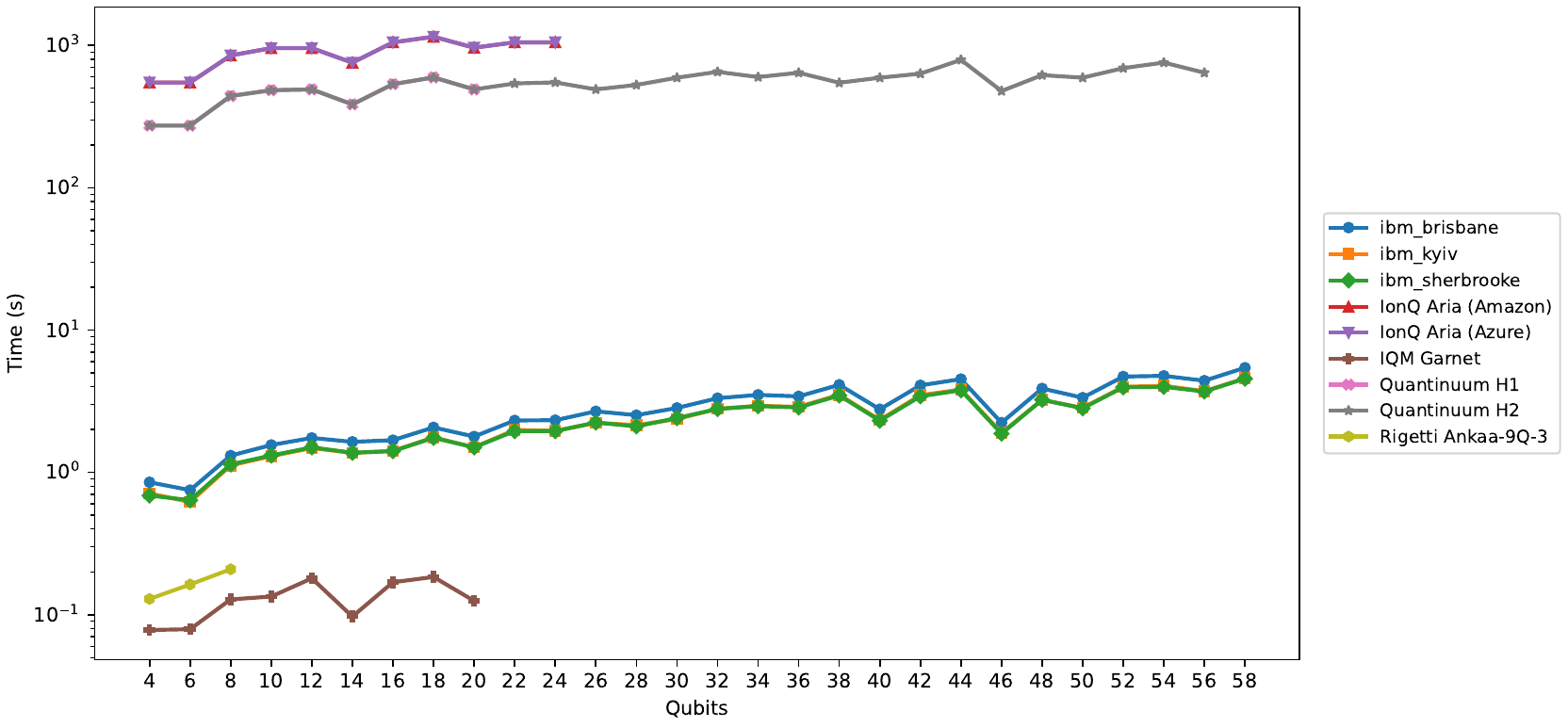}
        \caption{Estimated execution time.}
        \label{fig:reco_time}
    \end{subfigure}
    
    \vspace{0.5em}
    
    \begin{subfigure}{0.65\textwidth}
        \centering
        \includegraphics[width=\linewidth]{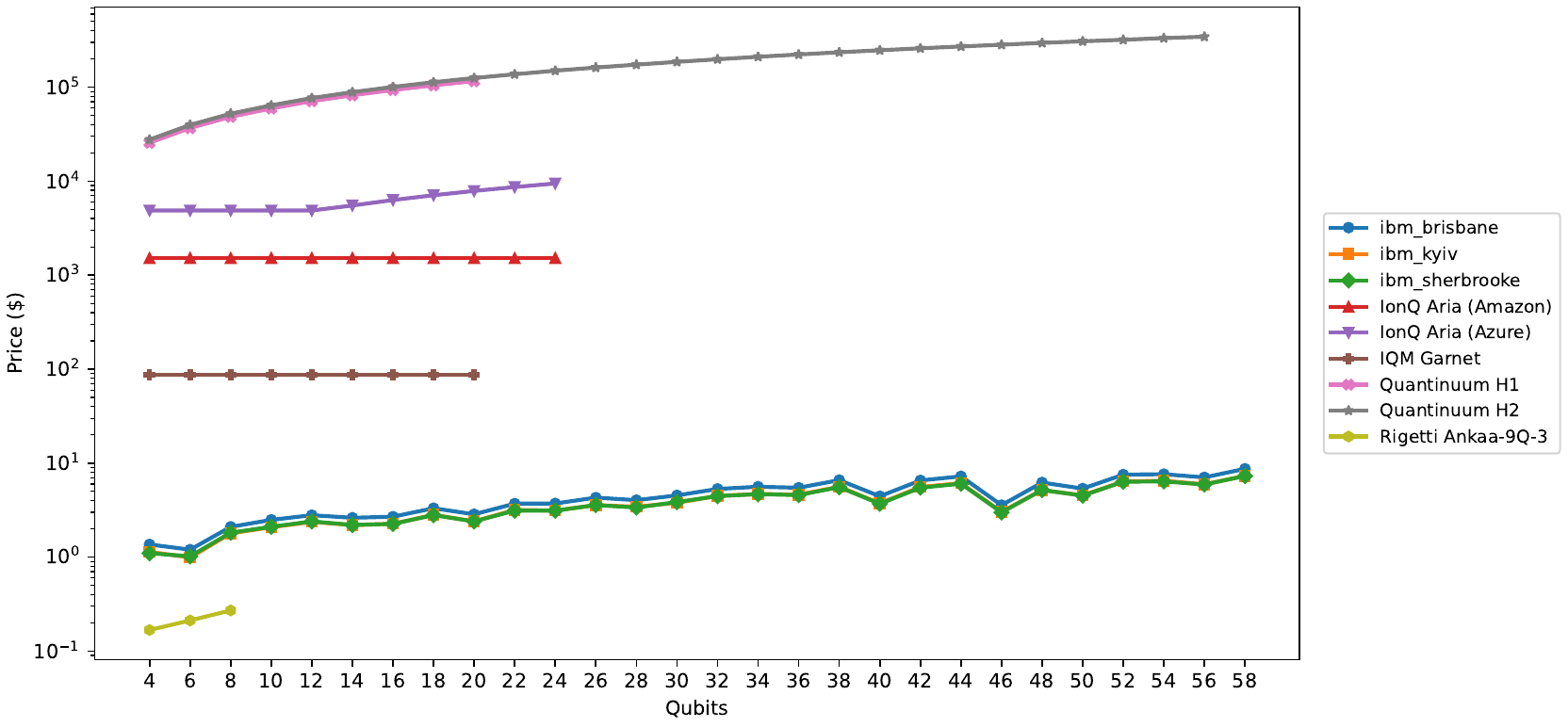}
        \caption{Estimated execution cost.}
        \label{fig:rec_prizes}
    \end{subfigure}

    \captionsetup{belowskip=0pt}
    \caption{Comparison of supported devices under QAOA workloads as a function of problem size in qubits: 
    (a) projected error rates, (b) execution time, and (c) execution cost. 
    Together, these metrics demonstrate the trade-offs that guide hardware recommendation.}
    \label{fig:recommender_metrics}
\end{figure}

Figure~\ref{fig:recommender_metrics} summarises the recommender’s evaluation of supported devices under QAOA workloads with respect to error rate, execution time, and cost, following the evaluation criteria defined in Section~\ref{Sec:Deployment}. Figure \ref{fig:recommender_metrics}(a) reports the projected error rates computed using the compound error model defined in Equation~\ref{eq:error}. The distinction between superconducting and trapped-ion devices is clear: trapped-ion platforms (Quantinuum, IonQ) scale with much lower error accumulation, whereas all superconducting devices (IBM, Rigetti, IQM) exceed the $\tau=0.5$ threshold beyond 12--14 qubits. For example, Quantinuum H1 maintains fidelity (defined as $F = 1 - \text{error rate}$) of approximately $90\%$ at 16 qubits, while \textit{ibm\_brisbane} falls below $35\%$. This gap directly impacts the feasibility of solving medium-to-large graph instances.

Figure \ref{fig:recommender_metrics}(b) compares execution times. Superconducting devices such as \textit{ibm\_brisbane} and \textit{ibm\_sherbrooke} achieved sub-second runtimes for small problems (e.g., $\sim$0.5--1.0\,s at 6 qubits), but degraded to 2--3\,s once circuits exceeded 30 qubits due to limited connectivity and the resulting SWAP/transpilation overheads~\cite{murali2019noiseadaptivecompilermappingsnoisy}. In contrast, trapped-ion devices like Quantinuum H1 showed much higher per-gate latency, already requiring $\sim$270\,s at 6 qubits and scaling steeply toward its 20-qubit limit. These differences reflect fundamental architectural trade-offs: superconducting qubits operate with nanosecond-scale gates~\cite{Kjaergaard_2020} but require additional operations to compensate for restricted connectivity, whereas trapped-ion systems perform gates more slowly (microseconds to milliseconds) but benefit from all-to-all connectivity that reduces circuit depth~\cite{Linke_2017}.

Figure \ref{fig:recommender_metrics}(c) presents the estimated execution cost. Here the contrast is most pronounced: trapped-ion devices, while highly accurate, incur prohibitively high costs---often thousands of dollars per 1000-shot experiment---making them quite impractical for routine academic use. By contrast, IBM superconducting devices remain relatively affordable (tens of dollars per experiment), though at the expense of fidelity. Rigetti Ankaa offers the lowest overall cost (\$0.09 at small scale), but its high error rates make it unsuitable for precision-sensitive workloads.

Using the default weights $(\lambda_1{=}0.6,\lambda_2{=}0.3,\lambda_3{=}0.1)$ in Algorithm~\ref{alg:recommender} and a fidelity threshold of $\tau{=}0.5$, circuit–device pairs with estimated fidelity below 50\% are filtered out, and the remaining candidates are ranked by their weighted scores. Table~\ref{tab:rec_device_choices} summarizes the selected devices across different qubit ranges. For small problems (4–8 qubits), the recommender selects \emph{Quantinuum H1}—although \emph{IonQ Aria} offers lower cost, H1's superior fidelity dominates under the default weighting. For medium-sized problems (10–20 qubits), H1 remains the top-ranked device. For larger problems (22–56 qubits), \emph{Quantinuum H2} is selected, as H1 exceeds its 20-qubit capacity, and although Aria remains feasible up to 25 qubits, H2 consistently yields a higher composite score. Beyond 56 qubits, no supported device satisfies the fidelity constraint. Full device-level scores and selection details are available in the replication package~\cite{ye_khan_c2q_dataset_2025}.
\begin{table}[h]
\centering
\caption{Devices selected by the recommender under default parameters 
$(\lambda_1{=}0.6,\;\lambda_2{=}0.3,\;\lambda_3{=}0.1,\;\tau{=}0.5)$,
respecting device qubit caps (H1: 20, H2: 56, Aria: 25), and considering qubit counts from 4 to 60 in steps of 2. 
\textit{Note: Results are based on the independent error model (Equation~\ref{eq:error}); while adequate as a first-order estimate, correlated and coherent errors on NISQ devices may lead to higher effective error rates in practice.}}
\label{tab:rec_device_choices}
\renewcommand{\arraystretch}{1.15}
\small
\resizebox{\textwidth}{!}{%
\begin{tabular}{c c p{9.2cm}}
\toprule
\textbf{Qubit Count} & \textbf{Selected Device} & \textbf{Rationale (aligned with error, time, and cost trends)} \\
\midrule
4--8   & Quantinuum H1 
& All devices except Ankaa-9Q-3 satisfy the fidelity threshold. Among the high-fidelity candidates, Aria offers lower cost but slower execution. Under the default weights $(\lambda_1{=}0.6)$, H1's lower error rate dominates the score, leading to its selection. \\

10--20  & Quantinuum H1 
& In this qubit range, trapped-ion devices significantly outperformed superconducting devices, as almost no superconducting backends maintain fidelity above 50\%. For $n{>}10$, the soft trapped-ion bias is active but applies uniformly to H1, H2, and Aria, preserving their relative ranking. H1 remains within its 20-qubit capacity and offers the lowest error among feasible devices, outweighing its higher latency and cost. \\

22--56 & Quantinuum H2 
& H1 exceeds its capacity ($>20$ qubits). H2 remains feasible up to 56 qubits and achieves the best score in this range; Aria is feasible only up to 25 qubits and is outscored by H2 for 22--25, while superconducting backends are filtered by the error model. \\

$>$56 & \emph{None (no feasible device)} 
& All supported devices exceed practical qubit feasibility (H2 max 56; Aria 25; H1 20). \\
\bottomrule
\end{tabular}}
\end{table}
\FloatBarrier
\begin{rqsummary}[title=Summary to RQ2 — Deployment evaluation]
Using the default weights $(\lambda_1{=}0.6,\lambda_2{=}0.3,\lambda_3{=}0.1,\tau{=}0.5)$ and actual device caps (H1:20, H2:56, Aria:25), the recommender selects \emph{Quantinuum H1} for both 4–8 and 10–20 qubits, and \emph{Quantinuum H2} for 22–56 qubits. No device is feasible beyond 56 qubits under the default threshold ($\tau{=}0.5$) and circuit depth ($p{=}1$). This behaviour follows the scoring function’s soft preference (Algorithm~\ref{alg:recommender}), which biases selection toward trapped-ion systems when fidelity dominates or qubit counts exceed 10. When user weights shift toward latency or cost, IBM superconducting devices become preferable at small sizes. Overall, selection is principled, capacity-aware, and tunable to user priorities.
\end{rqsummary}
\FloatBarrier

\subsection{Experiment 3: Full Workflow Evaluation (RQ3)} \label{ex:ex3}

The experimentation discussions in Sections~\ref{ex1} and \ref{sec:ex2} evaluated the individual modules (encoder, deployment) of the framework. In this section, we assess the complete end-to-end workflow of the proposed framework by jointly evaluating all modules in an integrated setting. All benchmark inputs, comprising 434 Python code snippets and 100 JSON specifications used in Experiment~1, were executed end to end on a quantum circuit simulator (Qiskit \textit{Aer}) to support scalability, reproducibility, and controlled evaluation. In addition, one representative instance of the Maximum Independent Set (MIS) problem was deployed on real quantum hardware on both Finland’s \textit{Helmi} processor~\cite{csc_helmi_specs} and IBM’s \textit{ibm\_brisbane} backend~\cite{ibm_quantum}, to validate the physical executability of the full pipeline under practical device constraints. 

Following the above setting, Table~\ref{tab:usability_problems} summarises the problem types, instance sizes (which determines the required number of qubits), and their respective categories, covering combinatorial optimisation, number-theoretic, and arithmetic tasks. The full dataset and a detailed report of all test cases have been publicly available in the replication package~\cite{ye_khan_c2q_dataset_2025}.

\begin{table}[h]
\centering
\caption{Benchmark problems and instance sizes used for evaluating full framework.}
\label{tab:usability_problems}
\sffamily
\begin{tabular}{l c c}
\toprule
\textbf{Problem Type} & \textbf{Instance Sizes} & \textbf{Category} \\
\midrule
MaxCut                      & 3--8 nodes                   & Combinatorial optimisation \\
Maximum Independent Set (MIS)       & 3--8 nodes                   & Combinatorial optimisation \\
Traveling Salesman (TSP)    & 3--6 nodes                   & Combinatorial optimisation \\
Clique                      & 3--6 nodes                   & Combinatorial optimisation \\
K-Coloring (KColor)         & 3--6 nodes                   & Combinatorial optimisation \\
Vertex Cover (VC)           & 3--6 nodes                   & Combinatorial optimisation \\
Integer Factorisation (Factor)       & 2--10 bit semi-primes        & Number-theoretic \\
Addition (ADD)              & 2--10 bit operands           & Arithmetic \\
Multiplication (MUL)        & 2--10 bit $\times$ 2--10 bit operands & Arithmetic \\
Subtraction (SUB)           & 2--10 bit operands           & Arithmetic \\
\bottomrule
\end{tabular}
\end{table}

\vspace{1em}




The workflow achieved a completion rate for Python inputs that was identical to the encoder completion rates reported in Table~\ref{table:parser_eval_results}. Here, the \emph{completion rate} refers to the percentage of cases where the entire pipeline—from classical specification through QCF translation, algorithm generation, deployment, and decoding—produced a valid output. Once the encoder successfully processed a problem, the subsequent stages proceeded without error. For JSON inputs, the completion rate was 100\%, consistent with the results of Section~\ref{ex1}, reflecting the reliability of structured specifications.

To illustrate the evaluation in practice, we present the Maximum Independent Set (MIS) problem as a representative case study. Since reporting the findings of all cases in detail is impractical, the complete execution results for all problem types are available in the replication package~\cite{ye_khan_c2q_dataset_2025}, while supporting artefacts for the MIS case are provided in Appendix~\ref{app:json}. MIS is a classical combinatorial optimisation problem that naturally maps to both QUBO and oracle encodings, making it well suited for validating the integration of problem parsing, QCF translation, algorithm generation, device recommendation, and execution. Listing~\ref{lst:mis_helmi} illustrates the use of the programming interface to solve a Maximum Independent Set (MIS) problem from a classical Python input (Listing~\ref{lst:mis_classical}). The encoder classifies the input as an MIS instance and extracts the corresponding problem data, as shown in Figure~\ref{fig:mis_case}a. The same MIS task can alternatively be specified using a structured JSON input (Listing~\ref{lst:json_mis}) and executed via the command-line interface, as demonstrated in Listing~\ref{lst:json_cli}.

    

From this experiment, the encoder successfully translated the input into the problem’s \textit{Quantum-Compatible Formats (QCFs)}: its oracle circuit (See Figure~\ref{fig:qcf_oracle}) and QUBO matrix representation (see Figure~\ref{fig:qubo_matrix}). These QCFs form the basis for the next stage, where the generator constructs executable quantum circuits. For the combinatorial optimisation problem MIS, the generator defaults to variational approaches such as QAOA and VQE, while also supporting Grover’s algorithm when an oracle formulation is available (see Table~\ref{table:problem_summary}).

The deployment stage transpiles the generated circuits (QAOA, VQE, and Grover). For demonstration, we follow the QAOA branch; an example QAOA circuit is shown in Figure~\ref{fig:qaoa_circuit_theta} and evaluates each circuit--device pair using the recommender’s weighted scoring algorithm (Algorithm~\ref{alg:recommender}). With the default parameters $(\lambda_1=0.6,\;\lambda_2=0.3,\;\lambda_3=0.1,\;\tau=0.5)$ (Table~\ref{tab:parameter_defaults}), fidelity is prioritised while latency and cost are considered moderately. Using the Maximum Independent Set (MIS) problem as a representative benchmark, this process selected the QAOA circuit on \textit{Quantinuum H1} as the optimal configuration (best score 0.311). Figure~\ref{fig:qaoa_comparison_app} highlights the trade-offs: (a) trapped-ion devices (Quantinuum, IonQ) achieve the lowest error rates, with Quantinuum H1 maintaining $96.10\%$ fidelity (equivalent to only $3.90\%$ total error); (b) superconducting devices provide much faster execution times but incur additional overheads from limited connectivity, while ion-trap platforms remain more accurate at larger qubit counts; (c) execution costs are lowest on superconducting platforms (e.g., IBM devices at $\sim\$2$ per run), whereas trapped-ion devices such as Quantinuum H1/H2 reach tens of thousands of dollars, creating several orders of magnitude difference.

Execution was performed on both simulators and real hardware backends. 
For the MIS instance in Figure~\ref{fig:mis_case}, the generated QAOA circuits were executed on Finland’s superconducting quantum computer \textit{Helmi} and IBM’s \textit{ibm\_brisbane}, in addition to simulator runs. The decoder then processed the raw measurement bitstrings and identified the most probable outcome, \texttt{1100} (little-endian), corresponding to the quantum state \(|1100\rangle\). This state indicates that vertices $\{2,3\}$ form the independent set. Figure~\ref{fig:mis_case}(b) presents this decoded output, illustrating how the decoder completes the workflow by mapping low-level quantum measurements into a human-readable solution.



\begin{rqsummary}[title=Summary to RQ3 — Generalisability and effectiveness]
The evaluation demonstrates that the framework generalises across all supported problem types and both input modalities (Python code and JSON). It successfully integrated variational, oracle-based, and arithmetic algorithms, and executed them on simulators as well as real hardware (\textit{Helmi} and \textit{ibm\_brisbane}). Across this test suite, the workflow achieved a completion rate of \textbf{93.8\%} (Python) and \textbf{100\%} (JSON), produced consistent results across problem types and encodings, and maintained performance within the limits of NISQ devices. A proxy-based usability analysis further indicates a substantial reduction in handwritten lines of code and explicit manual configuration steps across problem families, supporting the framework’s practical effectiveness alongside its technical generalisability.
\end{rqsummary}


To evaluate usability in terms of observable implementation workload, we adopt proxy-based measures that approximate the explicit work required to specify, configure, and execute quantum programs using the framework. Following ISO~9241-11~\cite{iso2018ergonomics}, which frames usability in terms of effectiveness and efficiency in a specified context of use, we interpret efficiency through measurable workflow-level indicators. In this study, implementation workload is operationalised using two concrete proxies: (i) handwritten lines of code (LoC), a widely used size-related indicator in software cost estimation models such as COCOMO~\cite{boehm1984software}, a classical empirical effort estimation model that relates development effort to program size; and (ii) the number of explicit manual configuration decisions required to obtain an executable quantum program. Importantly, the framework does not eliminate modelling or algorithmic decisions; rather, it encapsulates predefined configuration logic within its implementation, such that these decisions are no longer developer-visible configuration steps. To the best of our knowledge, there are currently no widely adopted effort estimation models specifically tailored to quantum software development, which motivates the use of established size-based proxies. While LoC does not directly measure cognitive effort and may be imperfect in contexts involving extensive code reuse, in quantum SDK workflows much of the handwritten code corresponds to explicit modelling, encoding, and backend configuration steps that must be specified and validated by the developer. In this context, LoC serves as a transparent and reproducible indicator of implementation workload.

The evaluation spans all supported problem types listed in Table~\ref{table:problem_summary}, with one representative instance selected per type. Each instance is implemented using a quantum algorithm supported by the framework and appropriate for the corresponding problem type (e.g., Grover’s algorithm for MIS), ensuring consistent problem-algorithm mapping across the study. For each instance, we compare a manual implementation using a mainstream quantum SDK (Qiskit)~\cite{javadiabhari2024quantumcomputingqiskit} with the corresponding implementation produced by \textit{C2$\ket{\rm{Q}}$}. Manual implementations explicitly perform all required development steps, including problem encoding, circuit construction, backend configuration, execution, and classical post-processing.

In this evaluation, \emph{manual configuration decisions} are defined at a coarse but consistent level of abstraction as explicit, developer-visible choices associated with a predefined list of high-level development steps (e.g., constraint encoding, ansatz configuration, backend and execution configuration), as shown in Table~\ref{tab:usability_comparison}, rather than individual API calls or low-level implementation details. The identification of manual configuration decisions was performed by the first author by mapping each manual implementation step to this predefined list of high-level development choices. This validation step was necessary because some manual SDK configuration decisions (e.g., encoding selections and parameter settings) are implicit and cannot be reliably reconstructed from the resulting circuit artifacts or outputs alone. To ensure technical correctness and fairness, the manual SDK implementations were first written by the first author and independently reviewed by the second author. For each problem instance, both the manual SDK implementation and the corresponding \textit{C2$\ket{\rm{Q}}$} workflow were executed on identical inputs, and their final classical outputs were directly compared. The second author independently verified both output correctness and the counting of configuration decisions, with disagreements resolved through discussion involving the fourth author. While this protocol strengthens the internal validity, we acknowledge that manual review may introduce residual human bias and does not substitute for fully automated large-scale verification. For replicability, all problem instances and manual SDK implementations have been made publicly available in the replication package~\cite{ye_khan_c2q_dataset_2025}.

As shown in Table~\ref{tab:usability_comparison}, the comparison reveals a consistent reduction in handwritten code volume and explicit configuration steps across all supported problem types when using \textit{C2$\ket{\rm{Q}}$} instead of manual quantum SDK implementations. Across the ten representative cases, manual implementations require on average approximately 70--75 lines of handwritten code (mean $\approx$73~LoC), whereas the corresponding \textit{C2$\ket{\rm{Q}}$} usage reduces the interaction to a small number of high-level pipeline statements. In particular, the core workflow consists of three statements for problem parsing, instantiation, and result reporting, as shown in Listing~\ref{lst:mis_helmi} (Appendix~\ref{app:json}). An additional one to two lines are required when explicitly linking to a user account and selecting a specific execution platform (e.g., IBM Quantum or \textit{Helmi}) for hardware execution. Excluding such optional platform-specific setup, this corresponds to an approximate $73{:}3$ reduction in handwritten code volume (i.e., roughly a \(20\times\)--\(25\times\) decrease) at the level of problem specification and execution orchestration. In addition, manual implementations involve on average around six explicit configuration decisions per problem instance. In contrast, these decisions are handled implicitly by the framework. Together, these quantitative differences provide a reproducible indication of reduced implementation workload in comparison to manual SDK workflows.

\begin{table}[htbp]
\centering
\caption{Indicative implementation workload metrics across supported problem types. Metrics include handwritten lines of code (LoC) and the number of explicit manual configuration decisions required to obtain an executable quantum program. Each instance is implemented using an appropriate quantum algorithm supported by the framework.}
\label{tab:usability_comparison}
\sffamily
\renewcommand{\arraystretch}{1.2}
\resizebox{\linewidth}{!}{
\begin{tabular}{l l c p{10cm}}
\toprule
\textbf{Problem Type} &
\textbf{Algorithm} &
\textbf{LoC} &
\textbf{Manual Configuration Decisions} \\
\midrule
ADD     & Full Adder      & 30  & operand encoding, register allocation, arithmetic circuit design, measurement mapping, backend and execution configuration, solution decoding \\
SUB     & Full Adder      & 33  & operand encoding, register allocation, arithmetic circuit design, measurement mapping, backend and execution configuration, solution decoding \\
MUL     & QFT Multiplier  & 43  & operand encoding, register allocation, QFT-based arithmetic construction, inverse QFT configuration, backend and execution configuration, solution decoding \\
Factor  & Grover         & 67  & candidate encoding, oracle construction, Grover parameter configuration, backend and execution configuration, solution decoding \\
MIS     & Grover         & 83  & constraint encoding, oracle construction, Grover parameter configuration, backend and execution configuration, solution decoding \\
MaxCut  & VQE            & 83  & constraint encoding, QUBO mapping, ansatz configuration, optimizer configuration, backend and execution configuration, solution decoding \\
TSP     & QAOA           & 81  & constraint encoding, QUBO mapping, circuit configuration, optimizer configuration, backend and execution configuration, solution decoding \\
Clique  & QAOA           & 100 & constraint encoding, QUBO mapping, circuit configuration, optimizer configuration, backend and execution configuration, solution decoding \\
KColor  & QAOA           & 115 & constraint encoding, variable expansion, QUBO mapping, circuit configuration, optimizer configuration, backend and execution configuration, solution decoding \\
VC      & VQE            & 92  & constraint encoding, QUBO mapping, ansatz configuration, optimizer configuration, backend and execution configuration, solution decoding \\
\midrule
\textbf{Average} & -- & \textbf{73} & \textbf{5.9 configuration decisions} \\
\bottomrule
\end{tabular}
}
\begin{flushleft}
\footnotesize
\textit{Note:} Manual configuration decisions are defined at a coarse but consistent granularity. Each listed item corresponds to a distinct, explicit choice that a developer must make in a manual quantum SDK workflow and that is handled implicitly by the \textit{C2$\ket{\rm{Q}}$} framework. Each decision subsumes multiple finer-grained tasks.
\end{flushleft}

\end{table}

\section{Discussion} \label{sec:discussion}
This section interprets the findings in Section~\ref{sec:evaluation}, and provides implications and logical understanding in relation to the research questions (RQ1--RQ3). 
Each subsection below addresses one RQ explicitly, linking observed evidence to the broader aims of the study.

\subsection{RQ1: Bridging classical and quantum programming}

The evaluation demonstrates that bridging classical and quantum programming is feasible through a modular encoder--deployment--decoder design. The encoder ensured that both Python code and structured JSON could be translated into quantum-compatible formats (QCFs), while the decoder reliably mapped quantum measurement outcomes back into classical solutions. This consistency across diverse input and output modalities is central for establishing confidence in the framework. It shows that the system can abstract away quantum-specific details without breaking the overall workflow. 

Another takeaway from the evaluation is modularity. Because encoding and decoding modules are separated from circuit generation and deployment, new problem types or algorithm families can be added without disrupting existing functionality. This separation of concerns strengthens maintainability and ensures that the framework can evolve alongside quantum hardware and algorithmic advances. Finally, the evaluation also revealed a trade-off between the expressiveness of user inputs and structural clarity required for reliable parsing. JSON-based problem specifications offer a high degree of structure, enabling consistent syntactic and semantic interpretation during data extraction and resulting in high reliability throughout the pipeline. In contrast, code snippets introduce variability in syntax, naming conventions, and control flow, but better reflect how developers naturally express problems. Supporting both formats allows the framework to balance robustness with flexibility, reinforcing the idea that a modular encoder–decoder pipeline is a practical strategy for bridging classical and quantum programming. From a design perspective, positioning the JSON interface as a lightweight, structured specification was essential for achieving reliable automation and reproducibility under current NISQ constraints.

In summary, the evaluation confirms that bridging classical and quantum programming is both technically feasible and practically valuable. The modular design ensures extensibility and maintainability, while support for both structured and unstructured inputs balances robustness with usability. Overall, modular encoder--decoder pipelines reduce the entry barrier for classical developers and provide a solid foundation for future hybrid classical--quantum software development.

\subsubsection{RQ1.1: Encoder effectiveness}

The parser and QCF translator form the foundation of the encoder. The evaluation shows that once the parser identified the correct problem type, subsequent data extraction and QCF translation succeeded reliably. Errors were largely confined to structurally similar graph problems such as \textsc{MaxCut} and \textsc{MIS}, showing that additional distinguishing features will be necessary when handling closely related domains.

Two lessons emerged during development. First, identifier normalisation proved essential: without it, the parser relied on problem-specific function names (e.g., \texttt{maxcut()}), which limited generalisation. Neutral identifiers forced the parser to learn structural patterns instead, leading to more robust behaviour. Second, the AST-based rule engine worked well on regular inputs such as explicit edge lists, but failed on dynamic constructs involving file I/O or helper functions. This exposed the limits of static-only parsing and suggests that hybrid parsing strategies could further improve the reliability of problem classification and data extraction. In contrast, JSON inputs achieved perfect accuracy, confirming the advantages of structured formats. Together, these findings confirm that the parser and QCF translator are effective for both unstructured and structured inputs, though with clear boundaries when inputs deviate from expected coding styles.

We also observed that several problem types, particularly \textsc{TSP} and arithmetic operations (\textsc{ADD}, \textsc{MUL}, \textsc{SUB}), achieved perfect classification scores. These results are a direct consequence of the framework’s modular design and the use of synthetic, well-structured datasets crafted to test baseline feasibility. By leveraging canonical representations and controlled syntax, the parser–translator pipeline was able to operate deterministically and with high precision. This outcome validates the soundness of the encoding process under ideal conditions and confirms that the framework can correctly and consistently handle a wide range of input formats when their structure is unambiguous. While these results demonstrate the robustness of the core design, we acknowledge that real-world code introduces variability and noise. To further test generalisability, future work will incorporate more diverse, developer-authored code snippets that reflect natural coding practices and less constrained input styles.

Overall, the parser and QCF translator are effective for mapping classical inputs into quantum-compatible formats. Identifier normalisation and rule-based extraction proved valuable design choices, though their limits suggest the need for hybrid parsing strategies. JSON inputs offered perfect accuracy, highlighting the strength of structured formats for reliability. While the parser does not support user-defined prompts during inference, we observed that its classification accuracy benefits from implicit guidance. For example, providing clearer variable names, avoiding deeply nested helper functions, and using regular structural patterns can improve recognizability. Alternatively, users can rely on the structured JSON input mode, which guarantees unambiguous interpretation and bypasses the variability inherent in free-form code.

\subsubsection{RQ1.2: Algorithm selection and configuration}

The generator consistently mapped QCFs to appropriate quantum algorithms. QUBO-based inputs invoked variational approaches such as QAOA or VQE, while oracle-based encodings triggered Grover. 
This behaviour was predictable and reproducible, showing that algorithm selection can be automated once a canonical QCF is available. For the framework, this means that developers do not need to intervene manually at the algorithm-selection stage, reducing potential points of failure and improving reproducibility of results. At the same time, the experiments revealed a boundary of the current design. The generator relies on fixed default parameters (e.g., QAOA depth $p=3$), chosen as hardware-agnostic settings to ensure robustness and reproducibility across problem families and NISQ devices, rather than to optimise algorithmic performance based on problem-specific QCF characteristics. While this sufficed to demonstrate correctness and workflow integration, achieving competitive performance on real hardware would require adaptive parameterisation or hybrid tuning strategies informed by the concrete input QCF. Thus, the findings illustrate both the strength of automated algorithm mapping and the limitations of the present implementation.

The findings of RQ1.2 confirm that algorithm selection can be automated once problems are expressed in QCFs, reducing manual effort and improving reproducibility. However, reliance on fixed default parameters limits performance on real hardware, indicating that future extensions should explore adaptive or hybrid tuning strategies.

\subsection{RQ2: Deployment evaluation}

The evaluation of QAOA circuits on the Max-Cut problem (see Experiment~2 in Section~\ref{sec:ex2}) shows that no single hardware platform dominates across fidelity, runtime, and cost. Superconducting devices such as \textit{ibm\_brisbane} and \textit{ibm\_sherbrooke} achieved sub-second runtimes at small scales (e.g., $\sim$0.5--1.0\,s at 6 qubits), but their error rates increased rapidly as problem size grew, making larger circuits infeasible. In contrast, trapped-ion platforms such as Quantinuum H1 sustained much lower error rates, maintaining fidelity above 80\% even beyond 15 qubits, but incurred runtimes of $\sim$100\,s at 6 qubits and scaling above $10^5$\,s by 60 qubits. IonQ Aria exhibited intermediate behaviour, with moderate fidelity and runtimes 
around $10^3$\,s for 10--30 qubits, but with execution costs exceeding \$1000 per run, rendering it practically unaffordable for routine use. These results highlight fundamental architectural trade-offs: superconducting systems offer speed but suffer from connectivity overheads, whereas trapped-ion systems trade latency for accuracy and scalability. In both cases, real execution on current NISQ devices remains limited by access restrictions and cost barriers.

In this context, the recommender provides a structured mechanism to navigate hardware heterogeneity. By combining device calibration data, pricing models, and user-defined weights into a single scoring function (see Algorithm~\ref{alg:recommender} in Section~\ref{sec:framework}), it ranks device--circuit pairs and filters infeasible options. This enables developers to select the quantum hardware without parsing low-level calibration tables or platform-specific policies. Although physical quantum devices are accessed through provider-specific platforms (e.g., IBM or IQM), the proposed framework remains hardware-agnostic at the architectural level. The workflow is explicitly separated into two phases: (i) a hardware-independent planning phase, which includes problem interpretation, algorithm selection, logical circuit construction, and hardware recommendation, and (ii) a backend-specific execution phase, where account authentication, access credentials, and provider-specific execution interfaces are handled. All planning and decision-making are performed on hardware-independent logical circuit representations, without committing to any specific quantum device. Platform-specific SDKs are invoked only at the final execution stage to transpile and execute circuits on concrete quantum hardware.

From a software engineering perspective, the implication is that hardware selection is no longer a manual bottleneck. Developers can focus on problem specification, while the \textit{C2$\ket{\rm{Q}}$} framework transparently manages the trade-offs between fidelity, latency, and cost. This automated abstraction of hardware-level decisions directly addresses RQ2 by demonstrating that automated recommendation not only hides platform-specific complexity but also provides interpretable, actionable guidance for deploying quantum programs under realistic NISQ constraints.

\subsection{RQ3: Generalisability and effectiveness}

The evaluation of the complete workflow shows that the \textit{C2$\ket{\rm{Q}}$} framework generalizes effectively across supported problem types, including combinatorial optimization number-theoretic, and arithmetic tasks. It also demonstrates strong generalisability across heterogeneous quantum hardware: the framework generated valid outputs on all supported quantum devices, and executed representative problems on real quantum processors including \textit{ibm\_brisbane} and \textit{Helmi}.
For Python code inputs, the completion rate matched the parser’s data extraction accuracy, while JSON inputs achieved 100\% success, reflecting the robustness of structured specifications. These results indicate that once problem data is correctly extracted, the downstream stages of QCF translation, algorithm generation, deployment, and decoding proceed reliably without manual intervention.

Generalisability is further supported by the breadth of algorithms tested---variational (QAOA, VQE), oracle-based (Grover), and arithmetic kernels. Once a problem is expressed in a standardised QCF, the generator consistently produces executable circuits that could be deployed across different quantum devices. Importantly, this behaviour held true not only in simulator-based tests but also in real-device executions on Finland’s \textit{Helmi} processor and IBM’s \textit{ibm\_brisbane}, 
demonstrating that the workflow remains stable under practical device constraints.

A key result of the proxy-based usability test (see Experiment~3 in Section~\ref{ex:ex3}) is that the \textit{C2$\ket{\rm{Q}}$} framework substantially reduces handwritten lines of code and explicit configuration steps compared to manual quantum SDK implementations. Across all supported problem families, the comparison reveals a consistent gap between concise \textit{C2$\ket{\rm{Q}}$} specifications and conventional SDK workflows, with the latter demanding significantly more handwritten code and developer-visible configuration choices. These results indicate that the proposed framework lowers the practical entry barrier to quantum programming by enabling developers to prototype quantum applications without directly managing low-level SDK modelling and configuration tasks.

Together, these findings confirm that the \textit{C2$\ket{\rm{Q}}$} framework is not only technically generalisable across problem classes and input modalities but also practically effective in reducing developers' effort and supporting execution on heterogeneous quantum devices. This combination of technical generality and practical usability addresses RQ3 by showing that an encoder–deployment–decoder pipeline can provide robust, scalable, and developer-friendly automation under realistic NISQ conditions.

\section{Implications} \label{sec:implications}
The research and industrial implications of this study are as follows:
\subsection{Research Implications}

An important implication lies in \textit{bridging the gap between classical and quantum software development}. The \textit{C2$\ket{\rm{Q}}$} framework demonstrates how classical specifications—expressed as Python code or structured JSON—can be systematically translated into executable quantum programs. This shows that familiar classical artefacts can serve as valid entry points into quantum workflows, reducing the conceptual distance between established programming practice and quantum execution. In doing so, the work provides an empirical foundation for future studies on how software abstractions can make quantum computing more accessible to the broader software engineering community.

In addition, the study advances \textit{quantum software engineering through modular design}. By classifying the workflow into distinct modules and sub-modules---\textit{encoder} (parser, QCF translator, generator), \textit{deployment} (transpiler, recommender, execution), and \textit{decoder}---the framework establishes a clear methodological principle for modular quantum software development. This modular design provides two key benefits. First, it offers an empirical foundation for analyzing how each stage impacts end-to-end reliability. Second, it enhances reproducibility, since researchers can benchmark or extend individual components---such as the parser or recommender---without reimplementing the entire workflow. This flexibility improves the scalability of the framework. In this way, the proposed approach contributes to a systematic, module-based paradigm for quantum software engineering, supporting its advancement as an emerging research discipline.

A further implication concerns \textit{enabling QCF translation from classical problem formulations}. The \textit{C2$\ket{\rm{Q}}$} framework demonstrates how quantum-compatible formats (QCFs) can be systematically derived from classical specifications, thereby providing a foundation for future developments in hybrid classical–quantum workflows where data and problem structures must be shared across paradigms. Moreover, the framework contributes to \textit{automating quantum circuit generation}. The generator module constructs executable circuits based on the characteristics of the input problem and the selected algorithms, reducing the reliance on manual circuit construction. This feature supports researchers in quantum algorithm selection and comparison, while also highlighting opportunities for systematic benchmarking of algorithm–circuit mappings.

Finally, we developed a hardware recommender that automatically suggests suitable quantum devices by evaluating factors such as error rates, latency, and cost. This recommender abstracts hardware-specific details and provides a systematic way to select quantum devices, reducing the complexity of cross-platform quantum software development. Its design opens research opportunities in hardware-aware circuit adaptation, quantum compilation, and performance optimization across heterogeneous platforms. By supporting portability, scalability, and reproducibility, the recommender also contributes to building a more robust foundation for empirical studies in quantum software engineering.

\subsection{Industrial Implications}

A key industrial implication is \textit{lowering the barrier for quantum adoption}. 
The proposed \textit{C2$\ket{\rm{Q}}$} framework allows software development organizations---particularly small and medium-sized enterprises (SMEs)---to explore quantum solutions without requiring in-depth quantum expertise. 
By abstracting away quantum circuit construction and device-level complexity, the framework enables non-specialist software engineers to experiment with quantum-enabled applications in a structured way. 
This democratizes access to quantum computing and supports adoption in domains where classical optimisation and algebraic problems already play a critical role, thereby fostering broader industrial engagement with quantum technologies.

Another important contribution concerns \textit{hardware abstraction and future scalability}. 
The framework integrates multiple quantum devices into a unified recommender module, allowing developers to compare heterogeneous platforms such as \textit{IBM}, \textit{IonQ}, \textit{Rigetti}, \textit{Quantinuum}, and Finland’s \textit{Helmi} on the basis of error, runtime, and cost. 
This abstraction ensures that applications developed today can transparently scale with future devices, protecting early investments while enabling cost- and performance-aware decision-making as hardware capabilities evolve.

Finally, the framework supports \textit{hybrid classical--quantum workflows}. By enabling suitable problem components to be translated into quantum circuits while leaving the rest of the application in a classical stack, the framework allows industries to experiment with quantum computing in an incremental way. Since the current infrastructure for pure quantum computing is still limited, the real-world benefits of quantum technologies can presently be realized only through hybridization. This staged adoption model therefore provides a practical path for companies to explore quantum technologies without requiring a complete redesign of existing enterprise systems.

\section{Threats to Validity} \label{sec:validity}
Following the guidelines of Wohlin et al.~\cite{wohlin2012experimentation}, we discuss the potential threats to the validity of this study in four dimensions: external, internal, construct, and conclusion validity.

\subsection{External Validity}
External validity refers to the extent to which the study findings can be generalised to other settings, scenarios or problems. The following threats were observed in our study: One potential threat relates to the \textit{limited coverage of supported problems}. The current implementation supports a subset of problem classes, including combinatorial optimisation problems as well as arithmetic and algebraic computations. Although this scope does not cover the full spectrum of quantum-relevant problems, the selected problems represent widely studied benchmarks that capture key domains in quantum optimization and computation. The modular design of the framework further ensures extensibility: new problem classes can be integrated incrementally by extending the parser, QCF encodings, and algorithm pool, without requiring changes to the core architecture.

A second threat concerns \textit{how optimisation objectives are specified in practice}. In this study, tasks are expressed using canonical problem families (e.g., MaxCut, MIS, TSP), rather than domain-specific formulations such as scheduling cost or resource utilisation that developers may naturally employ. While this choice enables controlled evaluation, reproducibility, and direct compatibility with quantum algorithms, it may not fully capture the diversity of real-world, domain-level objectives. This threat is partially mitigated by the support for classical code inputs, through which domain-specific objectives can be expressed implicitly via program logic (e.g., constraints or cost aggregation) without explicitly naming a canonical problem. By contrast, structured JSON inputs require objectives to be specified at the level of canonical problem families. Accordingly, the results should be interpreted as demonstrating feasibility at the level of problem-family abstraction rather than end-to-end domain-specific modelling. Nevertheless, the modular encoder design allows future extensions to introduce lightweight mapping layers that translate domain-level objectives into supported canonical representations without modifying the core pipeline.

Another threat concerns \textit{access to hardware}: cloud devices are scarce, heavily queued, and—on some platforms—expensive, which restricts broad real-device evaluation. Accordingly, our evaluation adopted a pragmatic strategy: we relied on simulators for scale, control, and reproducibility, and executed anchor running on Finland’s \textit{Helmi} and IBM’s \textit{ibm\_brisbane} to demonstrate physical executability despite multi-hour queues and trapped-ion pricing. This strategy provides a balanced evaluation: simulators verify correctness under controlled conditions, while targeted hardware runs confirm that the pipeline’s outputs are realizable on real quantum devices.

Finally, there is an issue of \textit{instance scale}. The results of Experiment~3 (see Section~\ref{ex:ex3}) demonstrate that error rates increase significantly with the increase of problem size on current quantum devices. In particular, superconducting qubit systems exhibit error rates exceeding 50\% once the required qubit count surpasses 12, largely due to accumulated gate errors in deeper circuits. This limitation reflects the state of NISQ hardware rather than the \textit{C2$\ket{\rm{Q}}$} framework itself and is consistent with prior studies showing that total error probability increases significantly with circuit depth and qubit count on current superconducting devices~\cite{aseguinolaza2024error}. As devices mature and coherence improves, the same workflow can naturally extend to larger instances without changes to the framework design.

\subsection{Internal Validity}
Internal validity refers to the degree to which factors within the study design or implementation may have influenced the results. The following are the main threats identified in our framework: A first threat lies in the \textit{rule-based data extraction}, which limits transferability of the extraction mechanism to more diverse or less structured code inputs.  
The AST-based extraction proved effective on well-structured Python code but was highly sensitive to syntactic variation, non-standard control flow, and diverse programming styles. This brittleness reduces transferability and may introduce vulnerabilities when the framework is applied to code outside the tested benchmarks. Nevertheless, adopting an explicit AST-rule strategy was an intentional design choice: it ensured transparency of the extraction process, facilitated reproducibility of our evaluation, and provided fine-grained control over how classical constructs were mapped to quantum-compatible formats (QCFs). In this sense, while the approach restricts generalisation, it offers clarity and interpretability that are essential for early-stage prototyping and for systematically studying the impact of parsing decisions on end-to-end workflow reliability.

A second concern is the use of \textit{fixed algorithm parameters}.  
The current implementation relies on default configuration settings for integrated algorithms such as QAOA, VQE, and Grover’s search. This constraint may reduce adaptability to diverse problem instances, as parameter choices (e.g., QAOA depth) strongly influence performance. However, using fixed defaults was a pragmatic choice to ensure consistency across experiments, support comparability of results, and reduce confounding factors in a proof-of-concept evaluation. While this limits performance optimisation, it provides a controlled setting to validate the correctness of the workflow. More adaptive parameterisation can be incorporated once baseline reliability of the full pipeline is established.

A third threat to internal validity arises from the \textit{Qiskit-centric design}. The framework currently relies on Qiskit for circuit construction and transpilation within a unified \textit{C2$\ket{\rm{Q}}$} pipeline. This design choice limits portability, as not all quantum platforms natively support Qiskit (e.g., Google’s Sycamore relies on Cirq). Nevertheless, Qiskit remains a widely adopted open-source quantum SDK with mature tooling, extensive community support, stable intermediate circuit representations, and broad support for quantum hardware backends through a unified transpilation and execution interface. In this work, these features make Qiskit a practical choice for prototyping and for demonstrating the feasibility of the proposed modular workflow, in which problem interpretation, algorithm selection, circuit generation, deployment, and result decoding are coordinated by the proposed framework. Although broader cross-platform coverage remains an avenue for future extension, the reliance on Qiskit provides a strong and widely compatible baseline for validating the framework’s system-level contributions.

\subsection{Construct Validity}
Construct validity concerns the extent to which the measures and evaluation methods capture the intended concepts. The following are the major construct validity threats: An important threat to construct validity is related to the \textit{bias from synthetic training data}. Because no large-scale, labelled corpora of quantum-relevant programs exist, the parser was trained on an LLM-generated dataset. This introduces risks of \emph{style homogeneity} and generator-specific artefacts that may reduce robustness on highly idiomatic human code. However, as described in Section~\ref{section:parser}, we mitigated these risks through several safeguards: prompts were designed to enforce coverage in the selected problem types, structural diversity, and identifier variation; syntactic validity was checked using Python’s \texttt{ast} module; near-duplicates were removed and classes were balanced; within-class structural and behavioural diversity of the generated programs was validated using software engineering metrics, algorithmic diversity, and structural instance diversity (Appendix~\ref{app:dataset_diversity}); and the parser was fine-tuned from a pre-trained model (CodeBERT) rather than trained from scratch. Moreover, the use of systematically validated synthetic corpora has been adopted in prior work on code and NLP tasks~\cite{wang2023selfinstructaligninglanguagemodels,taori2023stanford,xu2025kodcodediversechallengingverifiable,Nad__2025}, providing additional evidence of its methodological soundness. While future work should incorporate larger developer-authored datasets, the current setup ensures that our reported evaluation is grounded in widely accepted and carefully validated practices.

A second threat to construct validity concerns the \textit{size of input programs handled by the encoder}. The neural classification component is based on CodeBERT, which supports a maximum input length of 512 subword tokens produced by its tokenizer; inputs exceeding this limit are truncated during tokenisation. This design choice reflects the intended construct measured by the encoder, namely the identification of canonical problem formulations from concise problem-specification code rather than from large, monolithic program implementations. However, in practice, canonical problem structures such as graph constraints, objective definitions, or arithmetic formulations are typically expressed in short, self-contained code fragments that fall well within this token range. In our evaluation (Section~\ref{sec:evaluation}), all Python code snippets were observed to remain below the 512-token limit imposed by the CodeBERT tokenizer.

Another construct validity concern arises from the \textit{rule-based data extraction}. Although the AST-based extraction performed well on well-structured Python code, it proved brittle with dynamic constructs such as file I/O, helper functions, or object wrappers. This creates a gap between syntactic correctness and intended semantics, limiting construct validity. A promising direction is to augment static AST rules with semantic or dynamic analysis, or to adopt hybrid learning-based approaches, thereby improving robustness across diverse coding practices.

An additional construct validity threat relates to the \textit{manual identification and counting of configuration decisions used as a proxy for implementation workload}, as conducted in Experiment~3 (see Section~\ref{ex:ex3}). Although the process followed predefined criteria and involved independent review and arbitration among multiple authors, manual judgement introduces residual bias and potential interpretation differences. This can affect the exact counting of configuration decisions and therefore influence the precision of the reported workload comparison. To mitigate this risk, configuration decisions were defined at a consistent abstraction level and independently verified.

\subsection{Conclusion Validity}
Conclusion validity refers to the extent to which the study’s conclusions are supported by the collected evidence. In this study, the main concern for conclusion validity lies in the \textit{scope of evidence supporting our results}.  
Most evaluations were conducted on synthetic and small-to-medium benchmark instances, and many findings on hardware performance and recommender behaviour rely on simulator runs and provider-reported calibration data (e.g., error rates, gate times, coherence metrics). These choices reflect the current state of quantum hardware, where devices are constrained by limited qubit counts, restricted connectivity, gate errors, long queue delays, and high execution costs, making systematic large-scale experiments infeasible. Nevertheless, our evaluation also included representative executions on IBM’s \textit{ibm\_brisbane} and Finland’s \textit{Helmi} processor, which grounded the results in real-device behaviour beyond purely simulated tests. Taken together, these findings indicate that while broader hardware validation awaits more mature and accessible platforms, our conclusions are supported by a combination of simulated scalability studies, calibration-based performance modelling, and live hardware experiments.
\section{Conclusions and Future Work} \label{sec:conclusions}

This article has presented the design, implementation, and evaluation of \textit{C2$\ket{\rm{Q}}$}, a framework that automates the entire workflow from classical problem specification to quantum execution and result interpretation. The framework lowers the barrier for classical developers by abstracting away quantum-specific details, enabling them to obtain executable quantum programs from familiar inputs such as Python code or JSON specifications. The framework is publicly available as an open-source implementation under the Apache-2.0 license at \url{https://github.com/C2-Q/C2Q}. The repository also provides detailed instructions for reproducing the evaluation reported in this study. The latest release is also available as a Python package via PyPI at \url{https://pypi.org/project/c2q-framework/}, while installation and usage instructions are provided in the same repository.

The contributions of this work are threefold.  
First, we introduced a \textit{modular architecture} consisting of \textit{encoder}, \textit{deployment}, and \textit{decoder} modules. 
The encoder reliably classifies and translates diverse inputs into quantum-compatible formats (QCFs), achieving a workflow completion rate of 93.8\% on Python inputs and 100\% on structured JSON inputs. The deployment stage supports algorithm instantiation and device recommendation, showing that device selection can be automated by balancing error, runtime, and cost trade-offs across heterogeneous hardware. The decoder completes the workflow by mapping measurement outcomes into human-readable solutions, 
thus ensuring that outputs remain accessible to non-experts. Second, we provided a \textit{systematic evaluation} across three experiments: 
Experiment~1 validated the encoder through parser accuracy and QCF translation; Experiment~2 tested the deployment stage using QAOA circuits on Max-Cut; and Experiment~3 assessed the full workflow on 434 Python programs and 100 JSON inputs, executed on simulators and selectively on IBM and Helmi quantum hardware. The evaluation confirmed that the framework generalises across input modalities and problem domains, achieves workflow completion of \textbf{93.8\%} (Python) and \textbf{100\%} (JSON), and substantially reduces handwritten lines of code and explicit configuration steps compared to conventional quantum SDK implementations, as indicated by proxy-based workload measures. Third, we demonstrated that a \textit{principled engineering approach}—combining classical specification parsing, QCF abstractions, algorithm-device co-design, and cost modelling—can yield a full-stack quantum software workflow that is reproducible, extensible, and directly usable by classical software engineers. In summary, \textit{C2$\ket{\rm{Q}}$} demonstrates that a modular encoder–deployment–decoder pipeline can provide a practical bridge between classical software engineering and quantum computing. While current limitations of hardware access, training data, and supported problem scope constrain the applicability of our framework, the evaluation results and open-source implementation lay a foundation for further advances in quantum software engineering.

Looking forward, several directions remain open.  
First, the \textit{parser and QCF translator} can be extended to cover a broader set of canonical problem families and encodings, enabling the seamless integration of additional algorithms as the quantum software landscape evolves. In particular, future extensions will introduce mapping layers that translate application-domain optimisation objectives, such as scheduling costs or resource-utilisation objectives, into the canonical problem-family representations supported by the framework. In addition, the modular design allows new problem families to be incorporated in order to support emerging application domains with strong quantum potential, for example, materials- or chemistry-inspired optimisation problems, without modifying the core encoder-deployment-decoder pipeline. Second, the \textit{evaluation data} relied heavily on synthetic code generated with LLMs, which was useful for proof-of-concept but risks style bias. As the framework matures and is deployed in practical settings, where real-world feedback in the form of classical code from users becomes available, future work will incorporate developer-authored programs, mix real and synthetic data through semi-supervised learning, and expand support to additional languages such as C++. In the longer term, we plan to evolve our existing modular architecture, which is currently divided into separate encoder, generator, deployment, and decoder components, into a more intelligent and adaptive multi-agent system. This would build on the current separation of concerns by incorporating LLM-powered agents that can collaborate to improve generalisation across problem domains and input specifications, reduce brittleness of rule-based parsing in edge cases, and better handle diverse or unconventional input styles. While our rule-based parser remains valuable for transparency and fine-grained control in early-stage development, LLM-driven agents will provide greater flexibility in interpreting varied inputs and scalability for supporting new problem types and programming patterns as the system evolves.  
Third, the \textit{deployment stage} can be broadened. While the current framework supports major providers (IBM, Quantinuum, IonQ, Rigetti, and IQM), benchmarking was mainly conducted on simulators and calibration data due to cost and access barriers of real hardware. As access improves, systematic benchmarking across devices will allow the recommender to be validated more broadly and optimised beyond the default scoring function.



\appendix
\section*{APPENDICES}
\addcontentsline{toc}{section}{APPENDICES}

\setcounter{section}{0}
\renewcommand{\thesection}{\Alph{section}}
\section{Within-Class Dataset Diversity Analysis}
\label{app:dataset_diversity}

This appendix provides a multi-level quantitative characterisation of the diversity of the synthetic training dataset, complementing the dataset construction and validation described in Section~\ref{section:parser}. We analyse diversity at three complementary levels: implementation-level variation, algorithmic variation, and structural instance variation. The full implementation of this analysis and the corresponding results are provided in the replication package~\cite{ye_khan_c2q_dataset_2025}.

\subsection{Implementation-Level Diversity}
We refer to \emph{implementation-level diversity} as variation in how a canonical problem is expressed at the source-code level, independent of the underlying algorithmic strategy. We first compute static software metrics commonly used in empirical software engineering to quantify implementation-level diversity within each problem class
~\cite{ray2014large,nagappan2006mining}. For each problem type, we report the number of programs, lines of code (minimum, maximum, and mean), comment density (measured as the mean number of comment lines per program), control-flow construct usage (\texttt{for}, \texttt{while}, recursion), standard data structures (lists, dictionaries, sets, tuples), library usage (NetworkX), and mean cyclomatic complexity. These indicators capture variation in program size, structural logic, and implementation style.

Table~\ref{tab:dataset_diversity} summarises the resulting distributions. Across problem types, programs differ noticeably in size (e.g., KColor ranging from 10 to 44 LOC, TSP from 11 to 37 LOC), control-flow structure (e.g., MIS combining \texttt{for}, \texttt{while}, and recursion, whereas MaxCut relies almost exclusively on iterative loops), and structural complexity (e.g., CC$_{\mu}$ ranging from approximately 1.2 in arithmetic tasks to over 5.2 in MaxCut). These variations indicate that, within each canonical problem type, the dataset contains multiple implementation styles rather than near-duplicate solutions, providing meaningful diversity for evaluating learning-based program analysis.
\begin{table}[h!]
\centering
\caption{
Diversity statistics across problem types in the synthetic dataset. LOC$_{\min}$, LOC$_{\max}$, LOC$_{\mu}$: minimum, maximum, and mean lines of code. Comm$_{\mu}$: mean number of comment lines per program. NX: use of NetworkX. CC$_{\mu}$: mean cyclomatic complexity.
}
\label{tab:dataset_diversity}
\small
\renewcommand{\arraystretch}{1.15}
\begin{tabular}{lrrrrrrrrrrrrrrr}
\toprule
Family & \# & LOC$_{\min}$ & LOC$_{\max}$ & LOC$_{\mu}$ & Comm$_{\mu}$ &
\texttt{for} & \texttt{while} & Rec &
NX & List & Dict & Set & Tuple & CC$_{\mu}$ \\
\midrule
ADD     & 28 & 4  & 14 & 6.75 & 0.89 & 1  & 2  & 2  & 0 & 2  & 0 & 0 & 26 & 1.25 \\
SUB     & 28 & 4  & 13 & 6.68 & 0.96 & 1  & 2  & 2  & 0 & 2  & 0 & 0 & 25 & 1.29 \\
MUL     & 27 & 6  & 16 & 7.26 & 0.96 & 2  & 2  & 0  & 0 & 1  & 0 & 0 & 27 & 1.41 \\
Factor  & 31 & 10 & 31 & 15.61& 0.97 & 4  & 26 & 5  & 0 & 29 & 1 & 0 & 7  & 3.16 \\
MIS     & 70 & 6  & 21 & 12.36& 0.46 & 45 & 16 & 10 & 33& 70 & 0 & 2 & 70 & 3.00 \\
MaxCut  & 68 & 7  & 26 & 13.01& 0.57 & 54 & 4  & 0  & 27& 59 & 1 & 1 & 68 & 5.22 \\
TSP     & 44 & 11 & 37 & 17.41 & 0.59 & 14 & 31 & 0  & 7 & 43 & 2 & 0 & 43 & 2.59 \\
Clique  & 33 & 6  & 16 & 10.58& 0.39 & 26 & 0  & 7  & 6 & 29 & 0 & 0 & 33 & 4.03 \\
KColor  & 58 & 10 & 44 & 17.50& 0.53 & 48 & 9  & 16 & 11& 58 &10 & 0 & 52 & 4.22 \\
VC      & 47 & 11 & 35 & 13.30& 1.17 & 43 & 6  & 0  & 1 & 47 & 0 & 0 & 47 & 4.04 \\
\bottomrule
\end{tabular}
\end{table}
\subsection{Algorithmic and Structural Diversity}

Beyond implementation-level software metrics, we additionally assess diversity using algorithmic and structural measures designed to capture deeper variation in problem-solving strategies and instance structure.

Algorithmic diversity is measured via static analysis of each solution, implemented through AST-based feature extraction and rule-based classification into coarse-grained algorithm families (e.g., dynamic programming, brute-force enumeration, greedy heuristics, library-based approaches). For each problem type, we compute: (i) the number of detected algorithm families, (ii) Shannon entropy computed over the empirical algorithm-family distribution~\cite{shannon1948mathematical}, and (iii) the dominance ratio of the most frequent family.

Structural diversity is evaluated for graph-based problems using exact edge-set fingerprints (i.e., hashing sorted edge lists to detect identical graphs) and Weisfeiler--Lehman (WL) graph hashing~\cite{shervashidze2011weisfeiler}, with the implementation based on NetworkX~\cite{hagberg2008exploring}, providing a practical and scalable approximation for distinguishing exact duplicates and structurally distinct graph instances among generated instances. For non-graph tasks (ADD, SUB, MUL, Factor), Exact Unique is computed by hashing the extracted instance data (e.g., operand pairs for arithmetic and the integer to be factorised) to detect duplicate inputs.

Table~\ref{tab:diversity_overview} summarises the resulting algorithmic and structural diversity statistics. Graph-based combinatorial tasks such as Clique, MIS, MaxCut, and KColor exhibit moderate-to-high algorithmic diversity (entropy typically above 1.6, with MIS slightly lower at 1.35), with no single algorithm family exceeding approximately 55\% dominance. In contrast, TSP and VC are more algorithmically concentrated, reflecting the prevalence of dynamic programming and 2-approximation implementations, respectively. Arithmetic tasks (ADD and SUB) intentionally exhibit limited algorithmic variation, with diversity instead manifested through operand distributions and implementation-level differences. Regarding structural diversity, Exact Unique reports the fraction of instances with distinct extracted data under an exact fingerprint (e.g., identical edge sets counted once), whereas WL Unique reports distinct Weisfeiler–Lehman (WL) hashes as a scalable approximation of structural non-isomorphism. For graph-based problems, Exact Unique ratios range from 0.14 to 0.36 (e.g., 0.30 for Clique and 0.31 for MaxCut), while WL Unique ratios range from 0.14 to 0.22 (e.g., 0.18 for TSP and 0.22 for MaxCut). As expected, WL uniqueness ratios are lower because structurally equivalent graphs (e.g., differing only by node relabelling) collapse into the same WL class. The moderate WL uniqueness values are typical for bounded small-graph regimes (3--8 nodes), where the number of non-isomorphic structures is inherently limited.

\begin{table}[h!]
\centering
\caption{
Algorithmic and structural diversity across problem domains.
Count: number of syntactically valid programs. Exact Unique and WL Unique are computed over the subset of programs for which input instance data could be extracted (thus the denominator may be smaller than Count for some families).
\#Alg.\ Families: number of detected algorithm families.
Entropy: Shannon entropy over the family distribution.
Top Ratio: proportion of the most frequent algorithm family.
Exact Unique: ratio of structurally unique instances (exact fingerprint).
WL Unique: ratio of unique instances under Weisfeiler--Lehman (WL) graph hashing.
}
\label{tab:diversity_overview}
\small
\renewcommand{\arraystretch}{1.15}
\begin{tabular}{lrrrrrr}
\toprule
Family & Count & \#Alg.\ Families & Entropy & Top Ratio & Exact Unique & WL Unique \\
\midrule
ADD     & 28 & 2 & 0.37 & 0.93 & 1.00 & -- \\
SUB     & 28 & 2 & 0.22 & 0.96 & 0.96 & -- \\
MUL     & 27 & 3 & 0.73 & 0.85 & 0.84 & -- \\
Factor  & 31 & 3 & 1.46 & 0.48 & 0.81 & -- \\
MIS     & 70 & 4 & 1.35 & 0.47 & 0.27 & 0.14 \\
MaxCut  & 68 & 5 & 1.62 & 0.46 & 0.31 & 0.22 \\
TSP     & 44 & 3 & 0.78 & 0.82 & 0.25 & 0.18 \\
Clique  & 33 & 5 & 1.83 & 0.55 & 0.30 & 0.18 \\
KColor  & 58 & 5 & 1.90 & 0.50 & 0.14 & 0.14 \\
VC      & 47 & 7 & 1.31 & 0.77 & 0.36 & 0.21 \\
\bottomrule
\end{tabular}
\end{table}

\section{Programmatic Usage and Workflow Artefacts} \label{app:json}
This appendix provides representative programmatic usage examples and supporting workflow artefacts for the \textit{C2$\ket{\rm{Q}}$} framework. The materials complement the experimental discussion in Section~\ref{sec:evaluation} and are intended for reference.

Listings~\ref{lst:mis_helmi} and~\ref{lst:json_cli} illustrate the two supported programmatic usage modes of \textit{C2$\ket{\rm{Q}}$}: (i) an end-to-end workflow driven by a classical Python problem specification, and (ii) invocation via a command-line interface using a structured JSON specification. The first three lines in Listing~\ref{lst:mis_helmi} configure optional backend connections for executing the generated circuits on IQM’s \textit{Helmi} and IBM’s \textit{ibm\_brisbane} backends and are excluded from the line-count comparison in Section~\ref{ex:ex3}, which focuses on framework-level usage.

\begin{lstlisting}[language=Python, caption={Solving the Maximum Independent Set problem using the Helmi and IBM Brisbane quantum backends.}, label={lst:mis_helmi}]
# Representative MIS case study executed on Helmi and IBM Brisbane backends

# Set up quantum backends
helmi = IQMProvider(os.getenv("HELMI_CORTEX_URL")).get_backend()
service = QiskitRuntimeService()  # assumes account already saved
brisbane = service.backend("ibm_brisbane")

# Full C2|Q> pipeline for the MIS problem
problem_type, data = C2Q.parser.parse(is_snippet)
problem_instance = PROBLEMS[problem_type](data)

# Execute circuits and generate reports
problem_instance.report()
\end{lstlisting}

\begin{lstlisting}[
  language=bash,
  caption={Invoking \textit{C2$\ket{\rm Q}$} via the command-line interface using a structured JSON specification.},
  label={lst:json_cli}
]
# Run the C2|Q> pipeline from a JSON DSL specification
python3 -m json_engine \
    --input problem_input.json
\end{lstlisting}

Listings~\ref{lst:mis_classical} and~\ref{lst:json_mis} show the classical Python and JSON specification inputs used in Experiment~3 to initialise the MIS case study. Representative Python and JSON input examples for all supported canonical problem families, together with their corresponding generated artefacts, are provided in the replication package~\cite{ye_khan_c2q_dataset_2025}.

\begin{figure}[htbp]
  \centering
  \begin{minipage}[t]{0.53\textwidth}
    \begin{lstlisting}[language=Python, caption={Classical Python code for the Maximum Independent Set (MIS) task, used as \texttt{is\_snippet} in Listing~\ref{lst:mis_helmi}.}, label={lst:mis_classical}]
from itertools import combinations

def solve_graph_problem(n, edges):
    nodes = list(range(n))

    def is_independent(subset):
        return all(
            (u, v) not in edges and (v, u) not in edges
            for u, v in combinations(subset, 2)
        )

    best = set()
    for r in range(1, n + 1):
        for subset in combinations(nodes, r):
            if is_independent(subset):
                best = set(subset)
    return best

edges = [(0, 1), (0, 2), (1, 2), (1, 3)]
result = solve_graph_problem(4, edges)
print(result)  # Expected: {2, 3}
    \end{lstlisting}
  \end{minipage}
  \hfill
  \begin{minipage}[t]{0.43\textwidth}
    \begin{lstlisting}[language=json, caption={JSON input for the MIS task using an edge list, corresponding to \texttt{problem\_input.json} in Listing~\ref{lst:json_cli}.}, label={lst:json_mis}]
{
  "problem_type": "MIS",
  "goal": "find a maximum independent set",
  "instance": {
    "graph_rep": "edge_list",
    "graphs": {
      "G1": [
        [0, 1],
        [0, 2],
        [1, 2],
        [1, 3]
      ]
    }
  }
}
    \end{lstlisting}
  \end{minipage}
\end{figure}

\begin{figure}[htbp]
    \centering
    \begin{subfigure}{0.48\linewidth}
        \centering
        \begin{overpic}[width=\linewidth]{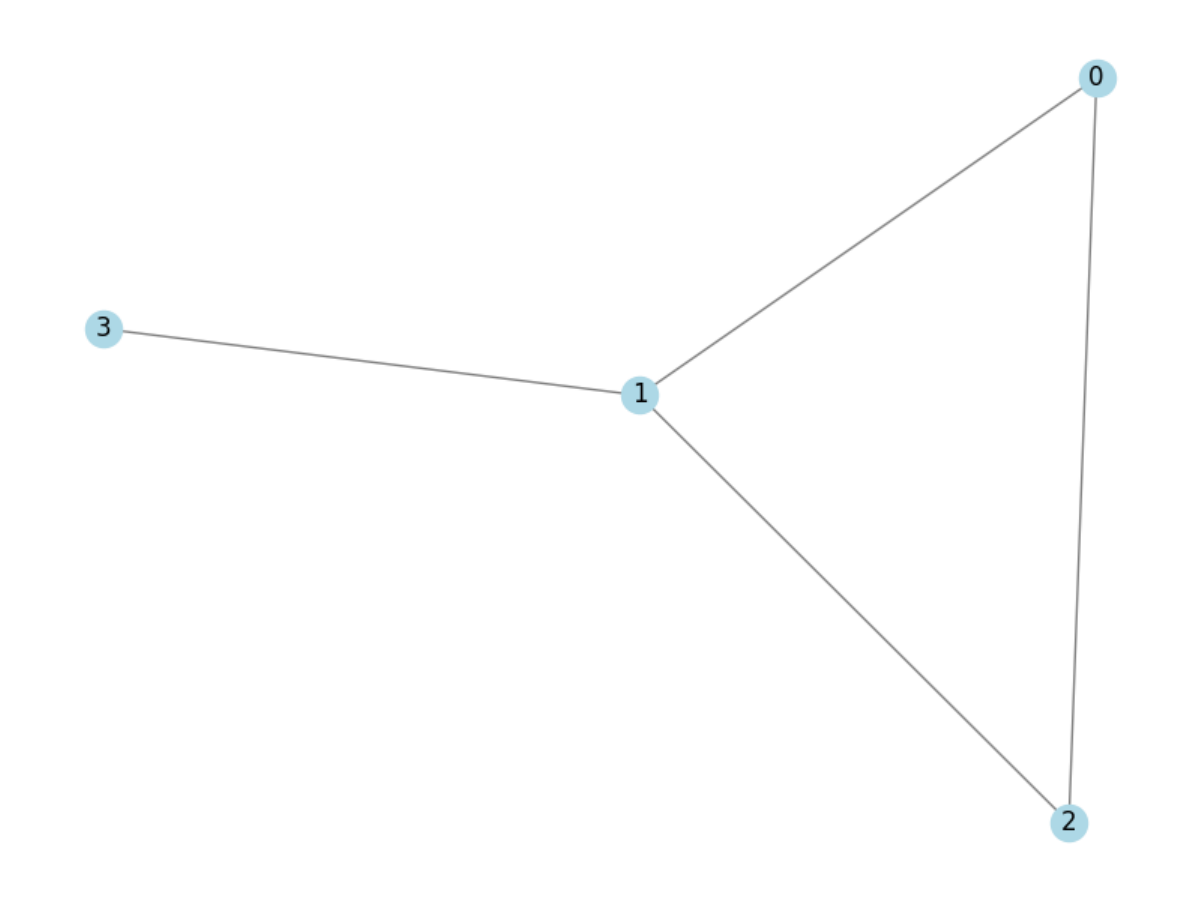}
            \put(5,65){\tiny \textbf{Nodes}: [0,1,2,3]}
            \put(5,60){\tiny \textbf{Edges}: [(0,1), (0,2), (1,2), (1,3)]}
        \end{overpic}
        \caption{Graph representation of the MIS problem extracted from input.}
        \label{fig:is_report_graph}
    \end{subfigure}
    \hfill
    \begin{subfigure}{0.48\linewidth}
        \centering
        \includegraphics[width=\linewidth]{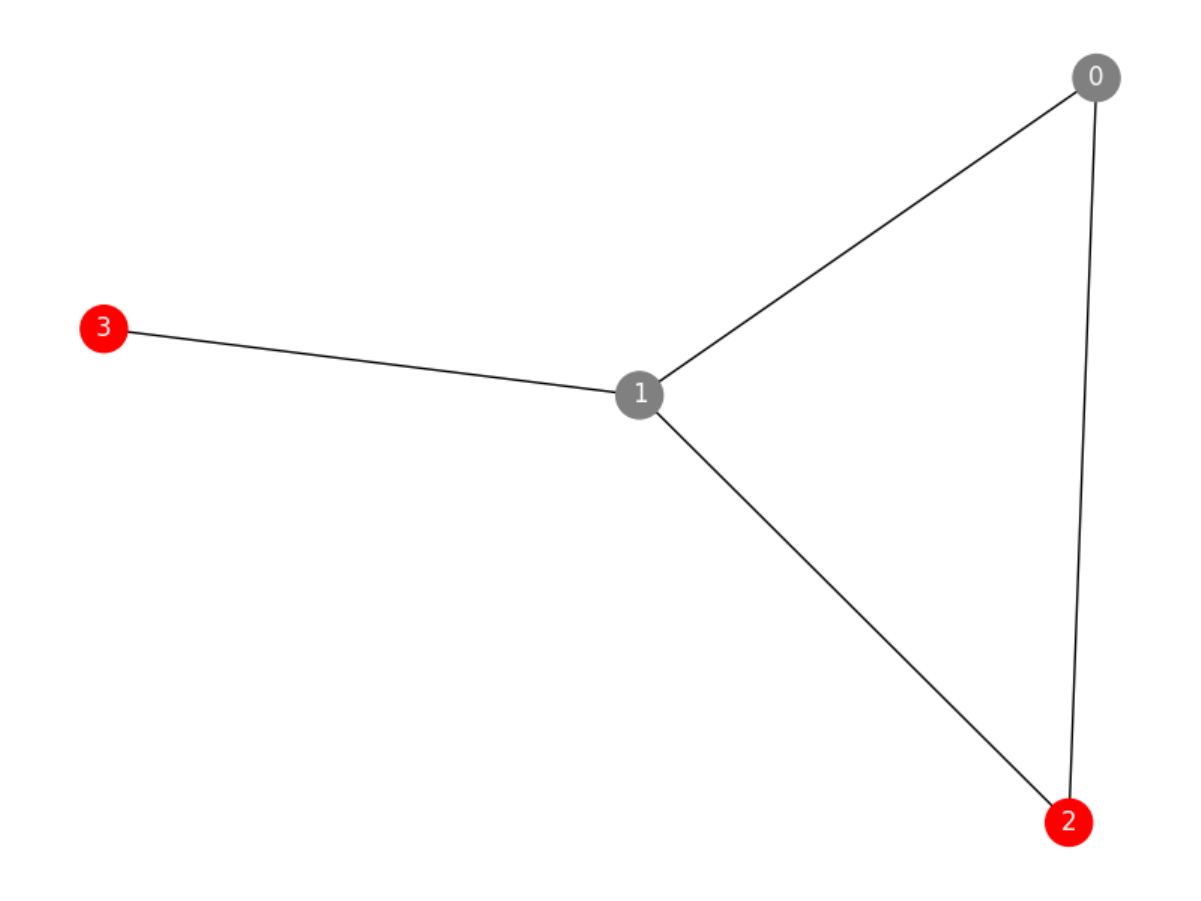}
        \caption{Measured output of QAOA execution. 
        The bitstring with highest probability (\texttt{1100}, little-endian) 
        corresponds to the state \(|1100\rangle\), indicating that vertices $\{2,3\}$ 
        form the independent set.}
        \label{fig:qaoa_result}
\end{subfigure}
    
    \caption{End-to-end demonstration of the MIS case study: (a) input graph extracted by the encoder; (b) execution result from QAOA, showing the selected independent set.}
    \label{fig:mis_case}
\end{figure}

Figures~\ref{fig:mis_case}--\ref{fig:bitstring_histogram} present visual artefacts from the same MIS case study, derived from the classical Python and JSON inputs shown in Listings~\ref{lst:mis_classical} and~\ref{lst:json_mis}. These artefacts include the extracted problem structure, quantum-compatible formats (QCFs), generated circuits, backend execution results, and decoded measurement outcomes. Together, they support the end-to-end workflow described in Section~\ref{ex:ex3} and are referenced there as needed.



\begin{figure}[htbp]
    \centering
    \scalebox{0.34}{
\Qcircuit @C=1.0em @R=0.2em @!R { \\
	 	\nghost{{q}_{0} :  } & \lstick{{q}_{0} :  } \barrier[0em]{10} & \qw & \qw & \qw \barrier[0em]{10} & \qw & \qw & \qw & \qw & \qw & \qw & \qw & \qw \barrier[0em]{10} & \qw & \qw & \qw & \qw & \qw & \qw & \qw & \qw \barrier[0em]{10} & \qw & \qw & \qw & \qw & \qw & \qw & \qw & \qw \barrier[0em]{10} & \qw & \qw & \qw & \qw & \qw & \qw & \qw & \qw \barrier[0em]{10} & \qw & \qw & \qw & \qw & \qw & \qw & \qw \barrier[0em]{10} & \qw & \qw & \qw & \qw & \qw & \qw & \qw \barrier[0em]{10} & \qw & \qw & \qw & \qw & \qw & \qw & \qw \barrier[0em]{10} & \qw & \qw & \qw & \qw & \qw & \qw & \qw \barrier[0em]{10} & \qw & \qw & \qw & \qw & \qw \barrier[0em]{10} & \qw & \qw & \qw\\
	 	\nghost{{q}_{1} :  } & \lstick{{q}_{1} :  } & \qw & \qw & \qw & \qw & \gate{\mathrm{X}} & \multigate{8}{\mathrm{or}}_<<<{0} & \gate{\mathrm{X}} & \qw & \qw & \qw & \qw & \qw & \gate{\mathrm{X}} & \multigate{8}{\mathrm{or}}_<<<{0} & \gate{\mathrm{X}} & \qw & \qw & \qw & \qw & \qw & \qw & \qw & \qw & \qw & \qw & \qw & \qw & \qw & \qw & \qw & \qw & \qw & \qw & \qw & \qw & \qw & \multigate{8}{\mathrm{or}}_<<<{2} & \qw & \qw & \qw & \qw & \qw & \qw & \multigate{8}{\mathrm{or}}_<<<{0} & \qw & \qw & \qw & \qw & \qw & \qw & \multigate{8}{\mathrm{or}}_<<<{0} & \qw & \qw & \qw & \qw & \qw & \qw & \qw & \qw & \qw & \qw & \qw & \qw & \qw & \qw & \qw & \qw & \qw & \qw & \qw & \qw\\
	 	\nghost{{q}_{2} :  } & \lstick{{q}_{2} :  } & \qw & \qw & \qw & \qw & \gate{\mathrm{X}} & \ghost{\mathrm{or}}_<<<{1} & \gate{\mathrm{X}} & \qw & \qw & \qw & \qw & \qw & \qw & \ghost{\mathrm{or}} & \qw & \qw & \qw & \qw & \qw & \qw & \gate{\mathrm{X}} & \multigate{7}{\mathrm{or}}_<<<{0} & \gate{\mathrm{X}} & \qw & \qw & \qw & \qw & \qw & \gate{\mathrm{X}} & \multigate{7}{\mathrm{or}}_<<<{0} & \gate{\mathrm{X}} & \qw & \qw & \qw & \qw & \qw & \ghost{\mathrm{or}}_<<<{0} & \qw & \qw & \qw & \qw & \qw & \qw & \ghost{\mathrm{or}}_<<<{3} & \qw & \qw & \qw & \qw & \qw & \qw & \ghost{\mathrm{or}}_<<<{1} & \qw & \qw & \qw & \qw & \qw & \qw & \multigate{7}{\mathrm{or}}_<<<{0} & \qw & \qw & \qw & \qw & \qw & \qw & \qw & \qw & \qw & \qw & \qw & \qw & \qw\\
	 	\nghost{{q}_{3} :  } & \lstick{{q}_{3} :  } & \qw & \qw & \qw & \qw & \qw & \ghost{\mathrm{or}} & \qw & \qw & \qw & \qw & \qw & \qw & \gate{\mathrm{X}} & \ghost{\mathrm{or}}_<<<{1} & \gate{\mathrm{X}} & \qw & \qw & \qw & \qw & \qw & \gate{\mathrm{X}} & \ghost{\mathrm{or}}_<<<{1} & \gate{\mathrm{X}} & \qw & \qw & \qw & \qw & \qw & \qw & \ghost{\mathrm{or}} & \qw & \qw & \qw & \qw & \qw & \qw & \ghost{\mathrm{or}}_<<<{1} & \qw & \qw & \qw & \qw & \qw & \qw & \ghost{\mathrm{or}}_<<<{1} & \qw & \qw & \qw & \qw & \qw & \qw & \ghost{\mathrm{or}}_<<<{2} & \qw & \qw & \qw & \qw & \qw & \qw & \ghost{\mathrm{or}} & \qw & \qw & \qw & \qw & \qw & \qw & \qw & \qw & \qw & \qw & \qw & \qw & \qw\\
	 	\nghost{{q}_{4} :  } & \lstick{{q}_{4} :  } & \qw & \qw & \qw & \qw & \qw & \ghost{\mathrm{or}} & \qw & \qw & \qw & \qw & \qw & \qw & \qw & \ghost{\mathrm{or}} & \qw & \qw & \qw & \qw & \qw & \qw & \qw & \ghost{\mathrm{or}} & \qw & \qw & \qw & \qw & \qw & \qw & \gate{\mathrm{X}} & \ghost{\mathrm{or}}_<<<{1} & \gate{\mathrm{X}} & \qw & \qw & \qw & \qw & \qw & \ghost{\mathrm{or}} & \qw & \qw & \qw & \qw & \qw & \qw & \ghost{\mathrm{or}}_<<<{2} & \qw & \qw & \qw & \qw & \qw & \qw & \ghost{\mathrm{or}} & \qw & \qw & \qw & \qw & \qw & \qw & \ghost{\mathrm{or}}_<<<{1} & \qw & \qw & \qw & \qw & \qw & \qw & \qw & \qw & \qw & \qw & \qw & \qw & \qw\\
	 	\nghost{{q}_{5} :  } & \lstick{{q}_{5} :  } & \qw & \qw & \qw & \qw & \qw & \ghost{\mathrm{or}} & \ctrl{1} & \ctrl{1} & \ctrl{1} & \targ & \qw & \qw & \qw & \ghost{\mathrm{or}} & \ctrl{1} & \ctrl{1} & \ctrl{1} & \targ & \qw & \qw & \qw & \ghost{\mathrm{or}} & \ctrl{1} & \ctrl{1} & \ctrl{1} & \targ & \qw & \qw & \qw & \ghost{\mathrm{or}} & \ctrl{1} & \ctrl{1} & \ctrl{1} & \targ & \qw & \qw & \ghost{\mathrm{or}} & \ctrl{1} & \ctrl{1} & \ctrl{1} & \targ & \qw & \qw & \ghost{\mathrm{or}} & \ctrl{1} & \ctrl{1} & \ctrl{1} & \targ & \qw & \qw & \ghost{\mathrm{or}} & \ctrl{1} & \ctrl{1} & \ctrl{1} & \targ & \qw & \qw & \ghost{\mathrm{or}} & \ctrl{1} & \ctrl{1} & \ctrl{1} & \targ & \qw & \qw & \gate{\mathrm{\left|0\right\rangle}} & \qw & \qw & \qw & \qw & \qw & \qw\\
	 	\nghost{{q}_{6} :  } & \lstick{{q}_{6} :  } & \qw & \qw & \qw & \qw & \qw & \ghost{\mathrm{or}} & \ctrl{1} & \ctrl{1} & \targ & \qw & \qw & \qw & \qw & \ghost{\mathrm{or}} & \ctrl{1} & \ctrl{1} & \targ & \qw & \qw & \qw & \qw & \ghost{\mathrm{or}} & \ctrl{1} & \ctrl{1} & \targ & \qw & \qw & \qw & \qw & \ghost{\mathrm{or}} & \ctrl{1} & \ctrl{1} & \targ & \qw & \qw & \qw & \ghost{\mathrm{or}} & \ctrl{1} & \ctrl{1} & \targ & \qw & \qw & \qw & \ghost{\mathrm{or}} & \ctrl{1} & \ctrl{1} & \targ & \qw & \qw & \qw & \ghost{\mathrm{or}} & \ctrl{1} & \ctrl{1} & \targ & \qw & \qw & \qw & \ghost{\mathrm{or}} & \ctrl{1} & \ctrl{1} & \targ & \qw & \qw & \qw & \gate{\mathrm{\left|0\right\rangle}} & \qw & \qw & \qw & \qw & \qw & \qw\\
	 	\nghost{{q}_{7} :  } & \lstick{{q}_{7} :  } & \qw & \qw & \qw & \qw & \qw & \ghost{\mathrm{or}} & \ctrl{1} & \targ & \qw & \qw & \qw & \qw & \qw & \ghost{\mathrm{or}} & \ctrl{1} & \targ & \qw & \qw & \qw & \qw & \qw & \ghost{\mathrm{or}} & \ctrl{1} & \targ & \qw & \qw & \qw & \qw & \qw & \ghost{\mathrm{or}} & \ctrl{1} & \targ & \qw & \qw & \qw & \qw & \ghost{\mathrm{or}} & \ctrl{1} & \targ & \qw & \qw & \qw & \qw & \ghost{\mathrm{or}} & \ctrl{1} & \targ & \qw & \qw & \qw & \qw & \ghost{\mathrm{or}} & \ctrl{1} & \targ & \qw & \qw & \qw & \qw & \ghost{\mathrm{or}} & \ctrl{1} & \targ & \qw & \qw & \qw & \qw & \gate{\mathrm{\left|0\right\rangle}} & \qw & \qw & \qw & \qw & \qw & \qw\\
	 	\nghost{{q}_{8} :  } & \lstick{{q}_{8} :  } & \qw & \qw & \qw & \qw & \qw & \ghost{\mathrm{or}} & \targ & \qw & \qw & \qw & \qw & \qw & \qw & \ghost{\mathrm{or}} & \targ & \qw & \qw & \qw & \qw & \qw & \qw & \ghost{\mathrm{or}} & \targ & \qw & \qw & \qw & \qw & \qw & \qw & \ghost{\mathrm{or}} & \targ & \qw & \qw & \qw & \qw & \qw & \ghost{\mathrm{or}} & \targ & \qw & \qw & \qw & \qw & \qw & \ghost{\mathrm{or}} & \targ & \qw & \qw & \qw & \qw & \qw & \ghost{\mathrm{or}} & \targ & \qw & \qw & \qw & \qw & \qw & \ghost{\mathrm{or}} & \targ & \qw & \qw & \qw & \qw & \qw & \ctrl{1} & \gate{\mathrm{\left|0\right\rangle}} & \qw & \qw & \qw & \qw & \qw\\
	 	\nghost{{q}_{9} :  } & \lstick{{q}_{9} :  } & \qw & \qw & \qw & \qw & \qw & \ghost{\mathrm{or}}_<<<{2} & \ctrl{-1} & \ctrl{-2} & \ctrl{-3} & \ctrl{-4} & \gate{\mathrm{\left|0\right\rangle}} & \qw & \qw & \ghost{\mathrm{or}}_<<<{2} & \ctrl{-1} & \ctrl{-2} & \ctrl{-3} & \ctrl{-4} & \gate{\mathrm{\left|0\right\rangle}} & \qw & \qw & \ghost{\mathrm{or}}_<<<{2} & \ctrl{-1} & \ctrl{-2} & \ctrl{-3} & \ctrl{-4} & \gate{\mathrm{\left|0\right\rangle}} & \qw & \qw & \ghost{\mathrm{or}}_<<<{2} & \ctrl{-1} & \ctrl{-2} & \ctrl{-3} & \ctrl{-4} & \gate{\mathrm{\left|0\right\rangle}} & \qw & \ghost{\mathrm{or}}_<<<{3} & \ctrl{-1} & \ctrl{-2} & \ctrl{-3} & \ctrl{-4} & \gate{\mathrm{\left|0\right\rangle}} & \qw & \ghost{\mathrm{or}}_<<<{4} & \ctrl{-1} & \ctrl{-2} & \ctrl{-3} & \ctrl{-4} & \gate{\mathrm{\left|0\right\rangle}} & \qw & \ghost{\mathrm{or}}_<<<{3} & \ctrl{-1} & \ctrl{-2} & \ctrl{-3} & \ctrl{-4} & \gate{\mathrm{\left|0\right\rangle}} & \qw & \ghost{\mathrm{or}}_<<<{2} & \ctrl{-1} & \ctrl{-2} & \ctrl{-3} & \ctrl{-4} & \gate{\mathrm{\left|0\right\rangle}} & \qw & \targ & \ctrl{1} & \gate{\mathrm{\left|0\right\rangle}} & \qw & \qw & \qw & \qw\\
	 	\nghost{{q}_{10} :  } & \lstick{{q}_{10} :  } & \qw & \gate{\mathrm{X}} & \gate{\mathrm{H}} & \qw & \qw & \qw & \qw & \qw & \qw & \qw & \qw & \qw & \qw & \qw & \qw & \qw & \qw & \qw & \qw & \qw & \qw & \qw & \qw & \qw & \qw & \qw & \qw & \qw & \qw & \qw & \qw & \qw & \qw & \qw & \qw & \qw & \qw & \qw & \qw & \qw & \qw & \qw & \qw & \qw & \qw & \qw & \qw & \qw & \qw & \qw & \qw & \qw & \qw & \qw & \qw & \qw & \qw & \qw & \qw & \qw & \qw & \qw & \qw & \qw & \qw & \targ & \gate{\mathrm{H}} & \gate{\mathrm{X}} & \qw & \qw & \qw\\
\\ }}
    \caption{Oracle circuit generated as a Quantum-Compatible Format (QCF) for the Maximum Independent Set (MIS) instance, encoding feasibility constraints for Grover’s algorithm.}
    \label{fig:qcf_oracle}
\end{figure}
\medskip

\begin{figure}[htbp]
\centering
\[
\begin{bmatrix}
-1.0 & 2.0 & 2.0 & 2.0 & 0.0 \\
0.0  & -1.0 & 2.0 & 2.0 & 0.0 \\
0.0  & 0.0  & -1.0 & 0.0 & 0.0 \\
0.0  & 0.0  & 0.0  & -1.0 & 0.0 \\
\end{bmatrix}
\]
\caption{The QUBO matrix generated as a QCF for the MIS instance and used as input to QAOA.}
\label{fig:qubo_matrix}
\end{figure}

\begin{figure}[htbp]
\centering
\scalebox{1}{
\Qcircuit @C=1.0em @R=0.2em @!R {
	\lstick{q_0} & \gate{H} & \multigate{3}{U_C(\gamma_1)} & \gate{R_X(\beta_1)} &
	\multigate{3}{U_C(\gamma_2)} & \gate{R_X(\beta_2)} &
	\multigate{3}{U_C(\gamma_3)} & \gate{R_X(\beta_3)} & \meter \\
	\lstick{q_1} & \gate{H} & \ghost{U_C(\gamma_1)} & \gate{R_X(\beta_1)} &
	\ghost{U_C(\gamma_2)} & \gate{R_X(\beta_2)} &
	\ghost{U_C(\gamma_3)} & \gate{R_X(\beta_3)} & \meter \\
	\lstick{q_2} & \gate{H} & \ghost{U_C(\gamma_1)} & \gate{R_X(\beta_1)} &
	\ghost{U_C(\gamma_2)} & \gate{R_X(\beta_2)} &
	\ghost{U_C(\gamma_3)} & \gate{R_X(\beta_3)} & \meter \\
	\lstick{q_3} & \gate{H} & \ghost{U_C(\gamma_1)} & \gate{R_X(\beta_1)} &
	\ghost{U_C(\gamma_2)} & \gate{R_X(\beta_2)} &
	\ghost{U_C(\gamma_3)} & \gate{R_X(\beta_3)} & \meter
}
}
\caption{Illustrative QAOA circuit generated for the MIS instance with $p=3$ layers. Cost and mixer parameters are interleaved as $(\gamma_k, \beta_k)$. The circuit is shown at the logical level prior to backend-specific transpilation.}
\label{fig:qaoa_circuit_theta}
\end{figure}


\begin{figure}[htbp]
    \centering

    \begin{subfigure}{0.48\linewidth}
        \centering
        \begin{tikzpicture}
        \begin{axis}[
            ybar,
            bar width=0.4cm,
            enlarge x limits=0.08,
            ymin=0, ymax=100,
            ylabel={Error (\%)},
            xtick=data,
            xticklabel style={rotate=45, anchor=east},
            symbolic x coords={
                Quantinuum H1, Quantinuum H2, IonQ Aria (Azure), 
                IonQ Aria (Amazon), IQM Garnet,
                ibm\_sherbrooke, ibm\_brisbane, Rigetti Ankaa-9Q-3, ibm\_kyiv
            },
            width=\linewidth,
            height=5cm,
            nodes near coords,
            every node near coord/.append style={font=\tiny, anchor=south}
        ]
        \addplot+[ybar, draw=black, fill=green!50!white] 
        coordinates {
            (Quantinuum H1, 3.90)
            (Quantinuum H2, 4.66)
            (IonQ Aria (Azure), 25.02)
            (IonQ Aria (Amazon), 25.02)
            (IQM Garnet, 31.58)
            (ibm\_sherbrooke, 32.31)
            (ibm\_brisbane, 32.67)
            (Rigetti Ankaa-9Q-3, 50.68)
            (ibm\_kyiv, 40.88)
        };
        \end{axis}
        \end{tikzpicture}
        \caption{Estimated error rates.}
        \label{fig:errors_qaoa}
    \end{subfigure}
    \hfill
    \begin{subfigure}{0.48\linewidth}
        \centering
        \begin{tikzpicture}
        \begin{axis}[
            ybar,
            bar width=0.4cm,
            enlarge x limits=0.08,
            ymin=0, ymax=1500, 
            ylabel={Execution Time (s)},
            xtick=data,
            xticklabel style={rotate=45, anchor=east},
            symbolic x coords={
                Quantinuum H1, Quantinuum H2, IonQ Aria (Azure), 
                IonQ Aria (Amazon), IQM Garnet,
                ibm\_sherbrooke, ibm\_brisbane, Rigetti Ankaa-9Q-3, ibm\_kyiv
            },
            width=\linewidth,
            height=5cm,
            nodes near coords,
            every node near coord/.append style={font=\tiny, anchor=south}
        ]
        \addplot+[ybar, draw=black, fill=blue!40!white] 
        coordinates {
            (Quantinuum H1, 712.95)
            (Quantinuum H2, 712.95)
            (IonQ Aria (Azure), 1327.5)
            (IonQ Aria (Amazon), 1327.5)
            (IQM Garnet, 0.183)
            (ibm\_sherbrooke, 1.331)
            (ibm\_brisbane, 1.590)
            (Rigetti Ankaa-9Q-3, 0.330)
            (ibm\_kyiv, 1.344)
        };
        \end{axis}
        \end{tikzpicture}
        \caption{Estimated execution times.}
        \label{fig:reco_time_single}
    \end{subfigure}

    \vspace{1em}
    \begin{subfigure}{0.7\linewidth}
    \centering
    \begin{tikzpicture}
    \begin{axis}[
        ybar,
        bar width=0.4cm,
        enlarge x limits=0.08,
        ymode=log, log basis y=10,
        ymin=1, ymax=1.5e5,
        ylabel={Cost (USD, log scale)},
        xtick=data,
        xticklabel style={rotate=45, anchor=east},
        symbolic x coords={
            Quantinuum H1, Quantinuum H2, IonQ Aria (Azure), 
            IonQ Aria (Amazon), IQM Garnet,
            ibm\_sherbrooke, ibm\_brisbane, Rigetti Ankaa-9Q-3, ibm\_kyiv
        },
        width=\linewidth,
        height=6cm,
        nodes near coords,
        every node near coord/.append style={
            font=\tiny, anchor=south, color=blue, yshift=1pt 
        }
    ]
    \addplot+[ybar, draw=black, fill=orange!50!white] 
    coordinates {
        (Quantinuum H1, 56625.0)
        (Quantinuum H2, 61155.0)
        (IonQ Aria (Azure), 4875.0)
        (IonQ Aria (Amazon), 1515.0)
        (IQM Garnet, 87.5)
        (ibm\_sherbrooke, 2.130)
        (ibm\_brisbane, 2.544)
        (Rigetti Ankaa-9Q-3, 0.429) 
        (ibm\_kyiv, 2.150)
    };
    \end{axis}
    \end{tikzpicture}
    \caption{Estimated execution costs (log scale).}
    \label{fig:rec_price_single}
\end{subfigure}

    \caption{Comparison of generated QAOA circuits across devices: 
(a) error rates, (b) execution times, and (c) execution costs. 
The execution costs (USD; with $\log_{10}$ values) are:
Quantinuum H1 = \$56{,}625 ($\log_{10}\!\approx 4.75$),
Quantinuum H2 = \$61{,}155 ($\log_{10}\!\approx 4.79$),
IonQ Aria (Azure) = \$4{,}875 ($\log_{10}\!\approx 3.69$),
IonQ Aria (Amazon) = \$1{,}515 ($\log_{10}\!\approx 3.18$),
IQM Garnet = \$87.50 ($\log_{10}\!\approx 1.94$),
ibm\_sherbrooke = \$2.13 ($\log_{10}\!\approx 0.33$),
ibm\_brisbane = \$2.54 ($\log_{10}\!\approx 0.41$),
ibm\_kyiv = \$2.15 ($\log_{10}\!\approx 0.33$),
and Rigetti Ankaa-9Q-3 = \$0.43 ($\log_{10}\!\approx -0.37$, omitted in (c) due to $y_{\min}=1$ on the log scale).}
    \label{fig:qaoa_comparison_app}
\end{figure}

\begin{figure}[h]
    \centering
    \begin{tikzpicture}
    \begin{axis}[
        ybar,
        bar width=8pt,
        width=0.9\textwidth,
        height=6cm,
        xlabel={Bitstring Outcome},
        ylabel={Counts},
        xtick=data,
        xticklabel style={rotate=45, anchor=east},
        symbolic x coords={
            0000, 0001, 0010, 0011, 0100, 0101, 0110, 0111,
            1000, 1001, 1010, 1011, 1100, 1101, 1110, 1111
        },
        ymin=0,
        enlarge x limits=0.02,
        nodes near coords,
        every node near coord/.append style={
            font=\small,
            anchor=south,
            yshift=2pt
        },
    ]
        \addplot+[fill=blue!30] coordinates {
            (0000,7) (0001,26) (0010,76) (0011,239)
            (0100,248) (0101,28) (0110,15) (0111,14)
            (1000,66) (1001,14) (1010,1) (1011,4)
            (1100,257) (1101,13) (1110,15) (1111,1)
        };
    \end{axis}
    \end{tikzpicture}
    \caption{Histogram of bitstring measurement outcomes over 1024 shots. Each bar shows the count of occurrences for a 4-qubit state.}
    \label{fig:bitstring_histogram}
\end{figure}

\FloatBarrier
\section{Quantum-Compatible Formats for Problem Encoding} \label{app:qcf_data}
We define \textit{Quantum-Compatible Formats (QCFs)} as intermediate representations that translate classical problem data into forms suitable for quantum circuit construction. In the \textit{C2$\ket{\rm{Q}}$} framework, the primary QCFs are \emph{Quadratic Unconstrained Binary Optimisation (QUBO)}, \emph{oracle-based encodings}, QCF-arithmetic encodings, and \emph{reversible-arithmetic encodings}. These formats serve as the bridge between classical specifications and executable quantum circuits (see Section~\ref{sec:framework}).

\subsection{Quadratic Unconstrained Binary Optimisation (QUBO) and Ising Model} \label{subapp:qubo}
The \textit{QUBO} model encodes combinatorial optimisation problems as the minimisation of a quadratic form over binary variables~\cite{glover2018tutorial}:
\begin{equation}
    \min x^T Q x = \sum_{i} Q_{ii} x_i + \sum_{i < j} Q_{ij} x_i x_j ,
\end{equation}
where \(x_i \in \{0, 1\}\). The upper-triangular matrix \(Q\) specifies linear coefficients (\(Q_{ii}\)) and pairwise couplings (\(Q_{ij}\)). QUBO is equivalent to the \textit{Ising model}~\cite{lucas2014} under the variable transformation \(s_i = 2x_i - 1\):
\begin{equation}
    H(s) = -\sum_{i < j} J_{ij} s_i s_j - \sum_{i} h_i s_i ,
\end{equation}
where \(s_i \in \{-1,+1\}\), \(J_{ij}\) denotes interaction strengths, and \(h_i\) denotes local fields. Mapping \(s_i \mapsto \sigma_z^i\) gives the corresponding quantum cost Hamiltonian:
\begin{equation}
    H_C = -\sum_{i < j} J_{ij} \sigma_z^i \sigma_z^j - \sum_{i} h_i \sigma_z^i ,
\end{equation}
which can be minimised by variational algorithms such as QAOA~\cite{farhi2014}.  

In the proposed \textit{C2$\ket{\rm{Q}}$} framework, QUBO is explicitly represented by the matrix \(Q\), which serves as the objective function for seamless integration with QAOA and VQE back ends.

\subsection{Oracle in the Quantum Query Model} \label{subapp:oracle}
A quantum \textit{oracle} is a black-box unitary operator \(O_f\) that marks solutions by flipping the phase of valid states, a central component in Grover’s search~\cite{grover1996}:
\begin{equation}
    O_f |x\rangle = (-1)^{f(x)} |x\rangle .
\end{equation}
In practice, such oracles are typically realised as quantum circuits parameterised by a Boolean function \(f(x)\). Within the proposed framework, oracle-based QCFs encode these functions into circuit representations, enabling Grover-style search for decision and combinatorial optimisation problems supported by \textit{C2$\ket{\rm{Q}}$}.

\subsection{Encoding the Maximum Independent Set (MIS)} \label{app:mis}
The Maximum Independent Set (MIS) problem is treated in its optimisation form, while oracle-based search internally relies on decision-style constraint encodings. Within the framework, MIS can be expressed in multiple Quantum-Compatible Formats (QCFs), enabling different algorithmic approaches within the framework. We support both oracle- and QUBO-based formulations.

The oracle encodes the \emph{independent set} and \emph{maximality} conditions of the input graph. For each edge \((u,v)\), we enforce the clause \(\lnot u \lor \lnot v\) to prevent adjacent vertices from both being selected. For each vertex \(v\), we ensure maximality by requiring \(v \lor v_1 \lor \dots \lor v_k\), where \(\{v_1,\dots,v_k\}\) are the neighbors of \(v\). The resulting Boolean formula \(\phi\) is checked during quantum execution via ancilla qubits, where each clause is mapped into a subcircuit using quantum equivalents of logical \textsc{AND}/\textsc{OR} operations. 
Ancilla qubits accumulate clause evaluations, and the oracle flips the phase only when all constraints are satisfied, following standard techniques for Boolean oracle construction~\cite{Barenco_1995}. This guarantees that measured solutions correspond to valid maximal independent sets. During results interpretation, we select the measurement outcome with the largest number of vertices, which provides an approximation of the MIS. This formulation directly supports Grover-style algorithms by embedding MIS constraints into the oracle unitary.

Alternatively, MIS can be encoded as a QUBO instance~\cite{lucas2014}. 
We define the matrix \(Q = Q_A + Q_B\), where:
\[
Q_A = A \sum_{(u,v)\in E} x_u x_v
\quad\quad
Q_B = -B \sum_{v} x_v .
\]
Here, \(x_v \in \{0,1\}\) indicates whether vertex \(v\) is selected. 
The penalty term \(Q_A\) prevents adjacent vertices from both being included, while the reward term \(Q_B\) maximises the size of the set. 
Following a standard practice~\cite{lucas2014}, we adopt \(A = 2B\) (here \(A=2.0, B=1.0\)), ensuring that adding a vertex is always beneficial when it does not violate independence constraints. This formulation maps naturally to the Ising Hamiltonian and is directly solvable with QAOA or VQE.

Together, these two QCF encodings illustrate the flexibility of the framework: the oracle supports search-style algorithms, while the QUBO encoding enables variational optimisation methods. Both can be generated automatically by the encoder, ensuring consistency between classical specifications and downstream circuit construction.

\section*{Acknowledgments}
The authors acknowledge the use of generative AI tools (ChatGPT) and a writing assistance tool (Grammarly) to improve the writing quality of this paper. Following the use of these tools, the authors thoroughly reviewed and revised the content, and they take full responsibility for the final version of the paper.

\section*{Funding}
This work has been supported by the following projects: \textit{Classi|Q⟩}, project number 24304728121, funded jointly by the City of Oulu and the University of Oulu, Finland; \textit{Securing the Quantum Software Stack (SeQuSoS)}, project number 24304955, funded by Business Finland; \textit{H2Future}, project number 352788, funded by the Research Council of Finland and the University of Oulu; and the National Natural Science Foundation of China (NSFC) under Grant No. 92582203.

\section*{Data Availability}
The code base of this work is publicly available via the C2\textbar Q\textgreater{} GitHub repository~\cite{C2Q_github_2025}, and the full replication package, including datasets and experimental results, is archived on Zenodo~\cite{ye_khan_c2q_dataset_2025}.

\bibliography{ref}{}
\bibliographystyle{ACM-Reference-Format}

\end{document}